\documentclass[aps,twocolumn,prd,superscriptaddress,floatfix,showpacs,showkeys]{revtex4-2}
\usepackage{newtxtext}
\usepackage[latin1]{inputenc}
\usepackage[T1]{fontenc}
\usepackage{lmodern}

\usepackage{empheq}
\usepackage{mathtools}
\usepackage{amsmath}

\makeatletter
\renewcommand{\p@subsubsection}{\thesection\,}
\makeatother

\usepackage{titlesec}

\titlespacing*{\subsubsection}{0pt}{2ex plus 1ex minus .2ex}{1ex plus .2ex}
\usepackage[titletoc]{appendix}

\usepackage{etoolbox}
\usepackage{amscd}
\usepackage{amsmath}
\usepackage{amssymb}
\usepackage{graphicx}\usepackage{scalerel}

\usepackage{dsfont}

\usepackage[ruled]{algorithm2e}
\usepackage{needspace}
\usepackage{newtxtext}
\usepackage{booktabs}
\usepackage{multirow}

\usepackage[colorlinks=true, pdfstartview=FitV, linkcolor=blue,
            citecolor=blue, urlcolor=blue]{hyperref} 
\usepackage{subcaption}
\usepackage{sidecap}
\usepackage{grffile}
\usepackage{xcolor}

\graphicspath{{./}}

\providecommand{\U}[1]{\protect\rule{.1in}{.1in}}
\newenvironment{proof}[1][Proof]{\noindent\textbf{#1.} }{\ \rule{0.5em}{0.5em}}

\definecolor{rred}{rgb}{0.7,0,0.1}
\definecolor{ccyan}{rgb}{0,.5,1}
\definecolor{greenrb}{rgb}{0.2,0.6,0.2}
\definecolor{faintpink}{rgb}{1.0, 0.9, 0.95}

\newcommand{\x}{\mathbf{x}}
\newcommand{\ub}{\mathbf{u}}

\def\kvec{\mathbf{k}}

\usepackage{mathrsfs}
\usepackage{bm}
\newcommand{\Action}{\mathscr{S}}

\def\be{\begin{equation}}
\def\ee{\end{equation}}
\def\bea{\begin{equation} \begin{aligned}}
\def\eea{\end{aligned} \end{equation}}
\def\beas{\begin{equation*} \begin{aligned}}
\def\eeas{\end{aligned} \end{equation*}}
\def\bes{\begin{equation*}}
\def\ees{\end{equation*}}

\def\d{\, \mathrm{d}}

\def\adots{
  \mathinner{\mkern1mu\raise1pt\hbox{.}\mkern2mu\raise4pt\hbox{.}
  \mkern2mu\raise7pt\vbox{\kern7pt\hbox{.}}\mkern1mu}}

\def\y{\boldsymbol{y}}

\newtheorem{thm}{Theorem}[section]
\newtheorem{lem}{Lemma}[section]
\newtheorem{defi}{Definition}[section]

\newtheorem{rem}{Remark}[section]
\newtheorem{cor}{Corollary}[section]
\def\bt{\begin{thm}}
\def\et{\end{thm}}
\def\bl{\begin{lem}}
\def\el{\end{lem}}
\def\bd{\begin{defi}}
\def\ed{\end{defi}}
\def\bc{\begin{cor}}
\def\ec{\end{cor}}
\def\bp{\begin{proof}}
\def\ep{\end{proof}}
\def\br{\begin{rem}}
\def\er{\end{rem}}
\def\bi{\begin{itemize}}
\def\ei{\end{itemize}}

\begin{document}

\title[]{The Nonlinear Physics and Statistical Field Theory of Twilight Clouds: Hovering Dynamics and Dynamic Inertia Emergence}

\author{Micka{\"e}l D. Chekroun}
\affiliation{Department of Earth and Planetary Sciences, Weizmann Institute, Rehovot 76100, Israel} 
\affiliation{Department of Atmospheric and Oceanic Sciences, University of California, Los Angeles, CA 90095-1565, USA}
\email{mchekroun@atmos.ucla.edu} 

\author{Ilan Koren}
\affiliation{Department of Earth and Planetary Sciences, Weizmann Institute, Rehovot 76100, Israel} 
\date{\today}

\begin{abstract}
Twilight clouds are small, optically faint warm clouds whose droplets fluctuate between haze and activated growth. We develop a nonequilibrium statistical field theory connecting this marginal activation regime to cloud-field organization, starting from K\"ohler microphysics, supersaturation transport, and turbulent mixing.

At parcel level, supersaturation rapidly tracks slow droplet growth with a finite dynamical lag. Mixing-induced supersaturation removal drives a saddle-node bifurcation, creating an edge state and shifting the activation threshold. The resulting geometry supports intermittent hovering between haze and activation under turbulent fluctuations, providing a mechanism for persistent transitional droplets in weak updrafts.

To connect this local mechanism to collective dynamics, we formulate a degenerate stochastic droplet-supersaturation reaction-diffusion system within the Martin-Siggia-Rose framework. 
A Gurari-type variational reduction drawn from polaron theory eliminates the fast supersaturation bath and reveals how it dresses the slow droplet-population field. Turbulence reshapes the activation landscape, shrinks the forbidden zone below the equilibrium K\"ohler curve, and raises the mixing threshold required to suppress activation. Each droplet population induces a screened supersaturation halo: its spatial footprint couples neighboring cloud elements, while its delayed adjustment generates dynamic inertia and temporally correlated noise.

The resulting nonlinear, non-Markovian pseudoparabolic stochastic equation retains a bounded frozen-background response at short wavelengths. Its long-wave coupling changes from smoothing to contrast amplification where the dressed slow manifold turns downward. Numerical comparisons with the parent two-field system show that a spatially correlated Markovian closure preserves evolving droplet-radius distributions and cloudlet organization. Near the activation fold, restoring residual-bath fluctuation memory improves recurrent-hovering frequency and patch extent without recalibrating the deterministic drift or fitting a correlation time.

These results connect microphysical transitions, collective spatial structure, and memory within a common analytical framework. More broadly, our reduction framework identifies how slow fields acquire interactions and inertia through the diffusive environments they modify, offering a route to interpretable closures for cloud physics and other environmentally mediated systems, including vegetation-water feedbacks and populations communicating through chemicals.
\end{abstract}



\maketitle

{\small 
\tableofcontents
}

\section{Introduction}

The coupling of cloud microphysics with climate dynamics remains one of the most challenging problems in atmospheric science. Failure to accurately parameterize clouds is responsible for the largest source of uncertainty in global climate projections \citep{zelinka2020}. Addressing this requires a more precise description of the nonlinear processes governing the onset and development of clouds, especially considering the vast spatial and temporal scale separation between global circulation and fine-scale turbulent motions.

Among all cloud types, the warm and smallest ones---often termed ``twilight clouds''---are particularly difficult to capture in climate models despite their abundance \citep{koren2007twilight,varnai2009modis,eytan2020,eytan2025net}. These clouds mark the transition from non-activating haze to fully activated cloud droplets. Their constituent particles reside in a precarious, marginal state highly sensitive to initial perturbations, with microphysical structures centered near the critical activation radius \citep{hirsch2014transition}. Complementary work has shown that these small clouds often form in buoyant humidity pockets, resulting in sub-Lifting Condensation Level (subLCL) clouds that oscillate between haze and cloud states \citep{hirsch2015properties,hirsch2017,altaratz2021environmental}.

Understanding the dynamics of droplet transitions in this marginal environment is the central goal of this study. To tackle this, we develop a theory that starts at the scale of an isolated moist parcel---where droplet growth is governed by K\"ohler thermodynamics, molecular diffusion, and mixing-induced supersaturation removal---and then lifts the resulting slow dynamics to a macroscopic cloud field through non-equilibrium statistical field theory. The unifying structure across these scales is the emergence of a slow manifold: a dynamical backbone on which the fast supersaturation environment is slaved to the slower evolution of the droplet-growth variable.

Classical K\"ohler theory \citep{kohler_1936} robustly explains the activation of individual cloud condensation nuclei \citep{khain_pinsky_2018}, but it is not by itself a theory of marginal clouds. In real clouds, populations of growing droplets deplete and reshape the surrounding supersaturation field \citep{Korolev2003,khain_pinsky_2018,pinsky2013}, while turbulent mixing and weak updrafts continuously perturb the thermodynamic balance. Detailed parcel models can incorporate these effects by coupling thermodynamics, droplet growth, and vertical motion \citep{hirsch2014transition,hirsch2015properties,hirsch2017,altaratz2021environmental}, but their complexity makes it difficult to isolate analytical mechanisms and scaling laws.

Here we focus on weak-updraft regimes, where supersaturation production is limited and droplets remain close to critical activation conditions. To expose the organizing dynamics of this regime, we adopt a reduced dynamical-systems approach \citep{LucariniChekroun2023,arabas2017,Santos_al24_SciAdv}, grounded in first-principles microphysics and closed by parameterizing unresolved turbulent exchange \citep{LucariniChekroun2023}. The starting point is the core K\"ohler diffusional-growth equation \citep{arabas2017,Santos_al24_SciAdv}, which is especially relevant near cloud base, where activation is initiated and where small, marginal or twilight clouds are prone to be encountered. This reduction is justified by a separation of timescales: supersaturation adjusts on microphysical timescales that are typically much shorter than the mesoscale evolution controlling the ascent. The vertical velocity may therefore be treated as quasi-steady at leading order, while retaining the essential competition between adiabatic cooling, condensational depletion of supersaturation, and unresolved mixing.  Measurements in newly forming shallow cumuli show substantial supersaturation fluctuations superposed on a quasi-steady background, emphasizing the competition between turbulent mixing and condensational relaxation \citep{siebert2017}.

In our reduced parcel model approach, environmental mixing enters as a net supersaturation removal rate $\lambda$, with turbulent variability represented by stochastic fluctuations of intensity $\sigma$ (Section \ref{Sec_stochastic}). The deterministic dynamics already reveal the central geometry: our analysis exposes that {supersaturation rapidly approaches and tracks its quasi-steady nullcline, with a lag quantified below}   (Section \ref{Sec_expo_attraction}), and mixing-induced removal drives a saddle-node bifurcation on this manifold (Section \ref{Sec_advection}). The saddle-node creates an edge state that replaces the classical K\"ohler activation radius by a dynamically shifted threshold $R^*>R_c$. This defines a forbidden zone: a region below the classical K\"ohler equilibrium curve where deterministic intuition would predict immediate return to haze, but where the slow-fast dynamics create a metastable transition layer.

The stochastic analysis then explains why this layer can persist. Using a small-noise reduction \citep{gardiner2009stochastic,freidlin2012random}, we derive in Section \ref{Sec_reducedSDE_main} an effective stochastic differential equation constrained to the slow manifold. A Lamperti transformation \citep{lamperti1962semi} casts this reduced dynamics into a gradient form \citep{berglund_2011,pavliotisbook2014}, exposing the potential landscape near the activation edge (Section \ref{Sec_Lamperti}). In the supercritical mixing regime $\lambda>\lambda_c$, the effective potential flattens near the saddle-node (Section
\ref{Sec_Lyapunov_exp}). Droplets need not immediately activate or evaporate; instead, they hover near the edge state for extended periods, producing intermittent transitions between haze and activation (Section \ref{Sec_Full_Parcel_Dynamics}). This mechanism gives a dynamical explanation for the fleeting, optically faint character of twilight clouds and recovers the observed scale of $1$--$2\mu{\rm m}$ droplet-size distributions in weak-updraft regimes \citep{hirsch2014transition}.

The 0D parcel description, however, is intrinsically local. It captures temporal hovering, but it cannot describe how neighboring cloud elements communicate through the supersaturation field, nor how hovering states organize across space. We therefore pass from isolated parcels to a spatially extended mean-field theory. 
Starting from the standard fluid-dynamical description of turbulent warm-cloud microphysics, and retaining K\"ohler activation physics at the level of a coarse-grained droplet population, we derive a {\it degenerate} stochastic droplet--supersaturation reaction--diffusion system; see Section \ref{Sec_Field_Theory} and {\it Supplementary Note 5}. The slow field is the mean-square-radius field $\phi(\x,t)=\langle R\rangle_{\x,t}$, representing the local coarse-grained droplet population, while the fast field $\chi(\x,t)$ is the resolved supersaturation bath sampled by that population. The system is degenerate in the sense that diffusion does not act on both fields symmetrically: the slow droplet field $\phi$ evolves locally through condensation and K\"ohler feedback, whereas diffusion acts only on the fast supersaturation field $\chi$, through the effective scalar diffusivity $D_\chi$, which includes the turbulent eddy-diffusivity contribution from the unresolved transport fluctuations. Thus, the model is not a standard two-field reaction--diffusion system with independent smoothing of both components; it is a partially parabolic slow--fast system in which neighboring droplet populations communicate indirectly through the diffusing supersaturation bath. This construction provides a leading-order closure that bypasses the need to resolve the full droplet-size distribution through bin microphysics \cite[e.g.][]{tzivion1987efficient,feingold1994explicit,khain2015spectral}, while still retaining the dominant nonlinear K\"ohler feedback mechanism inherited from the parcel level (Appendices \ref{App_Kohler_projection} and \ref{Sec_microphys_average}).

We then embed this stochastic reaction--diffusion system in the Martin--Siggia--Rose--Janssen--de Dominicis (MSR) path-integral formalism \citep{MartinSiggiaRose,janssen1976lagrangean,deDominicis1976,kamenev2011field,tauber2014critical,del2026field} (Section \ref{Sec_Field_Theory}). This step is not merely a change of notation: it recasts the droplet--supersaturation system as a response theory in which the fast supersaturation bath can be eliminated at the level of the action. The resulting construction reveals a useful structural analogy with the polaron problem in solid-state physics \citep{Devreese2009}. In the Fr\"ohlich--Gurari picture, a slow electron becomes dressed by a polarization envelope of a fast medium, thereby modifying its effective dynamics \citep{gurari1953,feynman1955slow}. In the cloud problem, the slow object is not an individual microscopic droplet, but the local coarse-grained droplet population represented by $\phi$. The corresponding fast medium is the supersaturation bath $\chi$, and the analogue of the polarization envelope is a statistically localized supersaturation halo generated by the turbulent bath response. The resulting object is therefore a collective ``cloud polaron'': a droplet field dragging a supersaturation response, rather than a single droplet carrying a microscopic vapor shell.

Using a Gurari-type variational shift, we integrate out the fast supersaturation bath around its turbulently dressed slow manifold. {As in Gurari's treatment of the polaron \citep{gurari1953}, the displacement of the fast field is a variational object, fixed by a stationarity condition; it is the retarded, spatially screened response of the bath to the droplet field, and its denominator has the same structure as Gurari's optimal phonon displacements, with the bath relaxation rate in place of the phonon frequency, the diffusive cost in place of the recoil energy, and the evolution of the droplet field in place of the motion of the electron.} The turbulent fluctuations enter this bath through the supersaturation equation itself: the unresolved mixing and exchange term with the dry air is decomposed into a mean removal rate and a fluctuating component, producing a multiplicative noise of the form $\sigma\chi\eta$ at the level of the fast field $\chi$, with $\eta$ denoting Gaussian noise that is white in time and correlated in space, as specified in Eq.~\eqref{Eq_noise_covariance} below. Thus, turbulence is not imposed as an additive random force on the droplet equation. It acts by modulating the supersaturation bath sampled by the local droplet population.

This distinction is essential in the MSR reduction aimed at a closed $\phi$-equation. Because the noise amplitude is state-dependent, $\Sigma(\chi)=\sigma\chi$, {and because the mixing-rate fluctuations have a finite correlation time, so that their white-noise limit is of Stratonovich type \citep{wong1965convergence},} the transformation from noise histories to supersaturation histories carries a nontrivial functional Jacobian  \citep{ZinnJustin2002,Aron_al2010,tauber2014critical,aron2016dynamical}. {After regularization at the Kolmogorov scale, converting the Stratonovich equation to It\^o form gives the relaxation-rate correction $-\sigma_S^2/\tau_K$. The equivalent midpoint formulation accounts for this correction through its prescription-dependent measure terms (Appendix~\ref{App_Jacobian_Regularization}); it is included only once. Here $\sigma_S^2$ is the dimensionless mixing-fluctuation strength and $\tau_K$ is the Kolmogorov turnover timescale.} The correction offsets part of the background supersaturation relaxation, so that turbulence dresses the K\"ohler slow manifold itself rather than merely adding fluctuations around it. At the parcel level, this turbulent dressing shifts the saddle-node threshold to a larger effective mixing rate and shrinks the forbidden zone. In physical terms, a stronger drying environment is required to suppress activation once the local coarse-grained droplet population is coupled to, and dressed by, its turbulent supersaturation bath.

The same reduction also determines how the local dressing {acquires} a spatial character. When the zero-diffusion bath limit is relaxed, the supersaturation response is no longer pointwise: neighboring coarse-grained droplet populations interact through the non-local operator
$ \mathcal L_\phi^{-1}=\big(\Gamma^*(\phi)-D_\chi\nabla^2\big)^{-1}$.
Here $\Gamma^*(\phi)$ is the turbulently dressed supersaturation relaxation rate toward the dressed slow manifold $\chi_0^*(\phi)$, while $D_\chi$ is the effective supersaturation diffusivity. The Bessel Green's function of $ \mathcal L_\phi^{-1}$ defines then the screening length,
$L_{\rm scrn}=\sqrt{D_\chi/\Gamma^*(\phi)}$,
over which the supersaturation halos attached to neighboring coarse-grained cloud elements interact effectively. Written in local form, this screened response yields the pseudo-parabolic operator
$-\nabla\cdot\big(\epsilon(\phi)\nabla\partial_t\phi\big)$,
with coefficient
$\epsilon(\phi)=D_\chi/\Gamma^*(\phi)=L_{\rm scrn}^2$,
which records the spatial footprint of the dragged supersaturation halo: a change of the droplet field at one point alters the growth tendency over the whole screening disk. When the finite relaxation time of the bath is retained at first order in the ratio of slow to fast rates, the halo also lags behind the droplet field, and this lag renormalizes the coefficient of $\partial_t\phi$. This is the cloud analogue of the polaron effective mass \citep{gurari1953,feynman1955slow}, and, as for the polaron, it is set by the ratio of the bath-mediated coupling rate to the bath relaxation rate. Together, the footprint and the lag form the dynamic-inertia operator, which confers on the macroscopic droplet field a memory of the supersaturation bath it drags (Fig.~\ref{Fig_MSR_schematic}).

{The same stationarity condition that fixes the Gurari displacement also yields the retarded, nonlocal bath response from which the screening and the lag both follow, so that the two components of the dynamic inertia are the spatial and temporal faces of a single dressed object.}

The resulting non-Markovian droplet-field closure is a nonlinear pseudo-parabolic stochastic partial differential equation (SPDE) for $\phi$ of Barenblatt--Zheltov--Kochina (BZK) type \cite{barenblatt1960basic,barenblast1960fundamental}, reminiscent of pseudo-parabolic models arising in heat conduction in materials with two distinct temperatures \cite{Chen_Gurtin1968theory}, fluid seepage through fissured rock formations \cite{barenblatt1960basic}, and radiative transfer theory \cite{sobolev1963treatise}. Its stochastic forcing is not imposed externally: it is the colored macroscopic noise left behind by the elimination of the supersaturation bath.

This structure also resonates with recent progress on the reduction of SPDEs forced by multiplicative noise acting in the unresolved, or orthogonal, component of the dynamics. In such settings, stochastic parameterization formulas show that fluctuations in the eliminated sector are transmitted to the resolved large-scale modes as non-Markovian stochastic coefficients in the reduced equation \cite{chekroun2023transitions,chekroun2025unravel}. These coefficients depend explicitly on the history of the noise path, and their memory content is determined self-consistently by the noise intensity and by the way the unresolved fluctuations interact with the resolved variables through the nonlinear terms of the SPDE. The present cloud reduction provides a physically explicit realization of this mechanism: multiplicative turbulent supersaturation fluctuations in the fast bath become, after elimination, colored and memory-bearing stochastic forcing in the slow droplet-field equation.

Other spatial terms arising in the non-Markovian closure are equally structured. {At long wavelengths, the halo-mediated coupling acts on the droplet field as
$\mathcal{K}_\phi=2D\nabla\cdot\big(\epsilon(\phi)\nabla\chi_0^*(\phi)\big)$,
where $\chi_0^*(\phi)$ is the turbulently dressed slow manifold, built on the mean-field projection $f(\phi)$ of the K\"ohler curve (Appendix \ref{App_Kohler_projection}). Its sign is set by the slope of the dressed slow manifold, which increases with $\phi$ in the haze regime and decreases beyond a maximum located close to the K\"ohler critical radius. In the haze regime the coupling acts diffusively, smoothing unactivated aerosol fluctuations (Section \ref{Sec_Phys_nonFickian}). Beyond the maximum, the same coupling becomes anti-diffusive, reinforcing spatial contrasts between cloudy and dry regions; the pseudo-parabolic operator keeps this backward diffusion well posed \citep{novick1991stable,padron2004effect}.} In this sense, the term has a ``rich-get-richer'' structure: activated coarse-grained droplet populations {draw supersaturation from their surroundings} and sharpen their contrast with neighboring haze-like regions. This provides a thermodynamic mechanism for edge sharpening and pattern formation, complemented by {a nonlinear squared-gradient term that emerges from the expansion of $\mathcal{K}_\phi$ and has the form that, for an interface height, generates Kardar--Parisi--Zhang (KPZ) growth} \cite{KPZ86,TGM97,corwin2012kardar}{; see} Section \ref{Sec_Phys_nonFickian}.

We test the spatial reduction directly in a regime where the bath's temporal memory is neglected but its spatial response is retained. The resulting Markovian closure evolves only the mean-square-radius field, with the same screened operator acting on drift and forcing (Section~\ref{Sec_Spatial_Markovian}). Starting from a uniform edge state, it reproduces the coupled droplet--supersaturation system's progression from hovering through connected haze regions to activated cloudlets (Fig.~\ref{Fig_closure_maps}). Agreement extends to the evolving distribution of local root-mean-square radius and the number and area of cloudlets (Fig.~\ref{Fig_closure_diagnostics}). These comparisons show that eliminating the bath can preserve both population statistics and spatial organization, providing a numerical test of the screened field theory.  A second comparison starts from uniform haze near the saddle-node threshold (Section~\ref{Sec_nearfold_memory}). Keeping the screened deterministic drift unchanged, we restore the residual bath's finite relaxation (making the closure non-Markovian) and find improved agreement with the parent system in the frequency and spatial extent of recurrent hovering (Figs.~\ref{Fig_nearfold_maps} and \ref{Fig_nearfold_diagnostics}). The bath sets the fluctuation-memory timescales without additional fitting. This experiment isolates the contribution of colored forcing to the non-Markovian reduction.

By uniting slow-fast parcel microphysics, turbulence modeling and non-equilibrium statistical field theory, this work provides a first-principles route from K\"ohler activation to a macroscopic stochastic cloud-field equation. The theory explains how turbulent mixing creates hovering near a saddle-node edge state, how turbulent dressing shrinks the forbidden zone, how screened supersaturation halos couple neighboring cloud elements, and how delayed bath relaxation generates dynamic inertia and colored noise. These mechanisms identify the operators through which subgrid supersaturation fluctuations organize the morphology and persistence of twilight clouds.

The reduction principle behind these results has a scope well beyond cloud physics. It transposes to fields the polaron mechanism by which a slow particle, by polarizing a fast medium, acquires a dressed energy and an effective mass. Its premise is generic: a slow field with little or no intrinsic transport, coupled to a relaxing, diffusive environment that it modifies. Eliminating that environment within the MSR formulation dresses the local dynamics, induces screened interactions through the environmental halo, renormalizes the inertia of the slow field, and colors its fluctuations. The same premise underlies dryland vegetation--water feedbacks, in which slowly spreading biomass redistributes soil water and organizes vegetation patterns \citep{gilad2007ecosystem,meron2016pattern}, and populations of oscillators or cells communicating through a rapidly diffusing chemical, whose adiabatic elimination yields a screened nonlocal coupling of the same form \citep{tanaka2003nonlocal}. It extends naturally to bacterial colonies whose growth and mobility depend on a diffusing nutrient \citep{kawasaki1997patterns} and to transport in heterogeneous media, where distributed exchange times generate memory \citep{dentz2003transport}. The polaron picture also has a classical counterpart in plasma physics, the dressed test particle with its shielding cloud \citep{rostoker1961fluctuations}, which links the reduction to dense and complex plasmas. In each setting, the slow-field closure framework described in this work offers a systematic route from the coupled dynamics to an effective equation for the slow field that retains the interactions, memory, and fluctuations inherited from its environment (Section~\ref{Sec_Discussion}).

The remainder of this paper is structured as follows. Sections~\ref{Sec_Diffusion_growth} and~\ref{Sec_advection} introduce at the parcel level, the reduced droplet-growth framework and the effect of mixing-induced supersaturation removal, notably in the creation of a saddle-node activation threshold. Section~\ref{Sec_expo_attraction} analyzes the slow manifold and the exponential supersaturation relaxation rate to it. Sections~\ref{Sec_stochastic} and~\ref{Sec_reducedSDE_main} derive the reduced stochastic dynamics  and use the Lamperti transformation to characterize hovering. Section~\ref{Sec_Field_Theory} then lifts the parcel theory to the MSR field level. In Section~\ref{Sec_MSR_spatio}, we construct the isolated slow-droplet action $\Action_\phi$, the isolated fast-bath action $\Action_\chi$, and the condensation coupling action $\Action_{\rm int}$. We then integrate over the fast supersaturation bath using a Gurari-type variational construction \citep{gurari1953} in Section~\ref{Sec_slow_effective_Gurari}. The first resulting closure, derived in Section~\ref{Sec_Markov_closure}, is obtained under two simplifying assumptions: a Markovian approximation, in which the bath relaxes instantaneously relative to the slow droplet field, and a zero-diffusion bath limit, in which the spatial redistribution of $\chi$ is neglected at leading order. These restrictions are relaxed in Section~\ref{Sec_Dynamic_Inertia}, where finite bath relaxation and spatial supersaturation redistribution generate the non-Markovian pseudo-parabolic droplet-field equation, including dynamic inertia, colored noise, and K\"ohler-mediated anti-diffusion.  Section~\ref{Sec_Closure_Results} tests the spatially correlated Markovian closure and the effect of residual-bath fluctuation memory against the coupled field system. Section~\ref{Sec_Discussion} discusses the implications for twilight clouds, subgrid cloud modeling, and response-based cloud closures.

\section{The Fleeting and Visually Faint Nature of Twilight Clouds}
Twilight clouds refer to the cloud-aerosol transition zone \cite{koren2007twilight}--a region of the sky that satellites often categorize as "clear," but which is actually populated by a complex mixture of humidified aerosols \cite{charlson2007climate, calbo2017thin, varnai2018satellite} and optically thin clouds that remain small relative to a satellite's spatial resolution. This zone is further characterized by 3D radiative effects, where larger clouds serve as secondary illumination sources for their surroundings \cite{marshak2006impact, wen20073}.

Because of these intricacies, detecting twilight clouds by deterministic means is notoriously challenging; their presence is typically revealed through statistical analysis. Studies have shown that the characteristic spatial scale of the twilight zone is $\sim 10 \text{ km}$ and that it covers approximately $60\%$ of the non-cloudy sky \cite{bar2011global, twohy2009effect}. Similar phenomena have been observed in cirrus clouds \cite{wollner2014signature}, with significant implications for both the shortwave and longwave climate energy balances \cite{eytan2020, eytan2025net}.

\begin{figure}[htbp]
    \centering
   \includegraphics[height=0.245\textwidth, width=.49\textwidth]{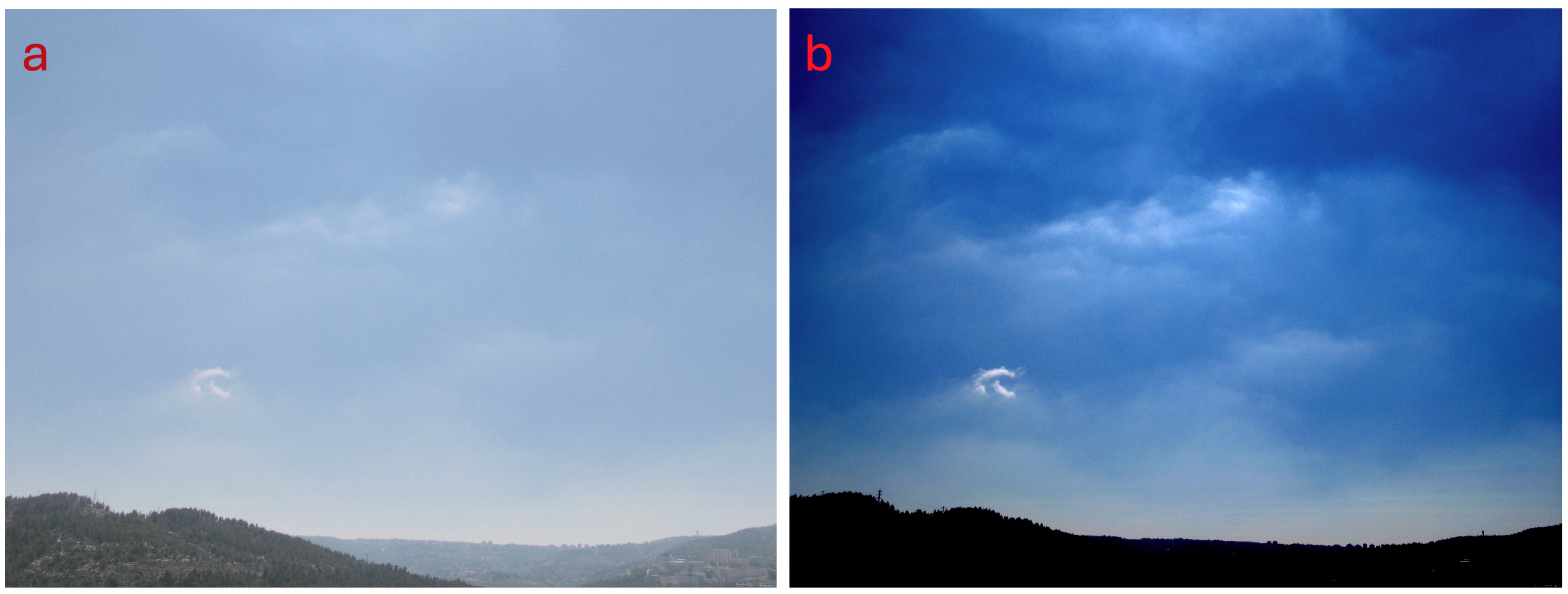}
    \caption{A typical twilight sky in the eastern Mediterranean summer. a) A true color snapshot showing mostly clear sky with a few small cloud segments. b) The same snapshot with an enhanced contrast. Note how all of the sky is filled with activated segments and haze.}
    \label{Fig_Twilight}
    \vspace{-.1cm}
\end{figure}

Figure \ref{Fig_Twilight} illustrates the elusive nature of this transition. Panel (a) shows a snapshot of the sky over the eastern Mediterranean during the summer of 2025. While only a few small cloud segments are visible to the naked eye, a contrast-enhanced version (Panel b) reveals that the entire domain is filled with a continuum of thin, small cloud elements and humidified (unactivated) aerosols, traditionally defined as haze.

The inherent difficulty in resolving these sub-grid cloud elements and the continuum of states between aerosol and cloud suggest that a deterministic framework may be insufficient for capturing the underlying dynamics. Given that these features are primarily identified through statistical signatures, we propose below that the transition from a stable haze state to an activated cloud state is best described as a stochastic process. By framing this transition as an escape problem in the presence of random fluctuations in mixing processes, we can model the emergence of twilight clouds as a probabilistic crossing of a critical threshold. This approach acknowledges the physical uncertainty inherent in the transition zone and allows for a more robust representation of the variability observed in the eastern Mediterranean's sky and beyond.

    \vspace{-.1cm}
\section{Diffusion Growth in Adiabatic Parcels}\label{Sec_Diffusion_growth}
\subsection{Background}\label{Sec_background_partI}
K\"ohler theory serves as a foundational framework in cloud physics for understanding how aerosol particles activate (at cloud's base) to become cloud droplets, a crucial step in cloud formation and precipitation. The central concept is the K\"{o}hler curve, which describes the equilibrium vapor pressure over a solution droplet's surface by considering two opposing effects: the Kelvin effect and Raoult's law. The Kelvin effect explains that the increased curvature of smaller droplets requires a higher surrounding vapor pressure to prevent evaporation. Conversely, Raoult's law states that the presence of dissolved solutes within the droplet lowers the equilibrium vapor pressure.
Mathematically, a standard approximate form for the  K\"ohler curve, $f(R)$, is given as \citep{kohler_1936}
\be\label{Eq_Kohler}
f(R)=\frac{A}{\sqrt{R}}- B R^{-3/2},
\ee
where $R$ is the square of the droplet's radius. The first term, $A/\sqrt{R}$, quantifies the Kelvin effect's contribution to the increased vapor pressure resulting from surface tension. The second term, $B R^{-3/2}$, represents the reduction in vapor pressure due to the presence of dissolved substances, as described by Raoult's law. The constant $B$ is related to the mass and solubility of the solute within the aerosol particle and can be expressed as $B=\kappa r_d^3$ \citep{kohler_1936,rogers1989}, where $\kappa$ is the hygroscopicity parameter \citep{petters2007single},
and $r_d$ is the dry radius of the aerosol particle. Using the squared radius $R$ simplifies subsequent mathematical derivations.

Under conditions of low supersaturation, the activation of aerosol particles into cloud droplets is a selective process, primarily affecting the largest and most hygroscopic particles. This activation hinges on the delicate equilibrium between the droplet's tendency to evaporate due to surface curvature (Kelvin effect) and its propensity to grow by water vapor condensation due to dissolved substances (Raoult effect) \citep{kohler_1936}. This competition results in two distinct thermodynamic states for aerosol particles \citep{khain_pinsky_2018}: stable haze particles, which remain in equilibrium with the surrounding humidity, and activated droplets, which undergo rapid growth by water vapor diffusion, theoretically growing indefinitely. However, on hazy conditions, when the thermodynamical conditions are relatively stable, the finite amount of available supersaturated water vapor in the atmosphere often leads to a coexistence of both haze particles and activated droplets within clouds, with individual particles potentially transitioning between these states, particularly in  relatively stable conditions that yield small, warm clouds \citep{hirsch2017, altaratz2021environmental}. Furthermore, under turbulent conditions, these microphysical regimes can give rise to intricate interactions, leading to non-linear phenomena such as oscillatory activation pulses at low particle concentrations \citep{yang_2024}.

Our modeling approach elaborates on this fundamental K\"ohler theory. 
As mentioned above, the K\"ohler curve marks the equilibrium supersaturation over drops of water containing a given amount of solute. To understand how droplets grow or evaporate, one typically uses the condensational growth equation, which describes the rate at which a droplet changes size due to condensation or evaporation. It is given by \cite[Eq.~(13.28)]{prupp}
\be\label{Eq_condens_eq}
\dot{R}= 2 D (S - f(R)),
\ee
where $D$  is the diffusional parameter  which encodes the resistance to vapor diffusion and latent-heat conduction in the microscopic boundary layer around an individual droplet; see {\it Supplementary Note 1}. It converts  a supersaturation excess into growth of $R=r^2$.
Here, $S$ denotes the ambient supersaturation, generally assumed to be a control parameter in Eq.~\eqref{Eq_condens_eq}; see e.g.~\citep{arabas2017}. 

In reality, the supersaturation is a time-dependent quantity that is coupled to the haze and activated particle size distribution \citep{ghan1993parameterization}. For warm clouds, its temporal evolution during adiabatic ascent, is often expressed,  using the quasi-hydrostatic approximation, as \citep[Eq.~(5.2.2)]{khain_pinsky_2018} 
\be\label{Eq_S_fundamental}
\dot{S}=A_1 w -A_2\dot{q}, 
\ee
where $q$ is the liquid water mixing ratio, $w$ is the updraft velocity of the air parcel, and the coefficients $A_1$ and $A_2$ are functions of temperature and pressure; see e.g.~\cite[Appendix 5.2.9]{khain_pinsky_2018}. 

The first term on the right-hand side (RHS) of Eq.~\eqref{Eq_S_fundamental} describes an increase in supersaturation due to adiabatic cooling of air during the parcel's ascent; the second term in the RHS describes the supersaturation sink caused by condensation of water vapor on drops' surface.

To find a closed model  for supersaturation from Eq.~\eqref{Eq_S_fundamental}, one typically relies on a well-known representation of the rate of changes in the liquid water mixing ratio due to condensation or evaporation of cloud droplets, that involves the normalized particle size distribution, $h$, according to \cite[Eq.~(5.2.12)]{khain_pinsky_2018}:
\be\label{Eq_general_q_rate}
\dot{q}=\frac{4 \pi N\rho_w}{\rho_a} \int_0^\infty h(R) \frac{\sqrt{R}}{2} \frac{\d R }{\d t} \d R,
\ee
where $N$ is the number concentration of the cloud droplets, while $\rho_w$ and $\rho_a$
denote the densities of liquid water and dry air, respectively. 

For a monodisperse population, the normalized particle size distribution $h(R)$ can be represented by a Dirac delta function centered at the single radius $R$: $h(r)=\delta(r-R).$ By using  Eq.~\eqref{Eq_condens_eq}, the expression of $\dot{q}$ then becomes:
\be\label{Eq_q_rate}
\dot{q}=\frac{4 \pi \rho_w N D}{\rho_a} \sqrt{R} (S-f(R)). 
\ee 
Substituting Eq.~\eqref{Eq_q_rate} into Eq.~\eqref{Eq_S_fundamental}, we arrive at the system 
\begin{subequations}\label{Eq_Korolev_rectif}
\begin{align}
&\dot{S} = \gamma-\alpha_N \sqrt{R}\Big(S -f(R)\Big), \label{Eq_Ka}\\
&\dot{R} = 2D\Big(S-f(R)\Big),\label{Eq_Kb}
\end{align}
\end{subequations}
where 
\bea\label{Eq_alphaN}
\gamma&=A_1 w,\\
\alpha_N &= 4\pi \rho_w \beta D N,
\eea
with with $\beta = A_2/\rho_a$.  The number concentration of the cloud droplets, $N$, appearing in Eq.~\eqref{Eq_alphaN}, is a key control parameter from a microphysical perspective as it quantifies  the air pollution: the higher $N$, the more polluted is the air. It will play a central role in our analysis below. Note that $[ \alpha_N]=[m^{-1}s^{-1}]$, so that the second term in the RHS of Eq.~\eqref{Eq_Ka} is of unit $[s^{-1}]$, making the model dimensionally consistent. 

Despite its simplicity, this model is a fundamental building block to clouds microphysics as it
describes the temporal evolution of a system containing $N$ identical particles, each characterized by a squared radius $R$ and sharing identical physical and chemical properties. In this monodisperse framework, $R$ represents the squared radius of a representative droplet whose growth is coupled to the surrounding environment.

A central feature of this system is the role of supersaturation. Here, $S$ denotes the bulk supersaturation of the air parcel. This $S$ tracks the available water vapor in the air parcel, which is consumed by the entire droplet population (represented by the $\alpha_N$ term in $\dot{S}$). In the evolution of the droplet radius, the driving force is the difference between this bulk value and the {\it equilibrium supersaturation} at the droplet's surface, $f(R)$, determined by the K\"ohler curve. By incorporating $f(R)$ directly into the growth law, the model effectively accounts for surface curvature and solute effects without requiring a separate variable for surface supersaturation, assuming that diffusive transport between the bulk and the surface is characterized by the effective constant $D$. While many variants of this system exist in the literature \citep{Korolev2003}, often disregarding K\"ohler effects when focusing purely on mass transfer in macroscopic cloud physics, our inclusion of $f(R)$ is essential for capturing the delicate stability of the haze state and the subsequent turbulent activation process discussed in this paper.

\subsection{Slow manifold and asymptotic power laws}\label{Sec_power_law}
We now investigate the slow manifold structure of the parcel model \eqref{Eq_Korolev_rectif}. In singular-perturbation theory, a slow manifold is an invariant, or approximately invariant, lower-dimensional set on which the fast variables have relaxed to a conditional equilibrium while the remaining variables evolve on a slower timescale \citep{tikhonov1952systems,Kuehn:2015tol}. Equivalently, it is the dynamical surface that organizes the long-time motion after the rapid transient has decayed. In the present parcel model, the supersaturation $S$ plays the fast thermodynamic variable, while the droplet-size variable $R$ evolves more slowly through condensational growth.

Assuming the parameters in Eq.~\eqref{Eq_Korolev_rectif} are strictly positive, one can prove that 
the solutions $(S(t),R(t))$ to this system, remains unbounded in the sense that $R(t)\rightarrow \infty$ as $t\rightarrow \infty$.  
{The goal of this section is to identify the leading quasi-steady balance governing this unbounded growth and extract the associated asymptotic power laws for $R(t)$ and $S(t)$. This balance approximates the slow dynamics; the curve itself is not invariant, and Section~\ref{Sec_expo_attraction} quantifies the tracking lag.}

The proof results from a simple argument. 
Since $S-f(R)=\dot{R}/ (2D)$ from Eq.~\eqref{Eq_Kb}, we observe that Eq.~\eqref{Eq_Ka} can be re-written as
\be
\dot{S}=\gamma-\frac{\alpha_N}{3 D} \frac{\d}{\d t} R^{3/2},
\ee
which leads to 
\be
S(t)=S(0)+\gamma t -\frac{\alpha_N}{3 D}\Big(R(t)^{3/2}-R(0)^{3/2}).
\ee
If we assume $R(t)$ is bounded by some $M>0$, we get for large $t$  that $S(t)\sim \gamma t$, which $S(t)$ grows linearly with time and tends to $\infty$ as $t\rightarrow \infty$, since $\gamma>0$ (adiabatic ascent). 

Now consider Eq.~\eqref{Eq_Kb}. Since $R(t)$ is assumed to be bounded, $f(R(t))$ is also bounded. Since $S(t)\sim \gamma t$ for large $t$, we infer that $\dot{R}\sim 2D \gamma t$ as $t\rightarrow \infty$.
This implies that $R(t)$ grows quadratically with time, which is incompatible with the assumption that $R(t)$ remains bounded. {Thus $R$ cannot remain uniformly bounded. On the growing branch, the large-time balance below gives $R(t)\to\infty$ and its leading growth law; unboundedness alone is not a proof of this limit.}  

Since $R(t)\rightarrow \infty$ as $t\rightarrow \infty$, the cloud droplet growth model (Eq.~\eqref{Eq_Korolev_rectif}) evolves towards a state where the condensation onto the growing droplet increasingly balances the continuous generation of supersaturation, leading to a gradual decrease in supersaturation while the droplet size continues to grow indefinitely. 

The balance equation from $\dot{S}\approx 0$, reduces to:
\bes
\gamma -\alpha_N \sqrt{R} S+ \alpha_N A -\alpha_N \frac{B}{R}\approx 0.
\ees
For large $R$, the term $\alpha_N B/ R$ approaches $0$. Thus, the balance equation simplifies to:
\bes
\gamma+ \alpha_N A \approx \alpha_N \sqrt{R} S.
\ees
From this, we can express $S$ in terms of $R$: 
\be\label{SR_Relation}
S\approx \frac{\gamma +\alpha_N A}{\alpha_N \sqrt{R}},
\ee
for $R$ sufficiently large.

{The leading terms of $S$ and $f(R)$ are both of order $R^{-1/2}$ and must be retained together. Their difference on the quasi-equilibrium curve is
\be
S-f(R)\sim\frac{\gamma}{\alpha_N\sqrt R},\qquad
\dot R\sim\frac{2D\gamma}{\alpha_N\sqrt R}.
\ee
Integration gives
\bea\label{Eq_Law_RS}
R(t)&\sim\left(\frac{3D\gamma}{\alpha_N}\right)^{2/3}t^{2/3},\\
S(t)&\sim\frac{\gamma+\alpha_N A}{\alpha_N^{2/3}(3D\gamma)^{1/3}}t^{-1/3}.
\eea
The same leading coefficient for $R$ follows from the exact balance $S+\alpha_NR^{3/2}/(3D)=\gamma t+\mathrm{const}$ when $S=o(t)$.}

The specific power law exponents (2/3 for $R$ and $-1/3$ for $S$) are indicative of a diffusion-limited growth process under a constant rate of supersaturation generation.
These power laws are consistent with those found in the literature for the aforementioned variants of Eq.~\eqref{Eq_Korolev_rectif}; see e.g.~\cite[Eq.~(4.3.26)]{sedunov1974physics}. 

The fact that the square of the droplet radius $R$ grows with time as $t^{2/3}$ indicates a growth process that is limited by the availability of water vapor diffusing towards the droplet. This  growth law is characteristic of droplet growth governed by diffusion in a supersaturated environment. The droplet grows continuously, but the rate of growth slows down over time.

The supersaturation $S(t)$ decreasing as $t^{-1/3}$ signifies that the growing cloud droplet is effectively consuming the excess water vapor from its surroundings. As the droplet gets larger, its surface area increases, allowing for more efficient condensation of water vapor onto it. This process leads to a gradual reduction in the relative humidity excess (supersaturation) in the air. The power law decay indicates a continuous depletion of the available supersaturation.

A stronger updraft $\gamma$ (due to factors like faster cooling of the air) leads to a faster overall growth of the droplet and a higher magnitude of supersaturation at any given time, even though $S(t)$ still decreases over time. A larger $D$ allows water vapor molecules to diffuse more readily towards the droplet surface, resulting in a faster growth rate of the droplet. A higher $\alpha_N$ (which depends on the number concentration of cloud condensation nuclei) implies more droplets are competing for the available water vapor. This would lead to slower growth for individual droplets compared to a situation with fewer droplets.

\begin{table}[h!]
    \centering
    \caption{Parameters for Eq.~\eqref{Eq_Korolev_rectif} used for Figure \ref{Fig_flow_tau=0}}
    \label{Table_RegimeA}
    \begin{tabular}{lc}
        \toprule
        Parameter & Value \\
        \midrule
        $\gamma \, [\% \, s^{-1}]$ & $1.2 \cdot 10^{-3}$ \\
        $D \, [\mu m^2 \, s^{-1}]$ & $50$ \\
        $A \, [\mu m]$ & $1.4 \cdot 10^{-3}$ \\
        $B \, [\mu m^3]$ & $3.5 \cdot 10^{-4}$ \\
        $\alpha_N \, [(\mu m \, s)^{-1}]$ & $0.011$ \\
        $N \, [\text{cm}^{-3}]$ & $50$ \\
        \bottomrule
    \end{tabular}
\end{table}

In terms of phase portrait, the dynamics along the $S$-direction is initially fast, leading to a quick adjustment of supersaturation relative to the slower growth of the droplet radius $R$.  The asymptotic laws describe  the state of the system as $ t\rightarrow \infty$ from which we can recover the system's slow manifold, where $R$ grows slowly and $S$ adjusts relatively quickly as $t\rightarrow \infty$.

\begin{figure}[htbp]
    \centering
   \includegraphics[height=0.3\textwidth, width=.45\textwidth]{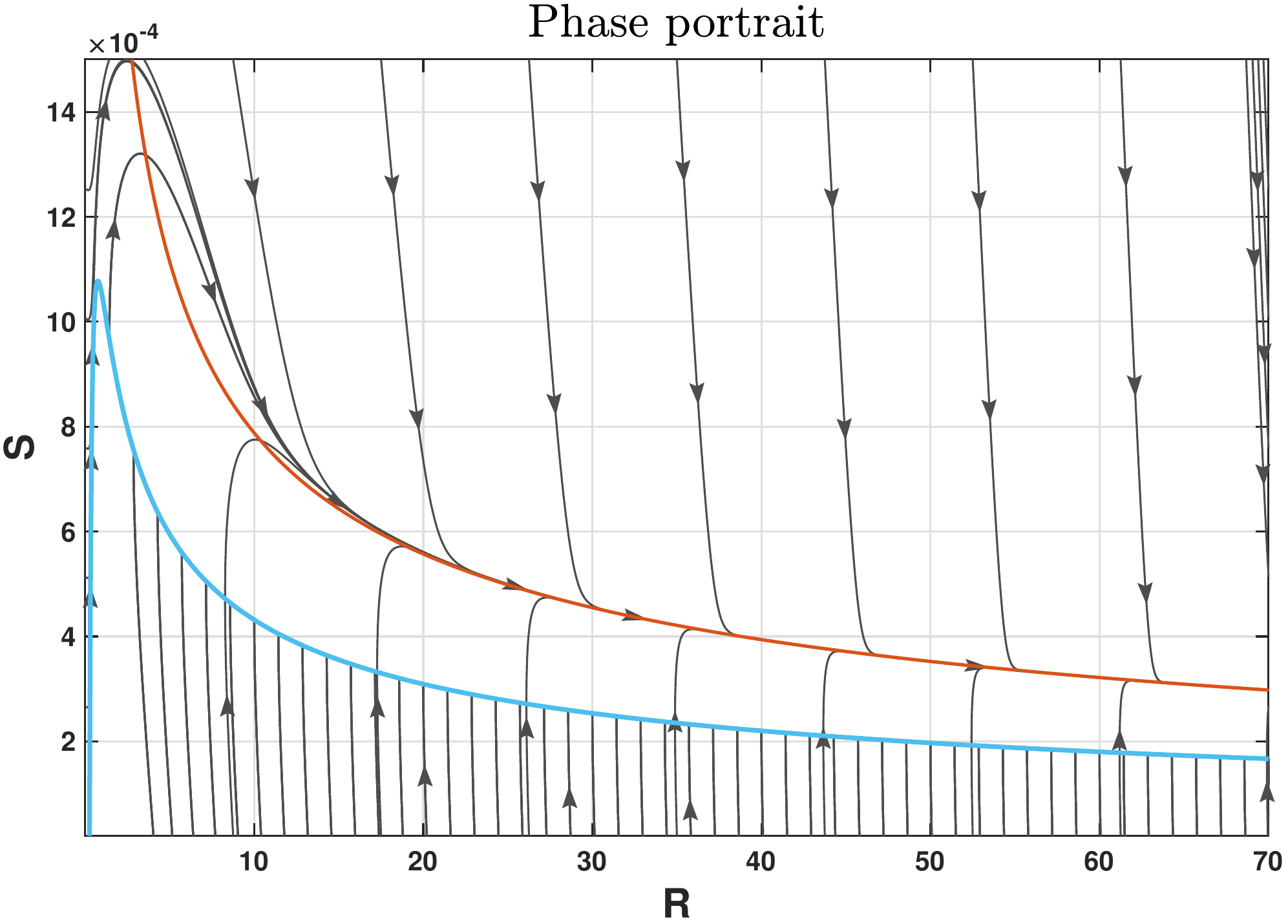}
    \caption{The K\"ohler curve is shown in cyan. The red curve shown is that from Eq.~\eqref{SR_Relation}, i.e.~$S=(\gamma +\alpha_N A)/(\alpha_N \sqrt{R})$. It provides an approximation of the system's slow manifold; see Text.}
    \label{Fig_flow_tau=0}
    \vspace{-.1cm}
\end{figure}

{Eliminating time from Eq.~\eqref{Eq_Law_RS} recovers $S\sim(\gamma+\alpha_NA)/(\alpha_N\sqrt R)$, consistent with Eq.~\eqref{SR_Relation}. Figure~\ref{Fig_flow_tau=0} illustrates this quasi-equilibrium curve.}

    \vspace{-.1cm}

\section{Mixing Net Effects and Critical Transitions}\label{Sec_advection}

\subsection{The parcel model}

In the previous section, the supersaturation equation was analyzed for an idealized adiabatic cloud parcel with fixed droplet concentration. In that setting, the only processes retained were the production of supersaturation by adiabatic cooling during an updraft and its depletion by condensational growth on droplets. Real cloud volumes are not closed thermodynamic boxes: supersaturation is also affected by transport and turbulent mixing. A more complete Eulerian balance, consistent with a density-weighted  transport formulation is discussed in \citep[e.g.][]{Clark1973}; \citep[Chap.~5.2]{khain_pinsky_2018}, and may be written compactly as
\be
\partial_t s = \kappa_\chi\nabla^2s + \gamma - A_2\dot{q}  - \frac{1}{\bar\rho} \partial_{x_i}\big(\bar\rho u_i s\big),
\ee
where $s$ is the local supersaturation, $\kappa_\chi$ is the molecular diffusivity, $\gamma$ represents the thermodynamic source associated with cooling, and $\dot{q}$ is the liquid-water production rate; see also {\it Supplementary Note 5, Step 4} and  \cite[Eq.~(5.2.49)]{khain_pinsky_2018}.

The link with a parcel description is obtained by expanding the conservative transport term:
\be
\frac{1}{\bar\rho} \partial_{x_i}\big(\bar\rho u_i s\big) = \ub\cdot\nabla s + s\,\frac{1}{\bar\rho} \partial_{x_i}\big(\bar\rho u_i\big).
\ee
Equivalently, introducing the material derivative following the local air motion,
\be
\frac{D^{\ub}s}{Dt} \equiv \partial_t s+\ub\cdot\nabla s,
\ee
the supersaturation balance becomes
\be\label{Eq_S_modified}
\frac{D^{\ub}s}{Dt} = \kappa_\chi\nabla^2s + \gamma - A_2\dot q_l - s\,\frac{1}{\bar\rho} \partial_{x_i}\big(\bar\rho u_i\big).
\ee
Thus, in a cloud element moving with the air, the resolved advective transport is absorbed into the parcel derivative. What remains locally is the production by cooling, the condensational sink, molecular or turbulent smoothing if retained, and a density-weighted transport term proportional to $s$.

At the present parcel level, we do not resolve the full Eulerian transport and filtering structure. Instead, we retain the leading bulk effect of environmental exchange by replacing the net density-weighted dilution, unresolved entrainment, and mixing-induced removal of supersaturated air by an effective linear relaxation term $-\lambda S$. The parameter $\lambda>0$ is therefore a bulk exchange rate, with units of inverse time, measuring how rapidly the co-moving parcel loses supersaturation to its environment.

This gives the reduced parcel model
\begin{subequations}
\label{Eq_growth_SR_lambda}
\begin{align}
\dot S &= \gamma - \alpha_N\sqrt{R}\big(S-f(R)\big) - \lambda S, \label{Eq_Ga} \\
\dot R &= 2D\big(S-f(R)\big). \label{Eq_Gb}
\end{align}
\end{subequations}
Here $R=r^2$ is the squared droplet radius, $f(R)$ is the K\"ohler equilibrium supersaturation, and $\alpha_N\sqrt{R}\big(S-f(R)\big)$ is the parcel-level condensational sink induced by the droplet population.  The term $-\lambda S$ is therefore not introduced as an independent microscopic force; it is the leading co-moving representation of density-weighted dilution and environmental exchange. It captures the net removal of supersaturation by mixing with the surrounding air, and $\lambda$ is {the corresponding effective removal rate in $[s^{-1}]$} due to the transport of supersaturated air out of the parcel or mixing with less saturated air.

The purpose of the present section is to isolate the effect of the mixing-induced removal term $-\lambda S$ on the long-time behavior of the parcel model. We first identify the natural threshold in $\lambda$ at which removal overwhelms sustained activation and the system is driven toward a haze-like state. This threshold provides the baseline around which the subsequent dynamical analysis is organized.

\subsection{Lower bound and critical slow down}\label{Sec_critical_slow_down}
The term $-\lambda S$ acts as a sink term in Eq.~\eqref{Eq_growth_SR_lambda}. Intuitively, if $\lambda$ is large enough, this  sink term can prevent the drops to activate, maintaining the overall bulk parcel into a haze state. 
We are after estimating a thresholding value of $\lambda$ for this to happen. 

The critical (squared) radius, given by $R_c = 3B/A$, discriminate between haze and activated states. 
For $R(t)$ to remain below $R_c$, a necessary condition is $\dot{R} \le 0$ as $R$ approaches $R_c$. Since $\dot{R} = 2D (S - f(R))$, this implies $S \le f(R)$ in a neighborhood of $R_c$.

Assuming a quasi-steady state for $S$, we have {$S \approx [\gamma + \alpha_N \sqrt{R} f(R)]/(\lambda + \alpha_N \sqrt{R})$}. Using this approximation, at $R = R_c$, the condition $S \le f(R)$, leads to 
$$ \gamma + \alpha_N \sqrt{R_c} f(R_c) \le f(R_c) (\lambda + \alpha_N \sqrt{R_c}),$$
that simplifies to $\gamma \le \lambda f(R_c)$.

\begin{figure}[htbp]
    \centering
   \includegraphics[height=0.25\textwidth, width=.5\textwidth]{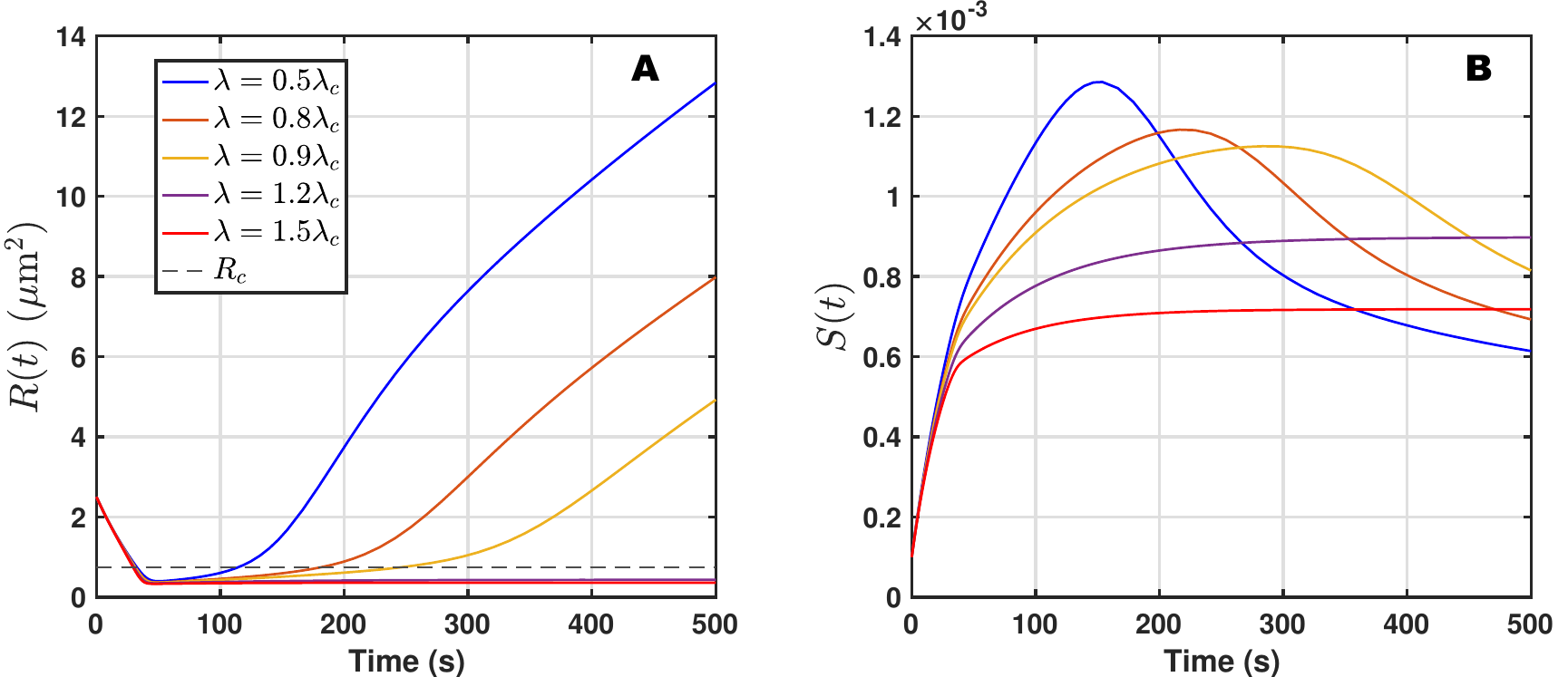}
    \caption{{\bf Lower bound verification, and critical slow down}.}
    \label{Fig_bound}
    \vspace{-.1cm}
\end{figure}

Since $f(R_c)= 2 A^{3/2}/(3 \sqrt{3B}),$ we infer that when 
\be\label{Eq_lower_bound}
\lambda \ge \lambda_c=\frac{3 \sqrt{3B} \gamma}{2 A^{3/2}},
\ee
the system \eqref{Eq_growth_SR_lambda} is likely to remain in a haze state. 
Note that the unit of the RHS in Eq.~\eqref{Eq_lower_bound} is in [$s^{-1}$], and is thus consistent to a rate. 
Figure \ref{Fig_bound}A illustrates the validity of this lower bound, where a few trajectories of $R(t)$, emanating from a common initial data,  are shown for different values of $\lambda$; three below $\lambda_c$ and two above. 
Figure \ref{Fig_bound}B shows the corresponding trajectories of the bulk supersaturation $S(t)$.
    \vspace{-.1cm}
Note that by  interpreting $\lambda$ as the inverse of a  timescale, $\tau_\lambda$, for homogenizing a parcel volume's supersaturation, the inequality \eqref{Eq_lower_bound} says in case of particles with radius $R>R_c$, that the suppression of activation is more effective as $\tau_\lambda$ gets smaller. 

\br\label{Rem_Mixing_Rate_Baker}
When $\lambda$ is near the lower bound, $\lambda_c$, the balance between the removal of supersaturation and the conditions allowing droplet growth past $R_c$ follows a delicate nonlinear relation.  A slight increase in $\lambda$ due to a perturbation would more effectively suppress supersaturation, hindering droplet growth and favoring the maintenance of a haze. Conversely, a slight decrease in $\lambda$ could allow supersaturation to build up enough for some droplets to potentially overcome the activation barrier at $R_c$. This is for instance manifested for the blue, orange and yellow curves shown in Fig.~\ref{Fig_bound}B presenting finite time intervals over which supersaturation grows, eventually leading to droplet activation. 

There is an interpretation from  the literature of $\lambda$ as $\lambda=\tau_m^{-1}=(\epsilon/L^2)^{1/3}$ \citep{baker1980} where $\epsilon$ is the kinetic energy dissipation rate and $L$ is the characteristic length of the air volume of the cloud parcel to which the equations are applicable. The timescale $\tau_m$ is understood as a characteristic timescale for homogenization of this cloud parcel due to mixing.

For small clouds in the transition zone,  easily switching to haze, if one takes $\epsilon=10^{-4}m^2 s^{-3}$ and $L=10m$, we find $\tau_m^{-1}=0.01 s^{-1}$. On the other hand, with the parameter values in Table \ref{Fig_flow_tau=0}, we find from   Eq.~\eqref{Eq_lower_bound} that $\lambda_c\approx 0.011 s^{-1}.$

Both approaches give a similar estimate of the critical threshold. It says that small clouds for which these numbers apply, are indeed susceptible to easily switching between a state allowing some droplet activation and a haze state in response to minor fluctuations in the turbulent mixing (and hence, in the effective $\lambda$). This aligns with the observation that these clouds are often unstable and sensitive to environmental conditions \citep{hirsch2014transition}.  Section \ref{Sec_reducedSDE_main} below elaborates on this aspect. 
\er
\subsection{Mixing-induced steady states}\label{Sec_steady_states}
To understand the conditions where the cloud system reaches a state of balance (steady states), we analyze when the rates of change of supersaturation and droplet size become both zero, i.e.~when $\gamma = \lambda f(R)$ and $S=f(R)$. This leads, due to Eq.~\eqref{Eq_Kohler}, to the following cubic equation for the droplet radius $r=\sqrt{R}$:
\begin{equation}\label{Eq_cubic}
P_{\lambda}(r)=\gamma r^3 - \lambda Ar^2 + \lambda B = 0.
\end{equation}
The solutions to this equation represent potential steady-state droplet sizes. The nature and number of these solutions depend on the discriminant $\Delta$ of this cubic equation ($\Delta  =\lambda^2 B(4\lambda^2 A^3 - 27\gamma^2 B)$) whose sign is determined by the parameter $\lambda$,  the removal rate of supersaturation. 

More precisely, a sign change of $\Delta$ occurs under a critical condition ($\lambda=\lambda_c$), where
\begin{equation}\label{Eq_discriminant}
\lambda_c^2 = \frac{27\gamma^2 B}{4A^3} = \bigg(\frac{\gamma}{f(R_c)}\bigg)^2,
\end{equation}
with $f(R_c)=2 A^{3/2}/(3 \sqrt{3B})$. 
Physically, this condition, also identified in Eq.~\eqref{Eq_lower_bound}, marks a threshold related to the balance between the updraft strength (proportional to $\gamma$) and the microphysical properties of the aerosols (related to $A$ and $B$). 
{For $\lambda<\lambda_c$, $\gamma/\lambda$ exceeds the maximum of $f$, so there is no positive equilibrium radius. The cubic has one real root, which is negative and physically inadmissible.}   

In the case  $\lambda>\lambda_c$, the situation becomes physically richer with the balance polynomial $P_{\lambda}$ possessing now three real roots {($\Delta>0$)}. 
These are found using the trigonometric form of Cardano's method, after transforming $P_{\lambda}$ into a depressed form $z^3 +p z +q$. 
It shows that when the removal rate $\lambda$ is sufficiently large, specifically when $\lambda>\lambda_c$, the three real roots are given by:
\begin{equation}\label{Eq_yk}
r_j = \frac{\lambda A}{3\gamma} \left(2 \cos\left(\frac{\phi  +2\pi j}{3}\right) + 1\right), \; j=0,1,2,
\end{equation}
 with $\phi=\arccos\left(1 - \frac{27\gamma^2 B}{2\lambda^2 A^3}\right)$. 
Only two of these roots  are physically  admissible (non-negative) steady droplet sizes. 
One steady state corresponds to very small, stable haze droplets (associated with the $r_2$-root).
The other steady state represents a larger droplet size ($R^{\ast}$ from the $r_0$-root) identified below as a saddle-point, i.e.~a linearly unstable steady state with mixed directions of stability. 

However, our analysis below reveals that the existence of this saddle point has a greater impact on the droplet size distribution than could be anticipated by a simple linear analysis. This will be particularly apparent when 
turbulence and its interaction with the system's inherent nonlinear effects are brought to the picture. 
For the moment, we analyze the consequences of this type of instability under these nonlinear effects.  

\subsection{Mixing-induced critical transition}\label{Sec_saddle_node_bif}
The critical slowdown pointed out in Figure \ref{Fig_bound} as $\lambda$ approaches $\lambda_c$ from below  signals the approach of a critical transition. 

To understand this transition, we analyze the stability of the emerging steady states for $\lambda>\lambda_c$. To do so,  we use the system's Jacobian matrix $J(\lambda)$
 \bes
 J(\lambda) =\begin{pmatrix} -\alpha_N \sqrt{R} - \lambda & \alpha_N \sqrt{R} f'(R) \\ 
2D & -2D f'(R) \end{pmatrix}.
\ees
 whose eigenvalues $\beta_{\pm}(\lambda)$ determine the stability. These eigenvalues depend on the trace and determinant of the Jacobian according to:
 \be\label{Eq_eig_gene}
\beta_{\pm}(\lambda) = \frac{\mbox{Tr}[J(\lambda)] \pm \sqrt{\mbox{Tr}[J(\lambda)]^2 - 4\mbox{Det}[J(\lambda)]}}{2}.
\ee

At the critical point $\lambda=\lambda_c$, when  $R=R_c$ and $S=S_c=f(R_c)$, since $f'(R_c)=0$, the Jacobian simplifies significantly ($ \mbox{Det}[J(\lambda_c)]=0$), leading to eigenvalues $\beta_+^*(\lambda_c)=0$ and $\beta_{-}^*(\lambda_c)=-\alpha_N \sqrt{R_c} - \lambda_c$. The zero eigenvalue is a hallmark of a {\it saddle-node bifurcation.}

    \vspace{-.1cm}
\begin{figure}[htbp]
    \centering
       \includegraphics[height=0.3\textwidth, width=.23\textwidth]{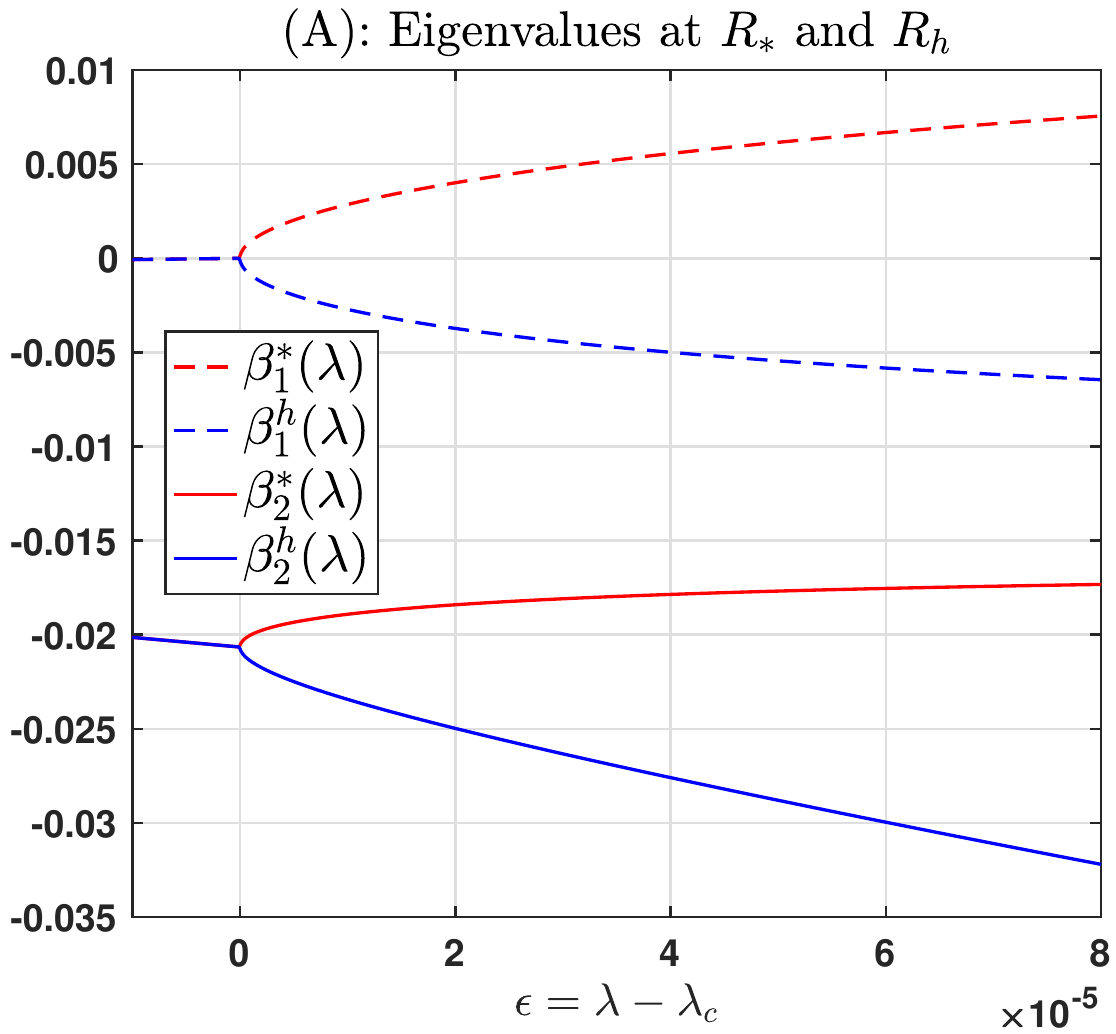}
   \includegraphics[height=0.3\textwidth, width=.23\textwidth]{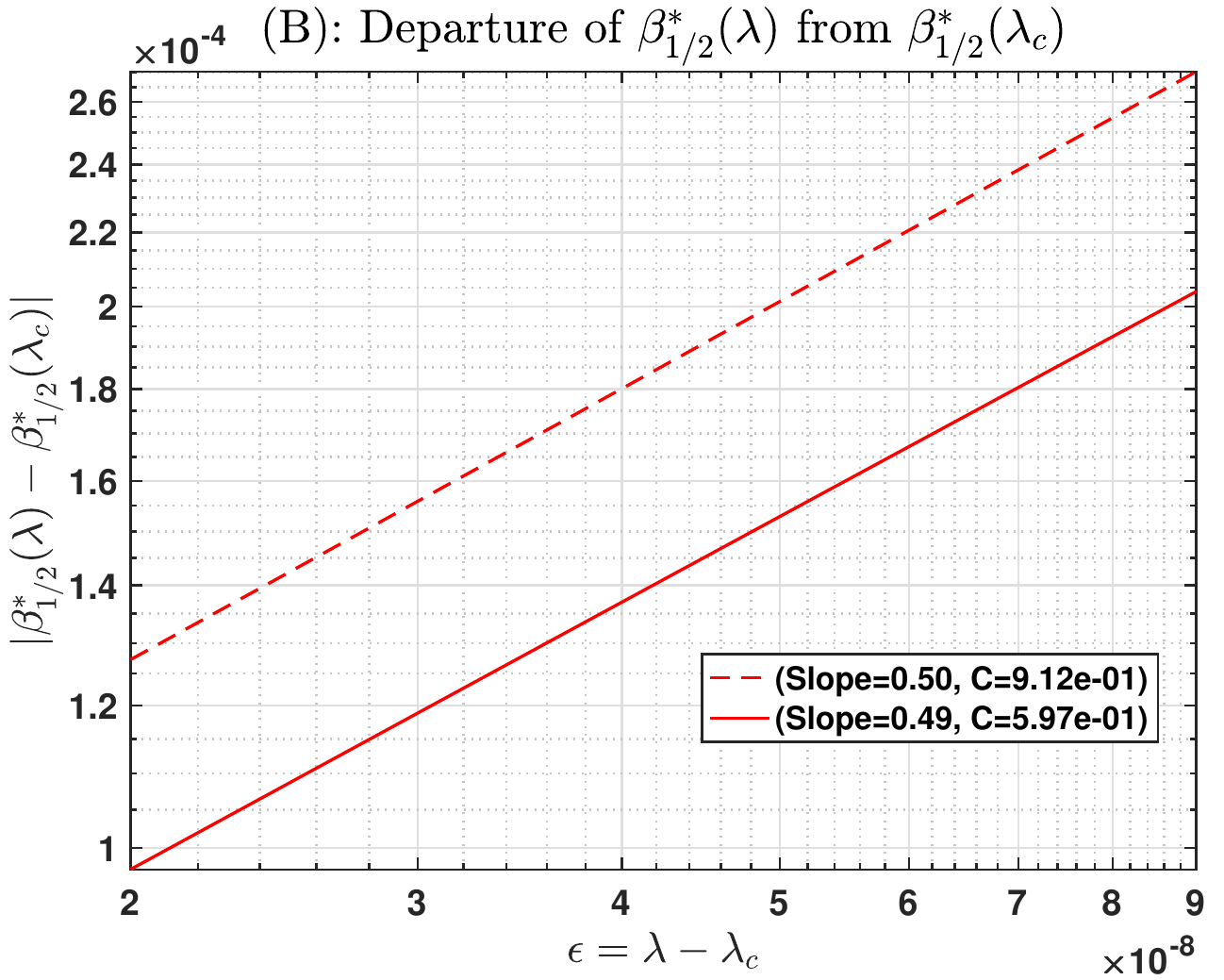}
    \caption{{\bf Eigenvalue dependence on $\epsilon$}. The eigenvalues are computed from Eq.~\eqref{Eq_eig_gene} for the parameter regime of Table \ref{Table_RegimeA}. The red curves (dashed and plain) shown in panel A are fitted as $C\epsilon^{\alpha}$ curves. The results are shown in panel B in a log-log plot, with the slope ($\alpha$) and constant $C$ as indicated in the legend.}
    \label{Fig_Eig_dep}
    \vspace{-.1cm}
\end{figure}

To explore the behavior near this critical point, we examined how the eigenvalues evolve for small deviations $\epsilon=\lambda-\lambda_c$. As shown in Figure \ref{Fig_Eig_dep}A, the eigenvalues $\beta_1^h$ and $\beta_2^h$ associated with the haze steady state $(R_h,S_h)$, as well as the eigenvalues $\beta_1^*$ and $\beta_2^*$ associated with the other steady state $(R^*,S^*)$, exhibit a scaling proportional to $\sqrt{\epsilon}$ as we move away from $\lambda_c$. This $\sqrt{\epsilon}$ scaling, clearly visible in the log-log plot in Figure \ref{Fig_Eig_dep}B, is a characteristic signature of a saddle-node bifurcation, as can be seen from the normal form $\dot{x} = - \mu + x^2$ where eigenvalues scale with $\sqrt{\mu}$ near the bifurcation ($r \sim \epsilon$) \citep{guckenheimer_book}.

Thus, when the removal of supersaturation is strong enough to overcome the tendency for the updraft to continuously increase supersaturation, a critical transition occurs in the system, corresponding to a saddle-node bifurcation. 
The presence of this bifurcation has a profound impact on the local system's behavior.  
This is furthermore enriched with another essential, more global, aspect of the system's nonlinear dynamics, its  slow manifold, that we analyze next. 

\section{Slow Manifold and Supersaturation's Rapid Return to Balance}\label{Sec_expo_attraction}

{In this section, we identify the supersaturation nullcline $\dot S=0$ as a leading quasi-equilibrium approximation to the slow manifold and quantify tracking of this moving balance. The nullcline itself is generally not invariant.} Before addressing the full model with mixing, it is instructive to consider the standard adiabatic equations recalled above (Eq.~\eqref{Eq_Korolev_rectif}). Even in this simplified classical framework, a dynamic balance is immediately identifiable. By setting $\dot{S}=0$ in Eq.~\eqref{Eq_Korolev_rectif}, we find that the supersaturation is tethered to a specific curve:
\be\label{Eq_balance_S_no_mixing}
S = g_0(R) = f(R) + \frac{\gamma}{\alpha_N \sqrt{R}},
\ee
whose approximation for $R$ sufficiently large is given by Eq.~\eqref{SR_Relation} derived above. 

It is remarkable that this curve, which we designate as the adiabatic slow manifold, has remained largely implicit in the literature despite the ubiquity of Eq.~\eqref{Eq_Korolev_rectif}. From Eq.~\eqref{Eq_balance_S_no_mixing}, it is obvious that $g_0(R) > f(R)$ for all $R>0$, confirming that the system inherently maintains a supersaturation level above the equilibrium (K\"ohler) curve to drive condensation. This gap, $\gamma/(\alpha_N \sqrt{R})$, represents the "excess" supersaturation required to offset the phase transition sink. 

The slow nature of this manifold is rooted in a sharp separation of timescales. While the radius $R(t)$ and the manifold $g_0(R(t))$ evolve according to the power laws derived in Eq.~\eqref{Eq_Law_RS} (specifically $R \sim t^{2/3}$ and $S \sim t^{-1/3}$), {at fixed $R$, supersaturation perturbations relax at rate $\eta_0=\alpha_N\sqrt R$. Along an evolving trajectory, motion of $g_0(R)$ supplies an additional tracking term, as derived below.} Using the asymptotic limit for $R(t)$, this relaxation rate scales as $\eta_0(t) \sim \alpha_N^{2/3} (3D \gamma)^{1/3} t^{1/3}$. Consequently, as the droplets grow, the "fast" relaxation toward the manifold becomes increasingly dominant over the "slow" power-law evolution of droplet growth, ensuring that the system remains tightly locked to the balance condition.

\begin{figure*}[t!]
    \centering
   \includegraphics[height=0.42\textwidth, width=.85\textwidth]{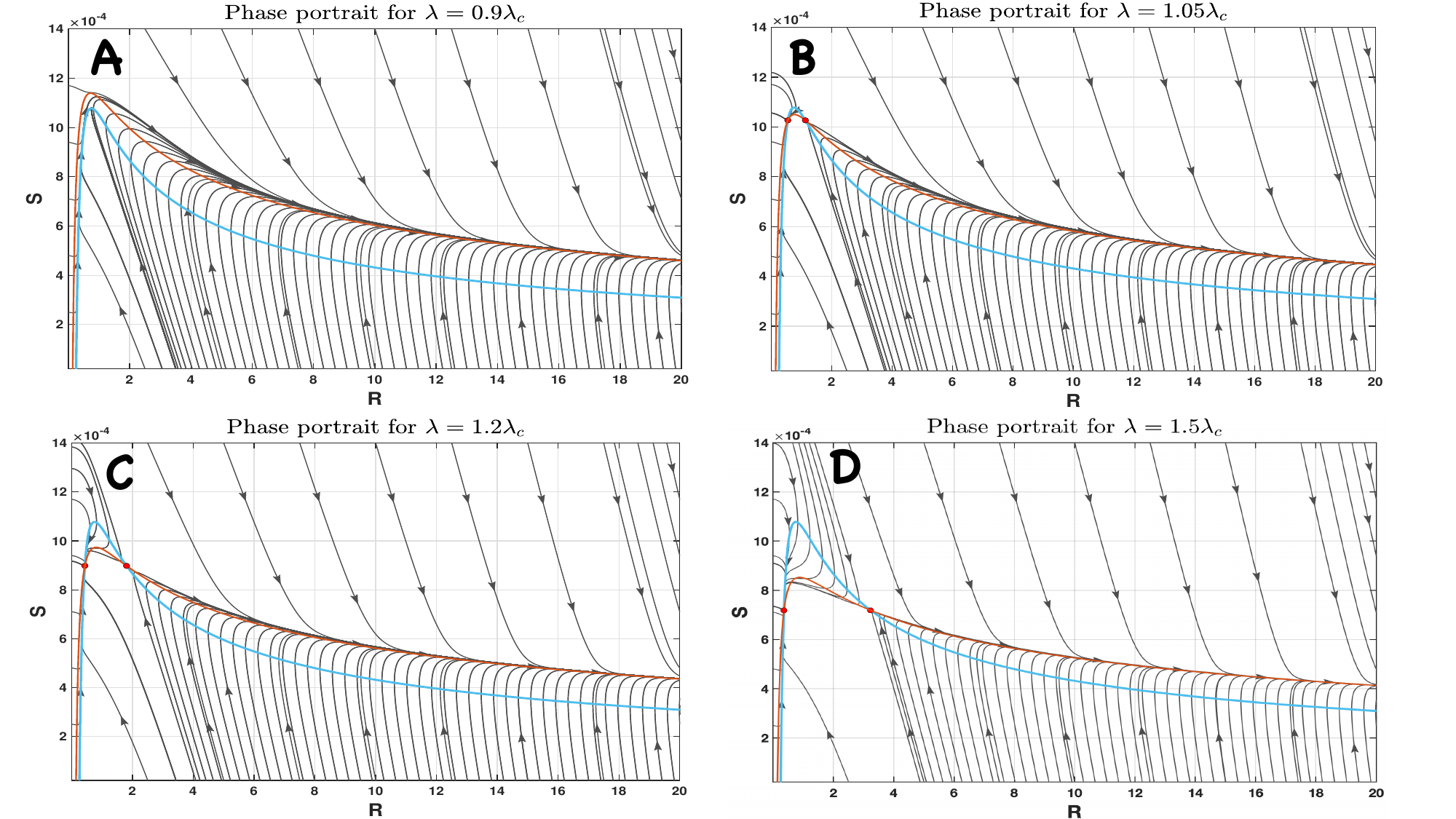}
    \caption{{\bf Phase portrait dependence on $\lambda$, and saddle-node bifurcation}. As $\lambda$ crosses from below $\lambda_c$, we observe the emergence of two steady states marked by red dots. These, correspond to a steady state ($R_h,S_h$) on the haze branch  with $S_h=f(R_h)$ and $R_h=y_2^2$, and a steady state ($R^\ast,S^\ast$), with $R^\ast=y_0^2$ and $S^\ast=f(R^\ast)$, where $y_0$ and $y_2$ are given by the analytic formulas Eq.~\eqref{Eq_yk} for $j=0$ and $j=2$, respectively.  In each panel, the red curve is given by the balance relation Eq.~\eqref{Eq_balance_S} (slow manifold $g(R)$), while the cyan curve is the K\"ohler curve (Eq.~\eqref{Eq_Kohler}). The parameter regime is that of Table \ref{Table_RegimeA}. Note the emergence of the "forbidden zone"---$R$-range for which $g(R)<f(R)$---as $\lambda$ moves away from $\lambda_c$, i.e.~for sufficiently strong mixing.}
    \label{Fig_PhasePortrait_dep}
    \vspace{-.1cm}
\end{figure*}

When we introduce the mixing rate $\lambda$, this fundamental structure is preserved but shifted.  From Eq.~\eqref{Eq_Ga}, we infer that this generalized balance is represented by the relationship:
\be\label{Eq_balance_S}
S=g(R)=\frac{\gamma+\alpha_N \sqrt{R}f(R)}{\alpha_N\sqrt{R} +\lambda}.
\ee
Comparing Eq.~\eqref{Eq_balance_S} to Eq.~\eqref{Eq_balance_S_no_mixing} reveals that mixing acts as an additional sink ($-\lambda S$) that drags the slow manifold $g(R)$ to lower values than its adiabatic counterpart $g_0(R)$. Specifically, as $\lambda$ increases, the manifold $g(R)$ is pushed down toward the equilibrium curve $f(R)$, and can even drop below it over some $R$-range corresponding to the K\"ohler's "forbidden zone,"  if the mixing with dry air is sufficiently strong; see Figure \ref{Fig_PhasePortrait_dep}.

\begin{figure}[htbp]
    \centering
       \includegraphics[height=0.48\textwidth, width=.43\textwidth]{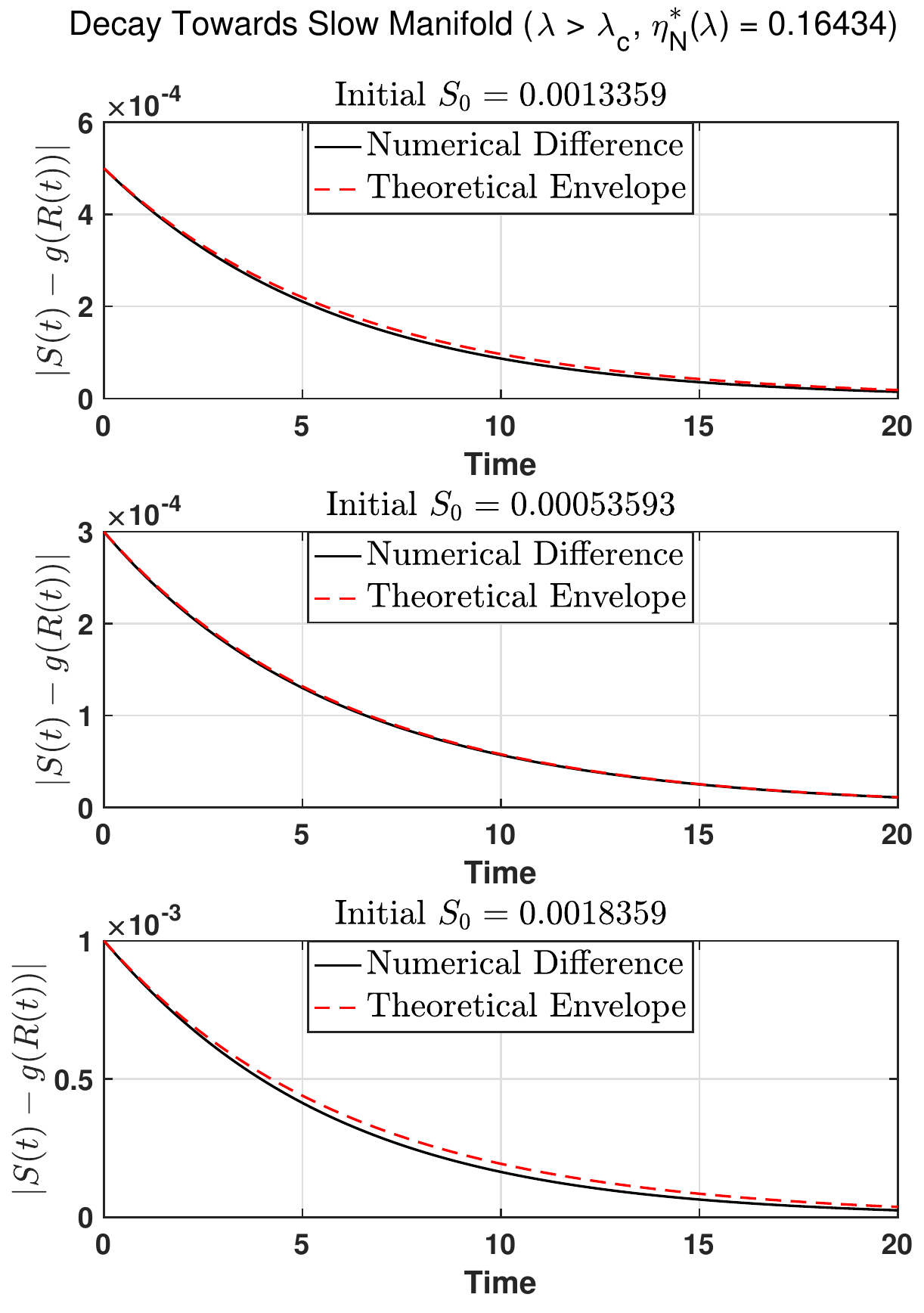}
    \caption{{\bf Decay rate towards the slow manifold}.  The initial data correspond, from top to bottom,  to $S_0=g(R_0)+5\cdot 10^{-4}$, $S_0=g(R_0)-3\cdot 10^{-4}$, and $S_0=g(R_0)+ 10^{-3}$, while in each case,  $R_0=3 R_c.$ The parameter regime is that of Table \ref{Table_RegimeA}, except $N=500 cm^{-3}${, $\gamma=1.5\cdot10^{-3}\,\%\,s^{-1}$,} and $\lambda=1.2 \lambda_c${, with $\lambda_c$ given by Eq.~\eqref{Eq_lower_bound} for this value of $\gamma$}. {The exponential guide uses $\eta_N^*(\lambda)$ from Eq.~\eqref{Eq_rate} with $\epsilon=0$ as a reference rate for the initial transient. It is not a global bound on the full tracking error, which also contains the moving-balance contribution in Eq.~\eqref{Eq_attraction}.}}
    \label{Fig_Decay_rate}
    \vspace{-.1cm}
\end{figure}

{The rapid adjustment toward $g(R)$ is a tracking property rather than exact attraction to an invariant graph. Set $E=S-g(R)$ and $q(R)=g(R)-f(R)$. The balance defining $g$ gives
\be\label{Eq_balance_gR}
\gamma-\alpha_N\sqrt R\,[g(R)-f(R)]-\lambda g(R)=0,
\ee
and therefore
\be\label{Eq_SE}
\dot S=-(\alpha_N\sqrt R+\lambda)E.
\ee
Differentiating $E$ along the coupled dynamics yields the exact identity
\be\label{Eq_Edot}
\dot E=-\eta(R)E+b_{\rm tr}(R),\quad
\eta(R)=\alpha_N\sqrt R+\lambda+2Dg'(R),
\ee
where $b_{\rm tr}(R)=-2Dg'(R)q(R)$ is the forcing due to motion of the quasi-equilibrium curve. In the growing regime with $g'<0$ and $q>0$, this forcing is positive; it cannot be dropped to obtain an upper decay bound.

For any trajectory segment on which $\eta(R(t))\ge\eta_{\min}>0$, variation of constants gives
\bea\label{Eq_ineqE}
E(t)&=e^{-\int_{t_0}^{t}\eta(R(u))\,\d u}E(t_0)\\
&\quad+\int_{t_0}^{t}e^{-\int_s^t\eta(R(u))\,\d u}b_{\rm tr}(R(s))\,\d s,
\eea
hence
\bea\label{Eq_attraction}
|E(t)|&\le e^{-\eta_{\min}(t-t_0)}|E(t_0)|\\
&\quad+\int_{t_0}^{t}e^{-\eta_{\min}(t-s)}|b_{\rm tr}(R(s))|\,\d s.
\eea
The initial mismatch decays exponentially, while the second term measures the persistent lag behind the moving balance. Since $g'(R)=\mathcal O(R^{-3/2})$ and $q(R)=\mathcal O(R^{-1/2})$ at large $R$, $b_{\rm tr}=\mathcal O(R^{-2})$, whereas $\eta\sim\alpha_N\sqrt R$. The leading tracking offset is consequently $b_{\rm tr}/\eta=\mathcal O(R^{-5/2})$ under slow variation of these coefficients.

For sufficiently large $R$, one may use the lower rate estimate
\be\label{Eq_rate}
\eta(R)\ge\eta_N^*(\lambda)=\alpha_N\sqrt{R^*}+\lambda(1-\epsilon),
\ee
for any fixed small $\epsilon>0$. This asymptotic estimate does not assert a uniform bound from the initial time for every $R(0)>R^*$. Figure~\ref{Fig_Decay_rate} illustrates rapid initial relaxation; its exponential guide describes the homogeneous transient in Eq.~\eqref{Eq_attraction}, not the full tracking error.}

    \vspace{-.1cm}

\section{Turbulent Mixing Effects: Stochastic Model}\label{Sec_stochastic}
{The rapid tracking of the quasi-equilibrium curve,} with a rate, $\eta_N^\ast(\lambda)$, that is sensitive to aerosol number concentration ($N$) and the supersaturation removal rate ($\lambda$), generally promotes a stable relationship between supersaturation and droplet size. This allows the gradual process of droplet growth to dominate the system's evolution. Nevertheless, the presence of a saddle equilibrium $(R^*,S^*)$ introduces a crucial element of instability, particularly when considering the turbulent nature of clouds and the inherent fluctuations in mixing. The interplay between these slow-fast processes and the saddle point determines the system's response to such perturbations.

To better understand this interplay under the presence of turbulent fluctuations, we consider the following stochastic version of Eq.~\eqref{Eq_growth_SR_lambda}:
\begin{subequations}\label{Eq_growth_SR_lambda_stoch}
\begin{align}
&\dot{S} = \gamma-\alpha_N \sqrt{R}\Big(S -f(R)\Big)-\lambda S +\sigma S {\circ}\dot{W}_t\label{Eq_Gas}\\
&\dot{R} = 2D\Big(S-f(R)\Big),\label{Eq_Gbs}
\end{align}
\end{subequations}
where  $\dot{W}_t$ denotes a white noise process{, and $\circ$ denotes the Stratonovich interpretation of the stochastic term \citep{gardiner2009stochastic}; the equivalent It\^o form is given in Eq.~\eqref{Eq_Gas_Ito} below}.

In the deterministic parcel model (Eq.~\eqref{Eq_growth_SR_lambda}), the net effect of environmental exchange on the bulk supersaturation was represented by the linear removal term $-\lambda S$. {We interpret it as a reduced environmental-exchange law. The density-weighted divergence in Eq.~\eqref{Eq_S_modified} vanishes under an anelastic constraint, $\nabla\cdot(\bar\rho\ub)=0$, and does not by itself derive this law. Exchange with an environment at supersaturation $s_{\rm env}$ would more generally give $-\lambda(S-s_{\rm env})$. Here its mean environmental contribution is absorbed into the prescribed net source, while only fluctuations proportional to $S$ are retained. Writing the unresolved exchange rate as a mean plus a rapidly fluctuating residual motivates $-\lambda S+\sigma S\circ\dot W_t$. This is a multiplicative exchange closure; fluctuations associated with a nonzero, variable $s_{\rm env}$ would require additional additive or correlated forcing.}

The fluctuations of the mixing rate are carried by turbulent eddies and therefore have a finite correlation time, of the order of the Kolmogorov  timescale $\tau_K$ (Appendix~\ref{Sec_Turb_Cascade}). Their white-noise limit is of Stratonovich type \citep{wong1965convergence}, which is the interpretation indicated in Eq.~\eqref{Eq_Gas} and adopted throughout the paper, including the field theory of Section~\ref{Sec_Field_Theory}. In It\^o form, Eq.~\eqref{Eq_Gas} reads
\bea\label{Eq_Gas_Ito}
\d S &= \Big[\gamma-\alpha_N \sqrt{R}\big(S -f(R)\big)-\lambda^* S\Big]\d t +\sigma S \,\d W_t,\\
\lambda^*&=\lambda-\frac{\sigma^2}{2},
\eea
so that the multiplicative fluctuations reduce the effective removal rate. The simulations of Sections~\ref{Sec_reducedSDE_main} and~\ref{Sec_Full_Parcel_Dynamics} integrate this It\^o form with the Euler--Maruyama scheme. The removal rate quoted there, and the threshold $\lambda_c$ of Section~\ref{Sec_advection} that it is compared with, therefore refer to the dressed rate $\lambda^*$; the corresponding bare rate is $\lambda^*+\sigma^2/2$. In particular, the bare mixing threshold for the saddle-node transition is $\lambda_c+\sigma^2/2$, which is the parcel counterpart of the field-level shift derived in Eq.~\eqref{Eq_lambda_critical_dressed} below, with $\sigma^2/2$ in place of $\sigma_S^2/\tau_K$. For the parameters of Table~\ref{Table_RegimeA}, $\lambda_c\approx 0.011\,{\rm s}^{-1}$, and $\sigma^2/2$ ranges from about $0.45\lambda_c$ for $\sigma=0.1$ to about $5.5\lambda_c$ for $\sigma=0.35$: in the regimes where hovering occurs, the turbulent dressing of the mixing threshold is not a small correction.

As will be clarified below, the multiplicative character of the noise disturbance is key to achieving realistic effects of turbulence within our simplified parcel model framework.
Physically, this type of noise can be interpreted as "parameter noise," encountered in many areas of physics, including turbulence theory (intermittency phenomena) \citep{Birnir13}, statistical physics \citep{schenzle1979multiplicative}, or non-equilibrium phase transitions \citep[e.g.][]{Cross_Hohenberg93,garcia1993effects,Hin00,SSG07,CLW15_vol1,CLW15_vol2,chekroun2023transitions}, where the intensity of fluctuations often depends on the state of the system.

It is noteworthy that the system given by Eq.~\eqref{Eq_growth_SR_lambda_stoch} is not a gradient system, and thus does not possess an "energy" landscape that could facilitate its analysis. Indeed, we observe that 
$$ \partial_R \dot{S} = \frac{\alpha_N}{2\sqrt{R}} (f(R) - S) + \alpha_N \sqrt{R} f'(R),$$ 
and $\partial_S \dot{R} = 2D$. Thus it cannot exist a function $V_{\lambda}$ such that $-\partial_S V_{\lambda} = \dot{S} $, and $ -\partial_R V_{\lambda} = \dot{R},$ as we have  $ \partial_{SR}^2 V_{\lambda}\neq \partial_{RS}^2 V_{\lambda}$. 

The slow manifold represents a quasi-equilibrium where the fast variable $S$ has adjusted to the slowly evolving variable $R$. Near the slow manifold, the dynamics are dominated by the attraction towards it (fast mode) and the motion along it (slow mode).

While the system exhibits a form of "potential" in the sense that the slow manifold attracts trajectories, the two-variable  system (Eq.~\eqref{Eq_growth_SR_lambda_stoch}) in the vicinity of the slow manifold is not expected to be a perfect gradient system, as the conditions for this to happen are not met for the original system. 

However, the evolution of $R$,  {\it on the slow manifold},  can be represented as a gradient flow in the one-dimensional space of $R$. The multiplicative nature of the noise, however, requires working a bit more to benefit from the interpretation of a Brownian particle evolving within an energy landscape. 

\section{Stochastic Reduced Dynamics on the Slow Manifold}\label{Sec_reducedSDE_main}

\subsection{Stochastic droplet growth reduced equation}\label{Sec_reducedSDE_der}

Formally, the droplet growth on the slow manifold is given by Eq.~\eqref{Eq_Gbs}, with $S$ replaced by $g(R)$, which leads,  using the expression for $g(R)-f(R)$ in the ``Useful Properties" paragraph of {\it Supplementary Note 1}, to the following model,
\be\label{Eq_SDE_master}   
\dot{R} = 2D \frac{\gamma - (\lambda + \sigma \dot{W}_t) f(R)}{\alpha_N\sqrt{R} +\lambda + \sigma \dot{W}_t}.
\ee
The stochastic noise in Eq.~\eqref{Eq_SDE_master}  is introduced as a parametric white noise, to reflect the uncertainties  in the removal rate  $\lambda$ caused by turbulent fluctuations and modeled  by the stochastic term in the  supersaturation equation  (Eq.~\eqref{Eq_Gas}) of the full model. 

We want to express this ODE with random coefficients into a standard SDE in its It\^o  form \cite{Oksendal_book,gardiner2009stochastic}:
\be\label{Eq_SDE_target} 
\d R = a(R) \d t + b(R)  \d W_t,
\ee
where $\d W_t$ represents an infinitesimal increment of a Wiener process (or Brownian motion) over an infinitesimal time interval $\d t$.  In Eq.~\eqref{Eq_SDE_target}, and in the language of stochastic analysis, $a(R)$ is called the drift coefficient and $b(R)$ denotes the diffusion coefficient. We use  
a standard small-noise expansion \citep{gardiner2009stochastic,freidlin2012random} to determine these terms from Eq.~\eqref{Eq_SDE_master}. 

Consider the denominator as $(\alpha_N\sqrt{R} +\lambda) + \sigma \dot{W}_t$. We can use the approximation $\frac{1}{X + \epsilon} \approx \frac{1}{X} - \frac{\epsilon}{X^2}$ for small $\epsilon$. Here, $X= \alpha_N\sqrt{R} +\lambda$ and $\epsilon = \sigma \dot{W}_t$, which gives 
    $$ \frac{1}{\alpha_N\sqrt{R} +\lambda + \sigma \dot{W}_t} \approx \frac{1}{\alpha_N\sqrt{R} +\lambda} - \frac{\sigma \dot{W}_t}{(\alpha_N\sqrt{R} +\lambda)^2}, $$
leading from Eq.~\eqref{Eq_SDE_master}  to 
    $$ \frac{\dot{R}}{2D} = [\gamma - (\lambda + \sigma \dot{W}_t] f(R)) \left[\frac{1}{\alpha_N\sqrt{R} +\lambda} - \frac{\sigma \dot{W}_t}{(\alpha_N\sqrt{R} +\lambda)^2} \right],$$
up to a small error. By expanding this product, we arrive at
 \beas
\frac{\dot{R}}{2D} =  &  \frac{\gamma - \lambda f(R)}{\alpha_N\sqrt{R} +\lambda} - \frac{\sigma \dot{W}_t \gamma}{(\alpha_N\sqrt{R} +\lambda)^2} \\
    &+ \frac{\sigma \dot{W}_t \lambda f(R)}{(\alpha_N\sqrt{R} +\lambda)^2} - \frac{\sigma \dot{W}_t f(R)}{\alpha_N\sqrt{R} +\lambda} + \frac{\sigma^2 (\dot{W}_t)^2 f(R)}{(\alpha_N\sqrt{R} +\lambda)^2}.
\eeas
We now keep terms up to the first order in $\sigma$ and neglect the term in $\sigma^2$, which gives: 
    $$   \frac{\dot{R}}{2D} =  \frac{\gamma - \lambda f(R)}{\alpha_N\sqrt{R} +\lambda} - \sigma \dot{W}_t \left[ \frac{\gamma - \lambda f(R)}{(\alpha_N\sqrt{R} +\lambda)^2} + \frac{f(R)}{\alpha_N\sqrt{R} +\lambda} \right].$$
By replacing $\dot{W}_t \d t$ with $ \d W_t$ and the expression of $g(R)-f(R)$ ({\it Supplementary Note 1}), we arrive at: 
\bes
\d R =2D  \frac{\gamma - \lambda f(R)}{\alpha_N\sqrt{R} +\lambda} \d t  -2D \sigma \frac{g(R)}{\alpha_N\sqrt{R} +\lambda}  \d W_t,
\ees
which can be written as
\be\label{Eq_reduced_SDE}
\hspace{-1.5ex}\boxed{\d R = 2D  (g(R)-f(R)) \d t   -2D \sigma  \frac{g(R)}{\alpha_N\sqrt{R} +\lambda}  \d W_t,\hspace{-0.5ex}}
\ee
providing the reduced SDE of droplet growth on the slow manifold.
{Here $\lambda$ stands for the dressed rate $\lambda^*$ of Eq.~\eqref{Eq_Gas_Ito}. The expansion fixes the drift at $\mathcal O(\sigma^0)$ and the noise at $\mathcal O(\sigma)$; the $\mathcal O(\sigma^2)$ part of the drift, which includes the contribution of the discarded $(\dot W_t)^2$ term, is not determined by it. We adopt the It\^o form \eqref{Eq_reduced_SDE}, consistent with the integration of the full model, and examine at the end of Section~\ref{Sec_Lyapunov_exp} which conclusions depend on this choice.}
 
 The drift term, $2D  (g(R)-f(R))$, accounts for the deterministic droplet growth on the slow manifold, driven by the difference between the quasi-equilibrium supersaturation, $g(R)$ (Eq.~\eqref{Eq_balance_S}), resulting from the slow-fast nonlinear dynamics, and the equilibrium supersaturation, $f(R)$,  governed by pure molecular diffusion.

The stochastic term  in  Eq.~\eqref{Eq_reduced_SDE} introduces random kicks to the droplet (squared) radius, $R$, at each infinitesimal time step. The magnitude of these kicks is proportional to the noise strength $\sigma$, the state of the supersaturation on the slow manifold ($g(R)$), and inversely proportional to the rate of adjustment to the slow manifold, $\alpha_N\sqrt{R} +\lambda$ (see Eq.~\eqref{Eq_SE}). The direction of the kick is random (positive or negative), reflecting the chaotic nature of turbulence. It represents  a way to mathematically represent the unpredictable, small-scale forcing that turbulence exerts on the cloud droplet's environment (here the bulk supersaturation $S$), leading to stochastic variations in its growth.

The term $\alpha_N\sqrt{R} +\lambda$  in the denominator of the diffusion term increases with $N$. This means that for a given $\sigma$, the magnitude of the stochastic fluctuations in $R$ (due to turbulence) becomes smaller in more polluted clouds.  Consequently, to achieve a similar level of stochastic forcing required to push droplets away from the haze state (small $R$) towards the unstable, precarious, equilibrium $R^*$ (near the activation barrier) in more polluted clouds, a larger value of  $\sigma$ (representing stronger turbulence) would be necessary. 

Thus, polluted clouds would require a higher level of turbulence for "twilight clouds" (with droplet radii near the activation barrier) to appear away from the haze state and to exhibit small activated islands as can be observed  in the lower left part of Fig.~\ref{Fig_Twilight}b ("mushroom-shaped" activated zone).
This highlights the complex interplay between aerosol concentration and turbulence in the formation of these elusive clouds.
{The local drift rates provide a quantitative diagnostic;} see Section \ref{Sec_Lyapunov_exp} below. To lead us to this analysis, we recast below the reduced SDE, Eq.~\eqref{Eq_reduced_SDE}, into a simpler form, more prone to analysis. 
\subsection{Edge state, Lamperti potential landscape, and hovering droplets}\label{Sec_Lamperti}
The reduced stochastic dynamics in Eq.~\eqref{Eq_reduced_SDE} involves state-dependent (multiplicative) noise, which complicates the direct intuition of the droplet's behavior as a motion within a potential well. To make the "twilight" dynamics more accessible, we map our system into the "Lamperti world." Mathematically, the Lamperti transformation \citep{lamperti1962semi} is a change of variables that effectively "unfolds" the coordinate space such that the diffusion coefficient becomes constant (specifically, equal to unity). Physically, this allows us to recast the complex interaction between nonlinear growth and turbulent fluctuations as a simple Brownian particle evolving within an energy landscape. {The invertible transformation preserves the stochastic process on such an interval, but its It\^o drift contains a correction. Potential extrema need not coincide with the deterministic parcel equilibria, nor preserve their number.}

In navigating this landscape, two critical radii serve as our primary coordinates. The first is the classical K\"ohler critical radius, $R_c$, which marks the peak of the equilibrium K\"ohler curve. The second, and more central to our stochastic analysis, is the bifurcating  state $R^*$. While $R_c$ is a fixed property of the aerosol's chemistry and size, $R^*$ is a threshold emerging from the nonlinear coupling of microphysics with the advective mixing rate ($\lambda$). {It is the deterministic edge state organizing the transition to sustained growth. With noise, crossing $R^*$ does not preclude a return to haze, and the maximum of the reduced Lamperti potential can be displaced from this physical reference.}

To gain insight into the droplet-radius dynamics generated by the reduced stochastic equation (Eq.~\eqref{Eq_reduced_SDE}), we use the Lamperti transformation, which converts a one-dimensional It\^o diffusion with state-dependent noise into an SDE with constant noise amplitude \citep[Chapter~3.6]{pavliotisbook2014}. Writing Eq.~\eqref{Eq_reduced_SDE} in the generic form
\be
\d R = a(R)\d t + b(R)\d W_t,
\ee
we define
\be
\hat R = \psi(R), \qquad \psi(R) = \int^R \frac{\d x}{b(x)}.
\ee
{We absorb the minus sign in Eq.~\eqref{Eq_reduced_SDE} into $W\mapsto-W$, which leaves the law unchanged, and use $b(R)>0$ on the physical interval containing the haze and edge states. Thus $\psi$ is increasing there. The transformation is local to intervals on which $b$ does not vanish.} Applying It\^o's formula gives
\beas
\d \hat R &= \psi'(R)\d R + \frac{1}{2}\psi''(R)(\d R)^2 \\
&= \left[ \frac{a(R)}{b(R)} - \frac{1}{2}b'(R) \right]\d t + \d W_t,
\eeas
where we used $\psi'(R)=1/b(R)$, $\psi''(R)=-b'(R)/b(R)^2$, and $(\d R)^2=b(R)^2\d t$.

Thus the Lamperti-transformed SDE takes the normalized form
\be
\label{Eq_Lamperti_SDE}
\d\hat R = \hat a(\hat R)\d t + \d W_t,
\ee
where
\be
\hat a(\hat R) = \left[ \frac{a(R)}{b(R)} - \frac{1}{2}b'(R) \right]_{R=\psi^{-1}(\hat R)}.
\ee
The transformed process has unit diffusion coefficient. It may therefore be interpreted as a Brownian particle evolving under the effective drift $\hat a(\hat R)$, or equivalently in the potential landscape $U(\hat R)$ defined, up to an additive constant, by $\partial_{\hat R}U = -\hat a$.
For the reduced SDE \eqref{Eq_reduced_SDE}, the drift and diffusion coefficients are
\be
a(R) = 2D\big(g(R)-f(R)\big), \qquad b(R) = \frac{2D\sigma g(R)}{\alpha_N\sqrt R+\lambda}.
\ee
It follows that
\bes
\frac{a(R)}{b(R)} = \frac{ \big(g(R)-f(R)\big) \big(\alpha_N\sqrt R+\lambda\big) } {\sigma g(R)}.
\ees
Moreover,
\bes
b'(R) = 2D\sigma \left[ \frac{g'(R)}{\alpha_N\sqrt R+\lambda} - \frac{\alpha_N g(R)} {2\sqrt R\big(\alpha_N\sqrt R+\lambda\big)^2} \right].
\ees
Hence, writing $\hat a(R)$ for the Lamperti drift before composition with $R=\psi^{-1}(\hat R)$, we obtain
\bea
\label{Eq_ahat}
\hat a(R) &= \frac{ \big(g(R)-f(R)\big) \big(\alpha_N\sqrt R+\lambda\big) } {\sigma g(R)}  \\
& - \, D\sigma \left[ \frac{g'(R)}{\alpha_N\sqrt R+\lambda} - \frac{\alpha_N g(R)} {2\sqrt R\big(\alpha_N\sqrt R+\lambda\big)^2} \right].
\eea
We now define the Lamperti potential $U(\hat R)$ by
\be
\hat a(\hat R) = -\frac{\d U}{\d\hat R}.
\ee
Equivalently,
\be
U(\hat R) = -\int^{\hat R} \hat a(y)\,\d y .
\ee
{For plotting and quadrature we express the potential as a function of $R$. Its slope in this coordinate must not be confused with the force or curvature in the Lamperti coordinate $\hat R$.} Using $\hat R=\psi(R)$ and
\be
\d\hat R = \psi'(R)\d R = \frac{1}{b(R)}\d R,
\ee
we find
\be
\label{Eq_Lamperti_potential_R}
U(\psi(R)) = -\int \hat a(R)\frac{1}{b(R)}\,\d R .
\ee
Since
\be
\frac{1}{b(R)} = \frac{\alpha_N\sqrt R+\lambda} {2D\sigma g(R)},
\ee
the potential may be written as
\be
\label{Eq_Lamperti_potential}
U_{\lambda,\sigma}(R) \equiv U(\psi(R)) = -\int \hat a(R) \frac{\alpha_N\sqrt R+\lambda} {2D\sigma g(R)} \,\d R ,
\ee
where $\hat a(R)$ is given by Eq.~\eqref{Eq_ahat}. We have made the dependence on $\lambda$ and $\sigma$ explicit because these parameters control the deformation of the effective potential landscape in the analysis below. The integral does not appear to admit a useful closed-form expression, so in practice we evaluate $U_{\lambda,\sigma}$ by high-precision numerical quadrature.
 
\begin{figure}[htbp]
    \centering
       \includegraphics[height=0.3\textwidth, width=.48\textwidth]{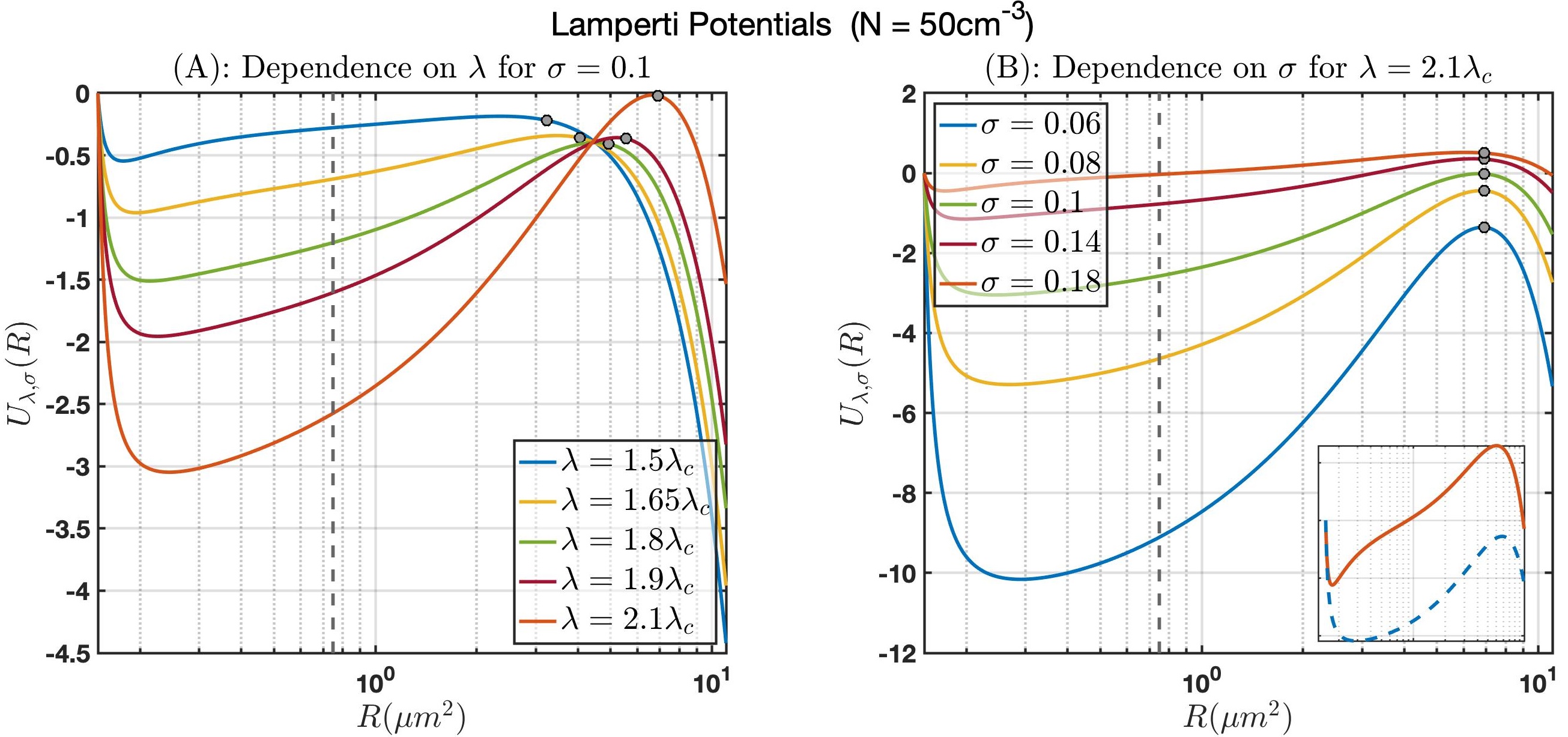}
              \includegraphics[height=0.3\textwidth, width=.48\textwidth]{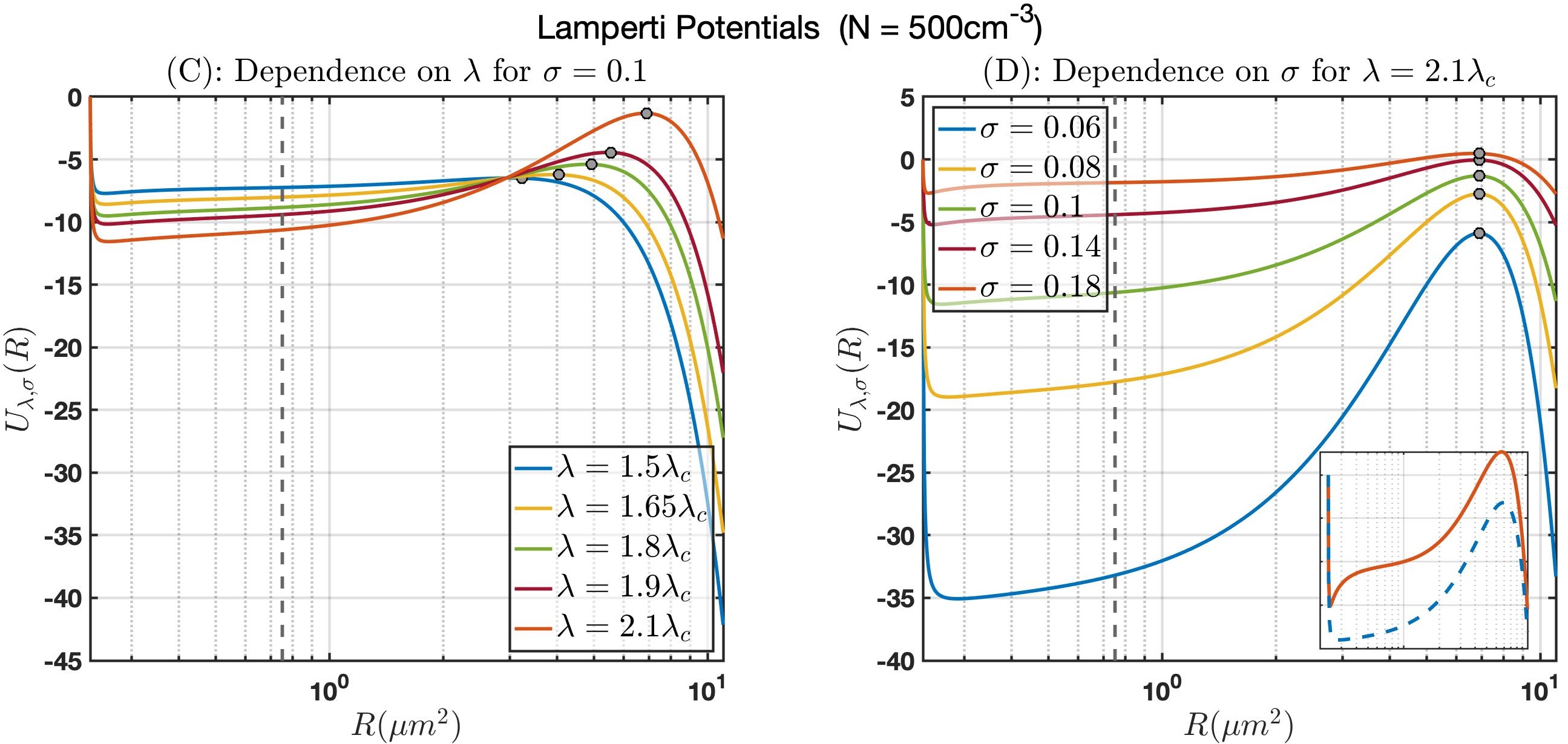}
    \caption{{\bf Lamperti potential's dependence on $\lambda$, $\sigma$, and $N$}. Panels A (resp.~C) and B (resp.~D) show, for $N=50 cm^{-3}$ (resp.~$N=500 cm^{-3}$),  the dependence  on  $\lambda$ and $\sigma$. 
    The other parameters values are those listed in  Table \ref{Table_RegimeA}. The insets show the Lamperti potential's shape in linear scale and standardized for $\sigma=0.18$ (orange curve) against a lower turbulent regime ($\sigma=0.06$, dashed-blue curve). The vertical dashed line corresponds to the K\"ohler critical radius $R_c = 3B/A$. The grey dots mark the location of $R^\ast$ on the respective Lamperti potentials.}
    \label{Fig_Lamperti_potential}
    \vspace{-.1cm}
\end{figure}

{Figure~\ref{Fig_Lamperti_potential} shows how the reduced potential depends on mixing, aerosol concentration and noise strength. Its haze-side minimum is denoted $\tilde R_h$, whereas the grey dots mark the deterministic edge $R^*$. At weak noise the relevant extrema approach the deterministic equilibria. At stronger noise they may move substantially, and additional extrema can appear. In particular, the principal barrier for Case 2 of Table~\ref{Table_energy_barrier} lies below the K\"ohler critical radius; the physical edge is then a reference point on the potential rather than its maximum. These distinctions are included in the local-rate comparison below.}

Melancholia states or edge states often appear as intermediate states in the transition between different stable attractors or regimes in the system's phase space, governing transient dynamics and dictating the pathways between them. In our context, near $R^*$, lies a critical transition zone between the stable haze droplet regime (Lamperti's potential minimum) and the regime of growing, activated cloud droplets ($R>R_c$). Indeed, $R^\ast$ is near the potential's peak location, 
and as such, corresponds to a landscape's region with a flatness gradation depending on both the smallness of slope of $U$ at $R^\ast$, and the smallness of the rate of change of this slope (curvature); see Table \ref{Table_Flatness}.

{The plotted potential becomes flatter over parts of the transition region as $\sigma$ increases, but its shape is shown in the original coordinate $R$. The dynamical force and local curvature must be evaluated in the Lamperti coordinate, as in Eq.~\eqref{Eq_local_Lamperti_rate}. In the It\^o-reduced closure, stronger noise can also sharpen the haze-side minimum and create additional extrema or inflection points. Those features depend on the order-$\sigma^2$ drift that the reduction does not determine. We therefore use them to describe this particular reduced model, while basing the physical hovering interpretation on the weak local edge response and the full-system trajectories. Returns remain possible after a noisy crossing of the deterministic edge.}

\begin{figure}[htbp]
    \centering
           \includegraphics[height=0.25\textwidth, width=.48\textwidth]{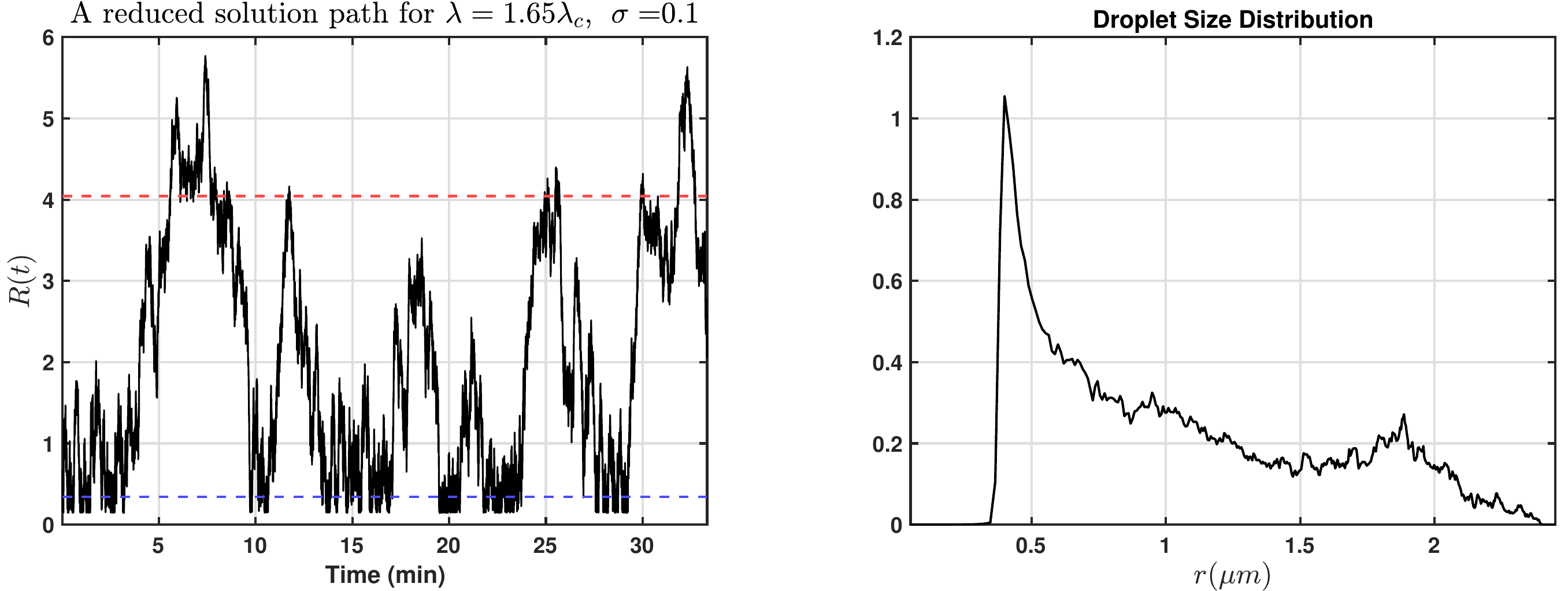}
        \caption{{\bf A transition path between the haze and edge states}. {\bf Left panel:} Here is shown a stochastic transition path connecting the edge state ($R*$) marked by the red dashed line and the haze state ($R_h$, blue dashed line), from integration of the slow manifold stochastic reduced equation (Eq.~\eqref{Eq_reduced_SDE}). The parameter regime is that of Table \ref{Table_RegimeA} for the values of $\lambda$ and $\sigma$ as indicated in the panel title. {\bf Right panel:} Corresponding droplet size distribution (DSD) in radius ($\mu m$).}
    \label{Fig_Transition_path}
    \vspace{-.1cm}
\end{figure}

\begin{figure}[htbp]
    \centering
              \includegraphics[height=0.25\textwidth, width=.48\textwidth]{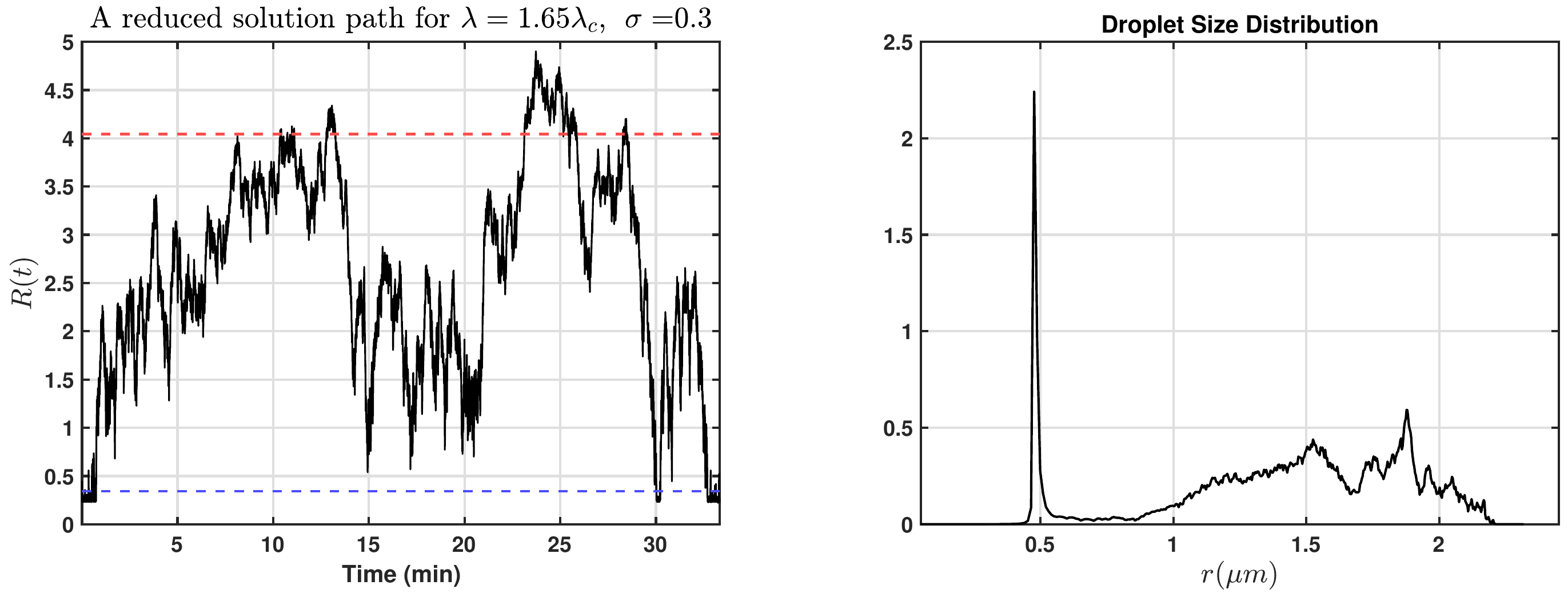}
        \caption{{\bf A transition path between the haze and edge states}. Same as Fig.~\ref{Fig_Transition_path} for which $N=500 cm^{-3}$ and $\sigma=0.3$.}
    \label{Fig_Transition_path2}
    \vspace{-.1cm}
\end{figure}

Figure \ref{Fig_Transition_path} illustrates this latter point, while Figure \ref{Fig_Transition_path2} shows a stochastic path hovering for a longer period of time in between the haze and edge states, for a more polluted and turbulent environment.
 We analyze in more details these cases in Section \ref{Sec_Lyapunov_exp}, below.

\begin{table}[h!]
    \centering
    \caption{\textbf{Flatness score $F$ at the edge state $R^\ast$.} Here, are considered  two different values of the supersaturation removal rate, $\lambda$, due to mixing. The flatness score ({\it Supplementary Note 2})
    increases with the noise strength $\sigma$ (turbulence) and decreases with the aerosol number concentration $N$ (pollution).}
    \label{Table_Flatness}
    \begin{tabular}{lcccc}
        \toprule
        & \multicolumn{2}{c}{\textbf{$\lambda=2.1\lambda_c$}} & \multicolumn{2}{c}{\textbf{$\lambda=1.65\lambda_c$}} \\
        \cmidrule(lr){2-3} \cmidrule(lr){4-5}
        \multirow{2}{*}{$\sigma$} & \multicolumn{2}{c}{$N$ (cm$^{-3}$)} & \multicolumn{2}{c}{$N$ (cm$^{-3}$)} \\
        \cmidrule(lr){2-3} \cmidrule(lr){4-5}
        & 50 & 500 & 50 & 500 \\
        \midrule
        0.06 & 0.7854 & 0.4754 & 0.8006 & 0.4725\\
        0.08 & 0.8543 & 0.5535 & 0.8582 & 0.5700\\
        0.1 & 0.8910 & 0.6341 & 0.8879 & 0.6524\\
        0.14 & 0.9258 & 0.7498 & 0.9155 & 0.7577\\
        0.18 & 0.9409 & 0.8154 & 0.9274 & 0.8135\\
        0.3 & 0.9562 & 0.8932 & 0.9393 & 0.8762\\
        \bottomrule
    \end{tabular}
\end{table}

Simultaneously, the supersaturation removal rate ($\lambda$) influences the position and height of the energy barrier at $R^*$, as shown in Figures \ref{Fig_Lamperti_potential}A and \ref{Fig_Lamperti_potential}C.
{The barrier and the local drift jointly influence excursions, but a noise-driven path has no conserved mechanical energy. Quantitative return or residence probabilities require first-passage calculations for the specified reduced diffusion or direct sampling of the full system.}

    \vspace{-.1cm}

    \vspace{-.1cm}

\subsection{{Local drift rates, energy barriers, and aerosol number concentration}}\label{Sec_Lyapunov_exp}
{We use the Lamperti potential to characterize the reduced dynamics, while keeping its closure dependence distinct from the full-system evidence for hovering. The potential is an effective stochastic landscape, not a thermodynamic energy in physical units. On an interval bounded by a haze-side minimum $\tilde R_h$ and a relevant maximum $\tilde R^*$, its barrier is
\be
U_{\rm barrier}=U_{\lambda,\sigma}(\tilde R^*)-U_{\lambda,\sigma}(\tilde R_h).
\ee
For the unit-noise Lamperti process, the Kramers exponential is $\exp(2U_{\rm barrier})$. The reported barriers are of order unity, so this is only a qualitative guide to escape, not a controlled rare-event estimate. A return probability additionally depends on the starting point and absorbing boundaries; it cannot be inferred from the barrier height alone.

Write $\hat a_R(R)=\hat a(\psi(R))$ for the transformed drift expressed as a function of the original variable. The local linearization rate in the Lamperti coordinate is
\be\label{Eq_local_Lamperti_rate}
\ell_L(R)=\left.\frac{\d\hat a}{\d\hat R}\right|_{\hat R=\psi(R)}
=b(R)\frac{\d\hat a_R}{\d R}
=-U''(\psi(R)).
\ee
It has units of inverse time. At a potential minimum it is negative; at a maximum it is positive. At $R^*$ it measures the local slope near the deterministic edge, whether or not that point is an extremum of the reduced potential. This is not a stochastic Lyapunov exponent: the latter requires a long-time tangent-growth average along stochastic trajectories, here $\lim_{T\to\infty}T^{-1}\int_0^T\ell_L(R_t)\,\d t$ when that limit exists \citep{arnold1998,Lyapunov2002}.

Table~\ref{Table_energy_barrier} reports the local rates obtained with Eq.~\eqref{Eq_local_Lamperti_rate}. The derivative is taken in the Lamperti coordinate; differentiating $\hat a_R$ with respect to $R$ alone omits the factor $b(R)$. The values use the parameters of Table~\ref{Table_RegimeA}, with $\alpha_N$ scaled proportionally to $N$.}
\begin{table*}[ht]
\centering
\caption{{Local Lamperti rates and effective barriers for the three illustrated regimes. Rates $\ell_L$ are in $\mathrm{s}^{-1}$ and refer to the It\^o-reduced closure. The haze rate is evaluated at its haze-side minimum; the edge rate is evaluated at the physical $R^*$. Barrier heights characterize the effective landscape. In Case 2, the relevant potential maximum lies below the K\"ohler critical radius.}}
{
\begin{tabular}{lcccccc}
\toprule
Regime & $\lambda^*$ & $N$ (cm$^{-3}$) & $\sigma$ & $\ell_L(\tilde R_h)$ & $\ell_L(R^*)$ & $U_{\rm barrier}$\\
\midrule
Case 1: Low Barrier & $1.65\lambda_c$ & 50 & 0.10 & $-1.77$ & $2.70\times10^{-3}$ & $0.6$\\
Case 2: Hovering & $1.65\lambda_c$ & 500 & 0.30 & $-8.29$ & $2.30\times10^{-4}$ & $1.16$\\
Case 3: Plateau Meandering & $2.10\lambda_c$ & 500 & 0.35 & $-8.72$ & $2.01\times10^{-4}$ & $1.28$\\
\bottomrule
\end{tabular}}
\label{Table_energy_barrier}
\end{table*}

\begin{figure}[h!]
\begin{center}
\includegraphics[height=0.3\textwidth, width=.45\textwidth]{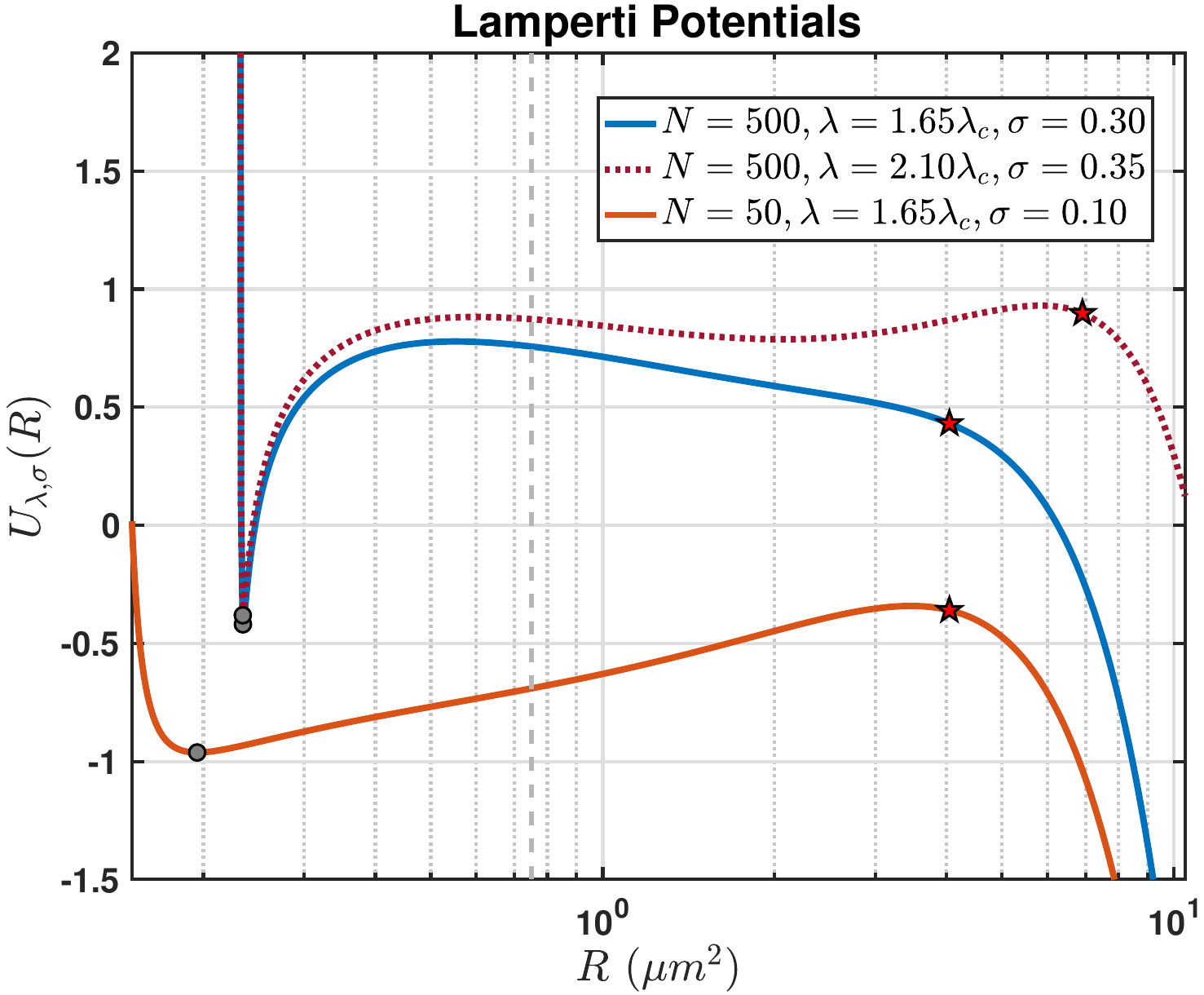}
\end{center}
\caption{{\bf Lamperti Potential Comparison.}. The potential in orange corresponds to the results shown in Figures \ref{Fig_Transition_path}  above and \ref{Fig_Hovering_N=50_lowsig} below. The potential in blue corresponds to the results shown in Figures \ref{Fig_Transition_path2}  above and \ref{Fig_Hovering_N=500_sig0dot3}  below. The potential shown as a dotted purple curve corresponds to the results shown in Figure \ref{Fig_Hovering_N=500_sig0dot35}  below. For each case, the red star marks the location of the edge state $R^*$, while the grey circle marks the location of the Lamperti haze state $\tilde{R}_h$. Note the presence of the inflection point in the case $N=500$, $\lambda=1.65 \lambda_c$, $\sigma=0.30${, in the It\^o reading of the reduced drift adopted here (see the end of Section~\ref{Sec_Lyapunov_exp})}, characterizing droplets transitioning from the haze to the edge state, that encounter a region of weaker driving force, subject to more erratic motion in this flatter potential region.} 
\label{Fig_Potential_comp}
\end{figure}

\begin{figure}[htbp]
    \centering
       \includegraphics[height=0.26\textwidth, width=.48\textwidth]{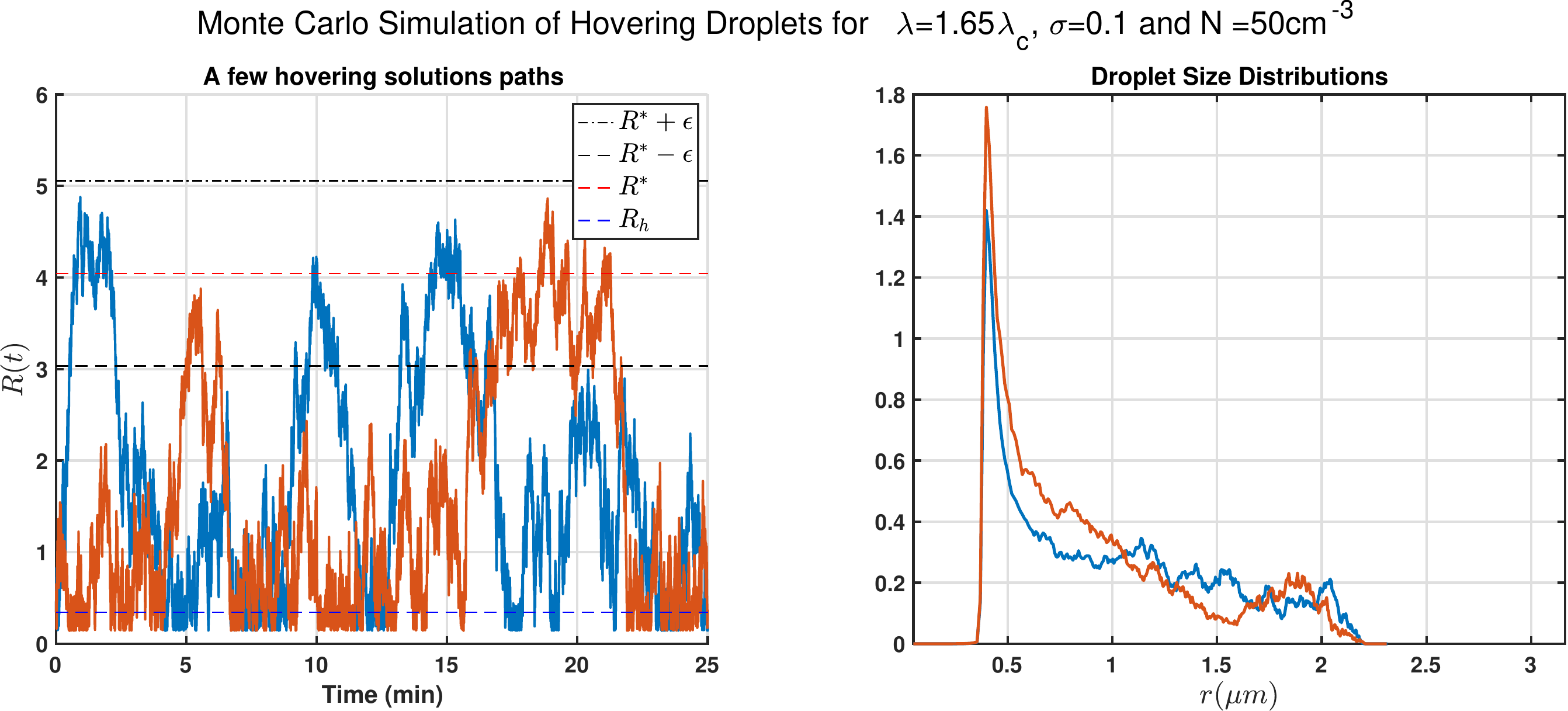}
    \caption{{\bf Hovering solution paths for $N=50cm^{-3}$.} The selection of these paths is performed through Monte Carlo's simulation.}
    \label{Fig_Hovering_N=50_lowsig}
    \vspace{-.1cm}
\end{figure}

\begin{figure}[htbp]
    \centering
       \includegraphics[height=0.26\textwidth, width=.48\textwidth]{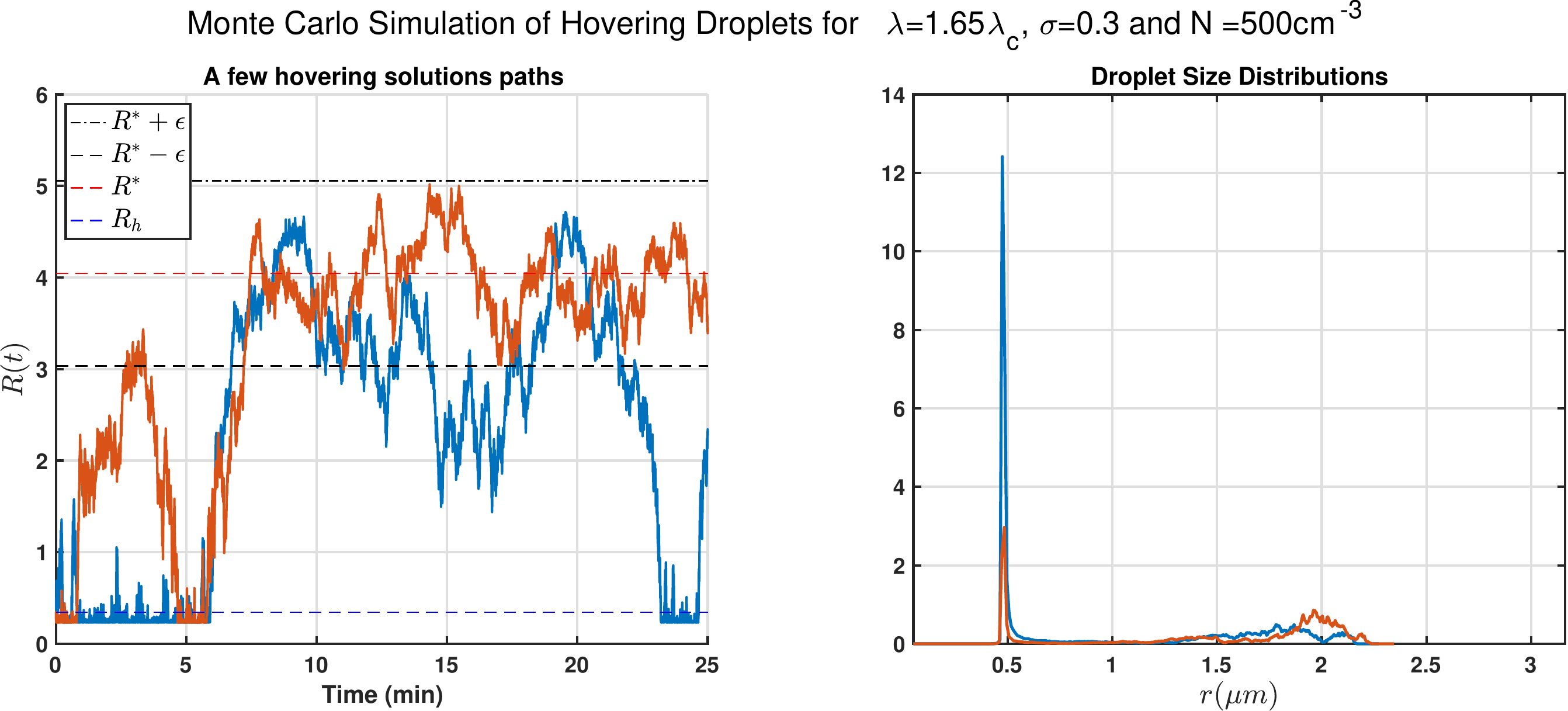}
    \caption{{\bf Hovering solution paths for $N=500cm^{-3}$ and $\sigma=0.3$.} The selection of these paths is performed through Monte Carlo's simulation. }
    \label{Fig_Hovering_N=500_sig0dot3}
    \vspace{-.1cm}
\end{figure}

\begin{figure}[htbp]
    \centering
       \includegraphics[height=0.26\textwidth, width=.48\textwidth]{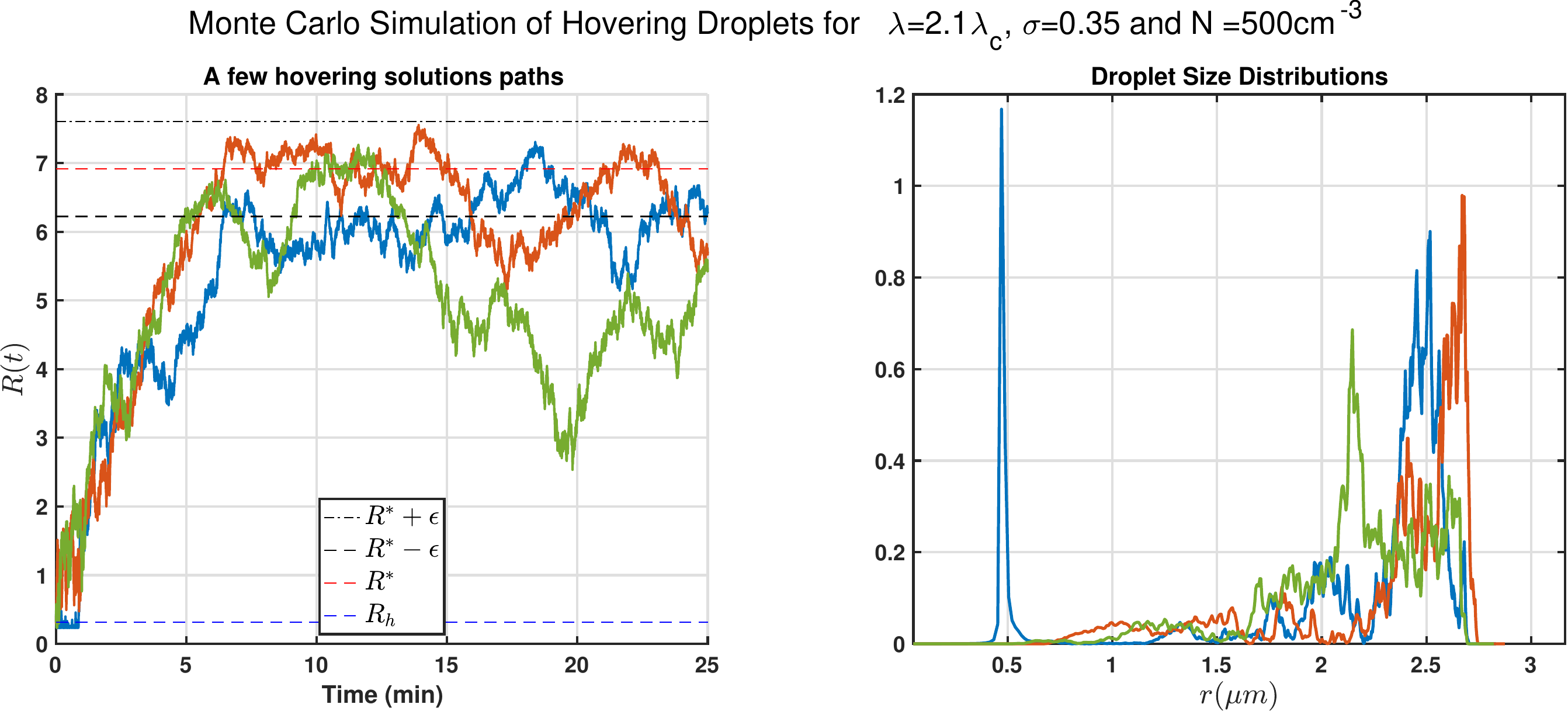}
    \caption{{\bf Hovering solution paths for $N=500cm^{-3}$.} The selection of these paths is performed through Monte Carlo simulations.}
    \label{Fig_Hovering_N=500_sig0dot35}
    \vspace{-.1cm}
\end{figure}

{The physical edge has a weak local drift slope in all three cases, especially for the higher aerosol concentration. The sample paths in Figures~\ref{Fig_Hovering_N=50_lowsig}--\ref{Fig_Hovering_N=500_sig0dot35} illustrate excursions toward this weakly repelling region, including extended wandering in Cases 2 and 3. These paths illustrate the mechanism; their selection does not estimate an unconditional hovering probability.

The reduced drift is determined only through order $\sigma^0$ by the formal expansion in Section~\ref{Sec_reducedSDE_der}. Reading the same coefficient pair in the Stratonovich sense adds $bb'/2$ to the It\^o drift and changes the pulled-back potential by $-\tfrac12\ln|b(R)|$, up to a constant. This correction is of order $\sigma^0$, whereas the leading potential is of order $\sigma^{-2}$. It can therefore substantially affect finite-noise landscape features.

For this alternative closure, the potential extrema coincide with the deterministic equilibria. The local edge rates for Cases 1--3 are $3.18\times10^{-3}$, $5.38\times10^{-4}$ and $2.57\times10^{-4}\,\mathrm{s}^{-1}$, compared with the values in Table~\ref{Table_energy_barrier}. Thus the small local edge rate and its reduction at larger $N$ occur in both closures. The haze rates, however, become $-0.302$, $-0.0905$ and $-0.151\,\mathrm{s}^{-1}$, and the reported alternative barriers are $0.73$, $0.30$ and $0.88$. Barrier ordering, haze confinement and additional extrema are consequently closure dependent. The stochastic potential supports an interpretation of the reduced trajectories; the evidence that hovering survives without this reduction comes from the full two-variable simulations in Section~\ref{Sec_Full_Parcel_Dynamics}.}

\subsection{Full Stochastic System Dynamics}\label{Sec_Full_Parcel_Dynamics}
 This section demonstrates the remarkable alignment between the insights gained from the analysis of the reduced one-dimensional SDE (Eq.~\eqref{Eq_reduced_SDE}) and the dynamics observed in the full two-variable stochastic system (Eq.~\eqref{Eq_growth_SR_lambda_stoch}). We analyze the three distinct regimes reported in Table \ref{Table_energy_barrier} discussed in Section \ref{Sec_Lyapunov_exp} for the reduced model.

\begin{figure}[htbp]
    \centering
       \includegraphics[height=0.35\textwidth, width=.48\textwidth]{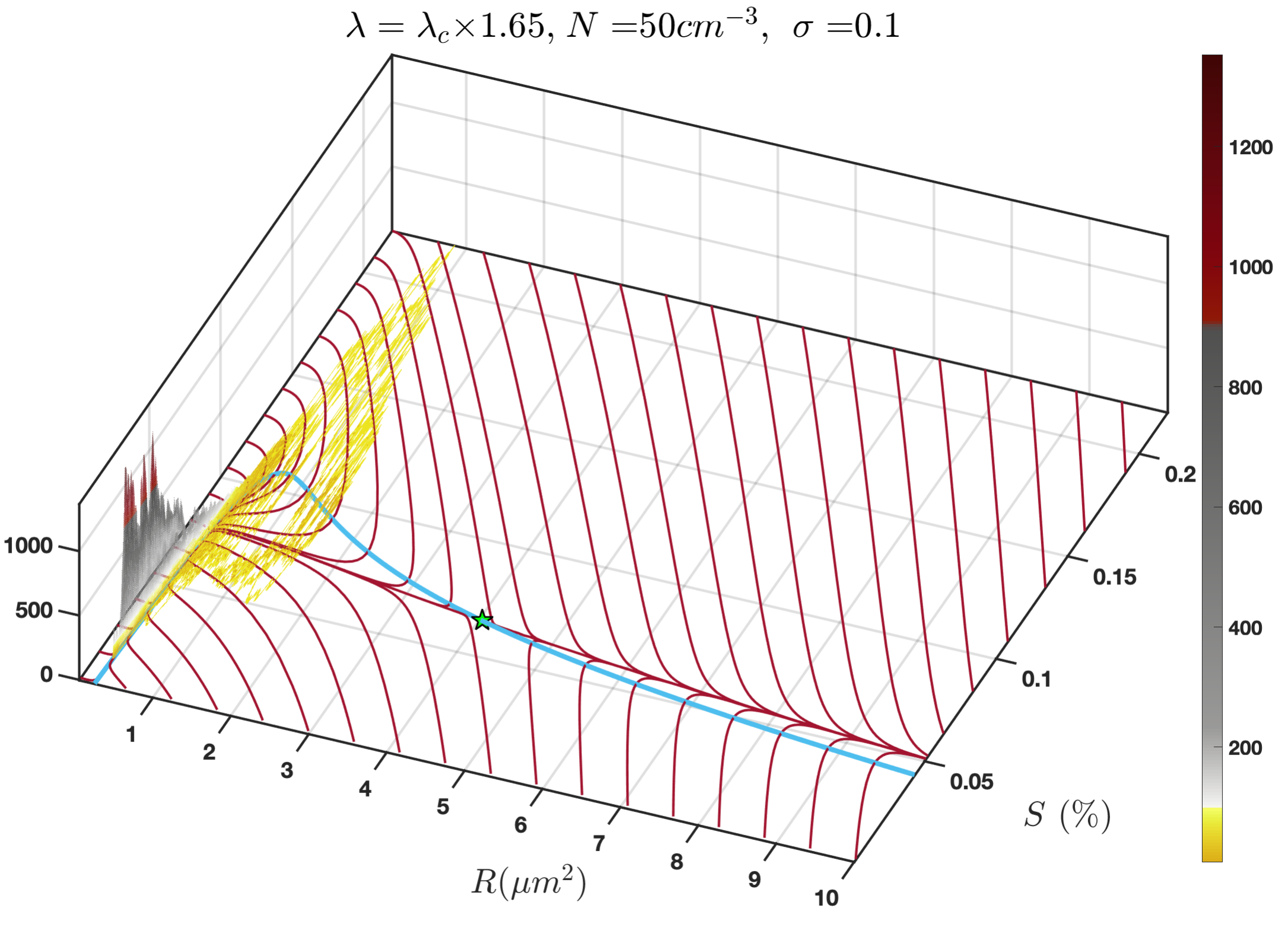}
    \caption{{\bf Full system's dynamics for Case 1.}
    Two-dimensional Probability Distribution Function (PDF) of the stochastic cloud droplet system (Eq.~\eqref{Eq_growth_SR_lambda_stoch}) in the $(R,S)$-phase space. The PDF is estimated from direct simulation using a colormap with yellow for low probability, and red for high probability. The cyan line represents the K\"{o}hler curve $(S=f(R), Eq.~\eqref{Eq_Kohler})$, which dictates the droplet's equilibrium growth. The red curves show the streamlines obtained by integrating the underlying  deterministic $SR$-system (Eq.~\eqref{Eq_growth_SR_lambda}). These illustrate how the haze and saddle structure the system's phase portrait. The green star marks the system's saddle equilibrium point, corresponding to the edge state ($R^\ast, S^\ast$) identified in Section \ref{Sec_advection}.}    
    \label{Fig_2Dcase1}
    \vspace{-.1cm}
\end{figure}

\begin{figure}[htbp]
    \centering
       \includegraphics[height=0.35\textwidth, width=.48\textwidth]{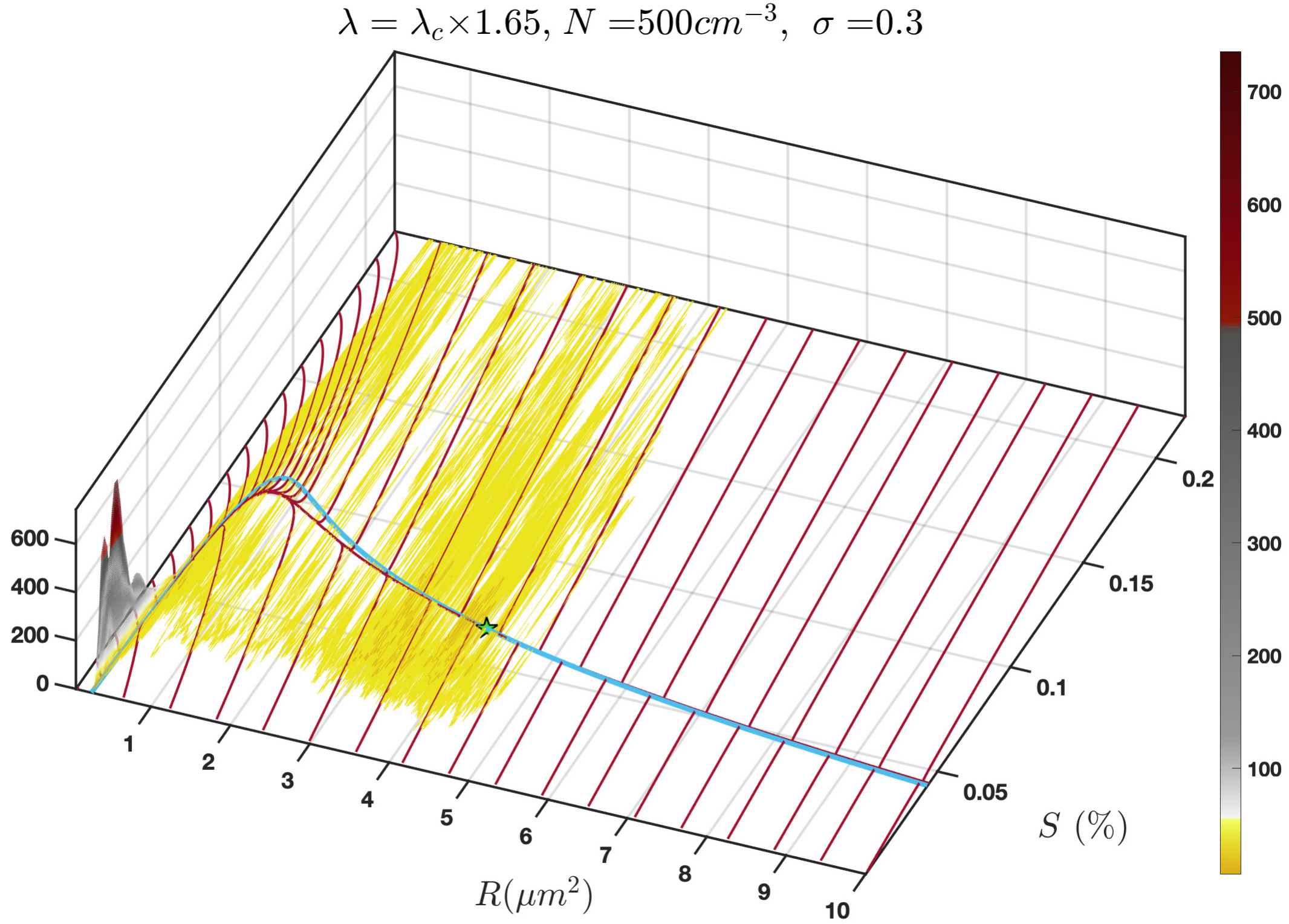}
    \caption{{\bf Full system's dynamics for Case 2.} Same as Fig.~\ref{Fig_2Dcase1}. The path selection  is here performed through Monte Carlo's simulation.}
    \label{Fig_2Dcase2}
    \vspace{-.1cm}
\end{figure}

\begin{figure}[htbp]
    \centering
       \includegraphics[height=0.35\textwidth, width=.48\textwidth]{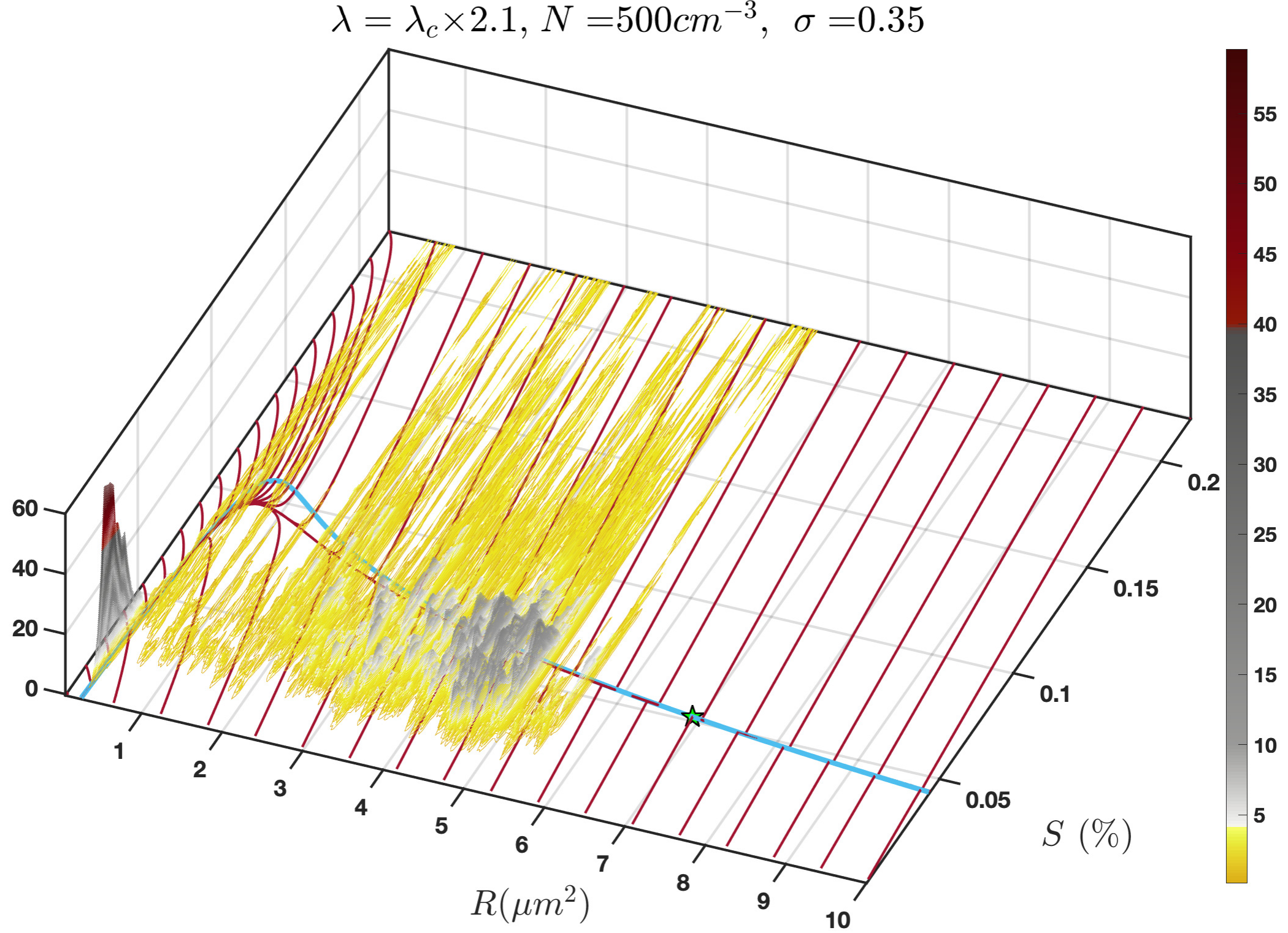}
    \caption{{\bf Full system's dynamics for Case 3.} Same as Fig.~\ref{Fig_2Dcase1}. The path selection is here performed through Monte Carlo's simulation.}
    \label{Fig_2Dcase3}
    \vspace{-.1cm}
\end{figure}

 For each case, the full stochastic system (Eq.~\eqref{Eq_growth_SR_lambda_stoch}) is integrated using an Euler-Maruyama scheme, starting from a dry radius under weak supersaturation ($S=0.01\%$). The runs are 25-minutes long, ensuring direct consistency with the reduced SDE analysis (Figures \ref{Fig_Hovering_N=50_lowsig}-\ref{Fig_Hovering_N=500_sig0dot35}). The corresponding two-dimensional Probability Distribution Functions (PDFs) are shown over the $(R,S)$-phase space in Figures \ref{Fig_2Dcase1}-\ref{Fig_2Dcase3}.

 In all three cases, the system's dominant mode remains confined to the haze branch---the stable segment of the K\"ohler curve. However, the qualitative distinction lies in the regions of the phase space visited by stochastic excursions. 
 In Case 1 (Low Barrier), excursions are tightly confined near the dominant haze mode (Figure \ref{Fig_2Dcase1}, yellow dots), indicating a rapid re-stabilization after stochastic disturbances. In Cases 2 and 3 (Hovering Regimes),  the excursions become significantly larger in both droplet radius ($R$) and supersaturation ($S$) fluctuations, a behavior directly analogous to the hovering phenomenon discussed for the reduced SDE.
 
\begin{figure}[htbp]
    \centering
       \includegraphics[height=0.26\textwidth, width=.48\textwidth]{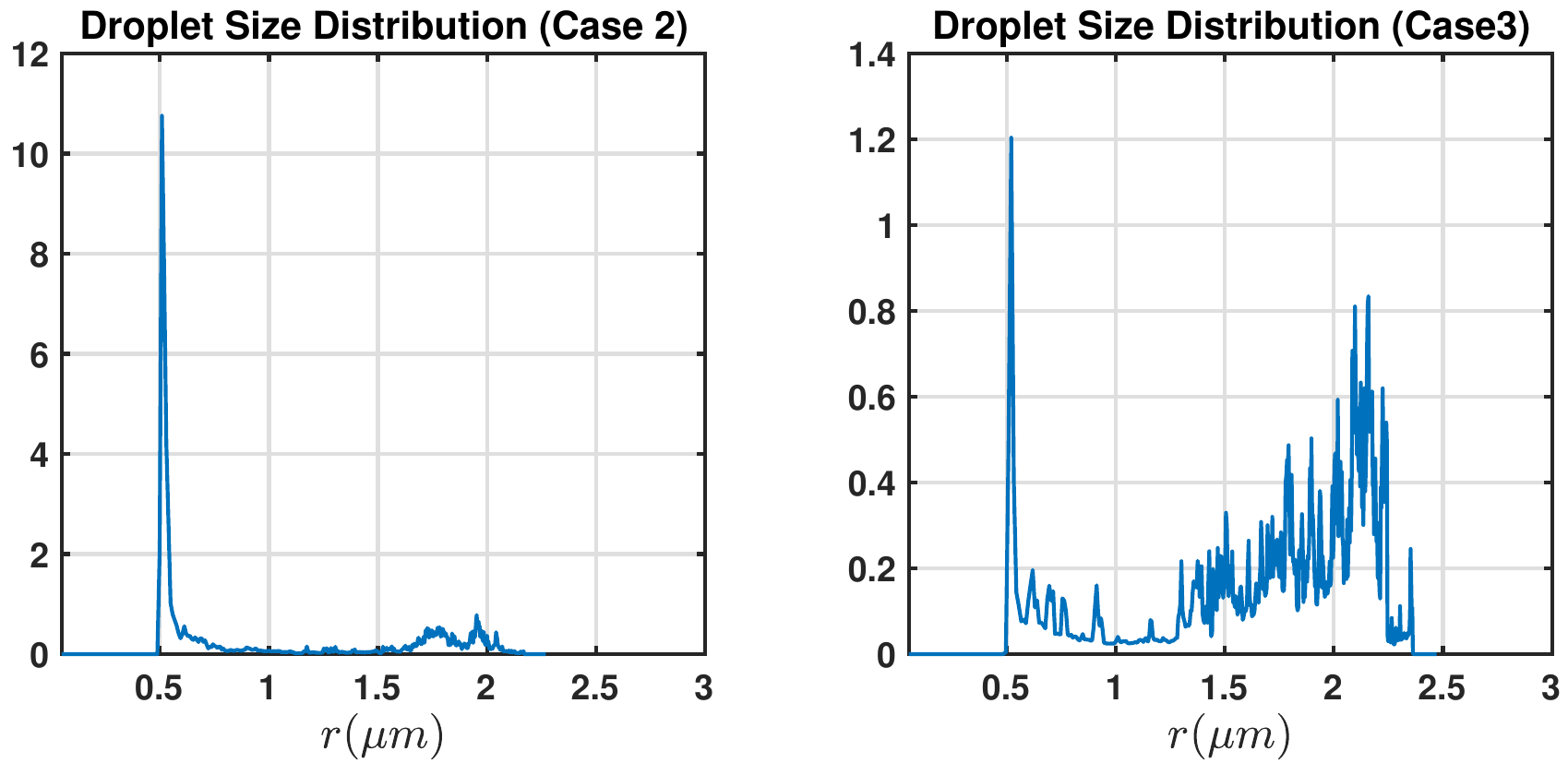}
    \caption{{\bf Droplet Size Distributions.} Here, the DSDs are obtained as 1D marginals of the two-dimensional PDFs shown in Figures \ref{Fig_2Dcase2} and \ref{Fig_2Dcase3}, respectively. The shapes of these DSDs are highly-consistent with those predicted by the reduced SDE (Eq.~\eqref{Eq_reduced_SDE}); compare with the blue curves in the right panels of Figures \ref{Fig_Hovering_N=500_sig0dot3} and \ref{Fig_Hovering_N=500_sig0dot35}, respectively.}
    \label{Fig_DSD_comp}
    \vspace{-.1cm}
\end{figure}

 A closer inspection of the PDF details reveals a hovering mode that is progressively more pronounced from Case 2 to Case 3. Crucially, the one-dimensional marginals in $R$ (Droplet Size Distributions, DSDs) from the full 2D system exhibit feature shapes that are strikingly reminiscent of those predicted by the reduced SDE; compare Figure \ref{Fig_DSD_comp} with the right panels of Figures \ref{Fig_Hovering_N=500_sig0dot3} and \ref{Fig_Hovering_N=500_sig0dot35}).

 {The agreement shows that the reduced SDE captures the dynamics of the full two-variable stochastic system that matter here: the confinement of the dominant mode to the haze branch and the hovering excursions toward the edge state, whose weak repulsion is the closure-independent feature identified in Section~\ref{Sec_Lyapunov_exp}. This is all the more notable given that the full system is not a stochastic gradient system, as pointed out in Section \ref{Sec_stochastic}. The full-system simulations involve no reduction and no closure of the reduced drift; they are exact, up to time discretization, for the parcel model \eqref{Eq_growth_SR_lambda_stoch}, with $\lambda$ the dressed removal rate of Eq.~\eqref{Eq_Gas_Ito}. The hovering mechanism and the droplet-size distributions reported below therefore do not depend on the reading of the reduced SDE.}

These results underscore the ability of our reduced model (Eq.~\eqref{Eq_reduced_SDE}) to capture the essential hovering dynamics of cloud droplets, highlighting the intricate and crucial roles that aerosol number concentration ($N$), mixing rate ($\lambda$), and turbulent fluctuations ($\sigma$), play in determining the pathway to the emergence of twilight clouds.

Notably, the simulated droplet size distributions span the $1$--$2 \mu m$ range, providing a robust physical match to the observations documented by \citep{hirsch2014transition}. Their field campaign, which utilized ground-based hyperspectral techniques to retrieve the properties of thin, warm convective clouds, confirmed the existence of these short-lived populations forming under weak updrafts. By recovering these specific droplet scales within our stochastic framework, we demonstrate that the "hovering" state near the activation barrier is not merely a mathematical curiosity of the slow manifold, but a fundamental mechanism governing the transition and persistence of the earliest stages of cloud formation.

\section{Statistical Field Theory of Cloud Physics}
\label{Sec_Field_Theory}

The preceding analysis of the 0D stochastic parcel model revealed a central nonlinear mechanism underlying twilight-cloud hovering tied to the ``forbidden zone'': droplet-growth variables are coupled to a fast supersaturation bath, and the coupled system rapidly collapses onto a slow manifold that defines an effective thermodynamic equilibrium distinct from the standard K\"ohler curve. This timescale separation, together with the emergence of a new thermodynamic balance (i.e.~the slow manifold) and the saddle-node structures induced by mixing, produces the forbidden zone and the intermittent switching between haze-like and activated states. However, the 0D model is intrinsically local. It treats the turbulent environment as a temporal stochastic forcing acting on a single representative parcel, and therefore cannot describe how hovering states persist, interact, or organize across an extended cloud layer.

We now pass from this localized parcel picture to a spatially extended field theory. The slow variable becomes the Mean-Square Radius Field $\phi(\x,t)$, while the fast variable becomes the coarse-grained supersaturation field $\chi(\x,t)$. The goal is to construct a dynamical action for these coupled fields, so that the fast environmental fluctuations can be treated systematically rather than appended as an external noise source. We pursue this goal within the Martin--Siggia--Rose--Janssen--de Dominicis (MSR) statistical field-theory framework \cite{MartinSiggiaRose,janssen1976lagrangean,deDominicis1976}, which converts stochastic evolution equations into a path-integral representation over fields and response fields \cite{tauber2014critical,kamenev2011field,del2026field}.

This formulation also gives a precise meaning to the analogy with the polaron problem in solid-state physics \cite{Devreese2009}. The slow droplet field plays the role of the heavy degree of freedom, while the fast, turbulently fluctuating supersaturation field plays the role of the environmental bath. Integrating out the fast field then produces an effective self-interaction for the slow field: the droplets become dynamically ``dressed'' by the supersaturation fluctuations they excite and deplete. In this sense, the cloud analogue of the polaron is not a literal quasiparticle, but a statistical-dynamical object: a slow droplet field carrying the memory and feedback of its fast vapor environment \cite{Devreese2009,gurari1953,feynman1955slow}.

A central advantage of the MSR framework is that it does not merely enable us to derive a deterministic mean-field closure. Once the fast supersaturation bath is integrated out, its turbulent fluctuations reappear in the reduced droplet dynamics as an emergent macroscopic stochastic forcing that goes beyond the small-noise expansion techniques used in Section \ref{Sec_reducedSDE_der} for the parcel model. In contrast with the parcel model, where we first reduced the deterministic parcel dynamics to its slow manifold and then derived a reduced stochastic equation by a small-noise expansion, the field-theoretic construction keeps the stochastic supersaturation bath inside the action before elimination. The effective noise in the $\phi$-equation is therefore inherited from the multiplicative turbulent fluctuations affecting the $\chi$-equation and transmitted through the nonlinear condensation coupling during the integration over the fast field.  We first derive  in Section \ref{Sec_Markov_closure} below, this dressed stochastic closure under two simplifying limits: a Markovian approximation, in which the bath relaxes instantaneously relative to the slow droplet field, and a zero-diffusion bath limit, in which the spatial redistribution of $\chi$ is neglected at leading order. 
 They are then relaxed in Section \ref{Sec_Dynamic_Inertia}, where the finite relaxation time and spatial propagation of the supersaturation bath generate a non-Markovian cloud droplet field equation with dynamic inertia.

The following sections make this construction explicit. In Section \ref{Sec_MSR_spatio}, we first build the isolated slow-droplet action $\Action_\phi$ and the isolated fast-bath action $\Action_\chi$, then introduce the condensation coupling action $\Action_{\rm int}$. We then use a variational integration over the fast bath, in the spirit of Gurari's method \cite{gurari1953} (Section \ref{Sec_slow_effective_Gurari}), to obtain the effective dressed dynamics of the slow droplet field (i.e.~its closure) within a Markovian approximation and zero-diffusion bath limit  (Section \ref{Sec_Markov_closure}).

\subsection{Basics of Statistical Field Theory}\label{Sec_Basics_SFT}

The goal of this section is to introduce, in minimal terms, the statistical field theory framework used below to pass from an isolated zero-dimensional parcel model to a macroscopic theory of droplet growth. As mentioned, the central tool is the MSR path-integral formalism \cite{MartinSiggiaRose,janssen1976lagrangean,deDominicis1976}. Its role is to convert a stochastic differential equation into a dynamical action principle, in the same way that equilibrium statistical mechanics converts a Hamiltonian into a Boltzmann weight. The essential difference is that equilibrium theory assigns weights to static configurations, whereas MSR assigns weights to entire time histories.

We first recall the equilibrium analogy. In classical statistical mechanics, the basic object is a state variable $q$, and the probability of observing a configuration is determined by the Boltzmann weight
\bes
P(q) \propto e^{-\beta H(q)},
\ees
where $H(q)$ is the Hamiltonian and $\beta$ is the inverse temperature. The partition function $\mathcal{Z} = \int dq e^{-\beta H(q)},$
normalizes this probability distribution and generates equilibrium averages. In this setting, the dimensionless quantity $\beta H(q)$ plays the role of a static action: configurations with large energetic cost are exponentially suppressed.

For a stochastic dynamical system, the elementary object is not a static configuration but a trajectory $q(t)$. Consider, for illustration, the Langevin equation
\be
\dot q(t) = f(q(t)) + \xi(t),
\ee
where $f(q)$ is an abstract deterministic drift (not to be confused with K\"ohler here) and $\xi(t)$ is a Gaussian white-noise forcing with covariance
\bes
\langle \xi(t)\xi(t') \rangle = 2D \delta(t-t'),  \; D>0.
\ees
The noise itself has the Gaussian path weight
\bes
P[\xi] \propto\exp\left[-\int dt \frac{\xi(t)^2}{4D} \right].
\ees
The MSR construction rewrites this probability as a functional integral over the trajectories $q(t)$ themselves. To do so, one enforces the Langevin equation by inserting the functional constraint $\delta\left[\dot q - f(q) - \xi\right].$
Using the Fourier representation of this delta functional introduces an auxiliary conjugate field $\bar q(t)$, called the response field. In the standard It\^{o} convention, the causal structure of the dynamics makes the associated functional Jacobian a constant, which can be absorbed into the path-integral measure. This point becomes important for the multiplicative-noise problem treated below whose corresponding regularization is discussed in Appendix~\ref{App_Jacobian_Regularization}. {There, as in the parcel model (Section~\ref{Sec_stochastic}), the noise is interpreted in the Stratonovich sense, the white-noise limit of mixing fluctuations with finite correlation time \citep{wong1965convergence}; {a direct MSR representation uses the midpoint discretization}, for which the Jacobian is field dependent \citep{Aron_al2010}. {Here we instead convert to the equivalent It\^o equation before constructing the dressed action, as detailed in Section~\ref{Sec_Jacobian_regularization}.}}

After integrating out the Gaussian noise, one obtains the MSR generating functional
\be
\mathcal{Z} =\int \mathcal{D}q \mathcal{D}\bar q e^{-\Action[q,\bar q]},
\ee
with dynamical action
\be
\Action[q,\bar q] =\int dt \left[\bar q(t)\big(\dot q(t)-f(q(t))\big) {-D \bar q(t)^2}\right].
\ee
{We use imaginary response fields and the weight $e^{-\Action}$ throughout; with this convention the quadratic response term has the minus sign displayed above.} The important point is that the stochastic equation has been rewritten as a dynamical field theory. The measure $\mathcal{D}q$ represents a sum over all possible time histories, while $\mathcal{D}\bar q$ introduces the conjugate response histories that enforce causality and generate linear response functions.

This construction is the dynamical analogue of the Boltzmann partition function. In equilibrium mechanics, the partition function sums over configurations weighted by $e^{-\beta H}$. In MSR theory, the generating functional sums over histories weighted by $e^{-\Action}$. The action $\Action[q,\bar q]$ therefore plays the role of a dynamical cost functional: it quantifies how the deterministic drift, stochastic forcing, and causal response structure combine to determine the statistical weight of a trajectory.

The response field $\bar q$ should not be interpreted as an additional physical degree of freedom, such as a new microphysical variable. Rather, it is an auxiliary field that records how the system responds to perturbations and enforces the Langevin dynamics inside the functional integral. This is what makes the MSR formalism more powerful than a direct path-probability description: it does not merely describe likely trajectories, but also generates correlation functions, response functions, and systematic fluctuation corrections.

The original Langevin equation can be recovered immediately from the unaveraged form of the construction. Before the Gaussian noise is integrated out, the action contains the term
\be
\Action_{\xi}=
\int dt 
\bar q(t)
\big(
\dot q(t)-f(q(t))-\xi(t)
\big).
\ee
Variation with respect to the response field imposes
\be
\dot q(t)=f(q(t))+\xi(t),
\ee
so the stochastic equation is not replaced by the field theory; it is embedded within it. In a saddle-point or weak-noise limit, the stationary configurations of the action identify the dominant histories. Away from that limit, however, the full functional integral retains the ensemble of fluctuating histories and is therefore suited to computing statistical observables rather than only a single most probable path.

This is also why the MSR formulation is particularly useful for reduction. In classical stochastic climate theory and stochastic parameterization, one often starts from a resolved--unresolved split and derives effective stochastic dynamics by averaging or homogenization under suitable scale-separation assumptions \citep[e.g.][]{MajdaTimofeyevVandenEijnden2001,FranzkeEtAl2015,GottwaldMelbourne2013}. When the scale separation is weak, substantial effort is instead devoted to retaining the memory and noise generated by the unresolved variables, for instance through Mori--Zwanzig projection techniques and related response-based methods \citep[e.g.][]{GottwaldCrommelinFranzke2017,MSM2015,SantosGutierrez2021,LucariniChekroun2023}. The MSR construction is complementary to these approaches. It keeps the fast degrees of freedom inside the action before they are eliminated. As a result, integrating out the fast sector directly produces not only an effective drift, but also the induced noise covariance, memory kernels, and response functions at the level of the action. This is essential for the present cloud problem, where eliminating the supersaturation bath must retain both the dressed deterministic drift and the stochastic correlations transmitted to the slow droplet field.

One should also not over-interpret the formal measure $\mathcal D q$ as requiring an explicit pointwise construction at this stage. It denotes integration over histories, made precise by a time-discretized or regularized version of the stochastic dynamics. What matters for the present derivation is that changes of variables inside this history integral are not innocuous. Transforming from the noise realization $\xi$ to the path $q$ generates a functional Jacobian, namely the determinant associated with the map from noise histories to stochastic trajectories \citep{ZinnJustin2002,tauber2014critical}. For additive noise this Jacobian often reduces to an irrelevant normalization, but for nonlinear or multiplicative stochastic dynamics it can become field-dependent and contribute directly to the effective action \citep{Aron_al2010,aron2016dynamical}. Thus the MSR functional integral does not merely rewrite the Langevin equation; it also keeps track of the measure corrections generated by the stochastic change of variables.

This observation is the bridge to the field-theoretic reduction used below. The advantage of the MSR reformulation becomes decisive when the stochastic dynamics are nonlinear, multiplicative, or spatially extended. In the present cloud-physics context, once the parcel variable is promoted to a field $q(\mathbf{x},t)$, the action becomes a functional of both $q(\mathbf{x},t)$ and its response partner $\bar q(\mathbf{x},t)$. This places cloud microphysics within the language of non-equilibrium statistical field theory. High-frequency or small-scale degrees of freedom can then be integrated out at the action level, producing effective large-scale drift, dissipation, noise covariances, memory kernels, and interaction terms. In field-theoretic language, the bare parcel-level dynamics become ``dressed'' by unresolved fluctuations.

This is the conceptual step used in the following sections. Starting from a stochastic parcel model of droplet growth, we construct its MSR action, promote the parcel variables to spatially dependent fields, and use the resulting statistical field theory to derive a macroscopic description of the turbulent droplet--supersaturation continuum. The point is not to replace the original stochastic dynamics, but to embed them in a framework where unresolved fluctuations can be eliminated systematically while retaining their induced drift, noise, and response structure.

\subsection{Dynamic Action for the Mean-Field Droplet--Supersaturation System}
\label{Sec_MSR_spatio}

Passing from a single-droplet variable $R$ to a continuous macroscopic field is not merely a change of notation. At a fixed macroscopic point $\x$, the coarse-grained cloud element contains a local population of droplets with a distribution of squared radii, and the exact evolution of its moments generally involves higher moments and spectral correlations. For example, the K\"ohler contribution depends on $\langle f(R)\rangle_{\x,t}$, while the vapor sink involves moments such as $\langle \sqrt{R}\big(s-f(R)\big)\rangle_{\x,t}$, where $s$ is the local microscopic supersaturation field. 
For the reduced droplet-sector closure pursued here, we project these unresolved  microphysical averages by functions of the resolved mean-square-radius field $\phi$.

 We define the slow macroscopic field $\phi$ as the local mean-square radius of the droplet population contained within that element (Appendix \ref{Sec_microphys_average}):
\be\label{Eq_Def_phi}
\phi(\x,t) = \langle R\rangle_{\x,t} = \langle r^2\rangle_{\x,t}.
\ee
Here, the bracket $\langle \cdot \rangle_{\x,t} \equiv \int \cdot \, h_{\x,t}(R)\,\d R$ denotes a microphysical average over the local normalized droplet-size distribution ($\int_0^\infty h_{\x,t}(R) \, \d R = 1$), not a spatial convolution over neighboring fluid elements.  More general droplet-sampled observables involving the local supersaturation are defined in Appendix \ref{Sec_microphys_average}.

In standard warm-cloud microphysics, the unknown spectrum $h_{\x,t}(R)$ is typically represented explicitly through bin or sectional schemes, or indirectly through bulk moment closures \citep[e.g.][]{tzivion1987efficient,feingold1994explicit,reisin1996rain,khain2015spectral}. Our reduced field theory takes a different route.

Rather than parameterizing the full droplet spectrum, we project its dominant effect onto the single resolved order parameter $\phi$. This projection keeps the K\"ohler nonlinearity and the leading condensation sink while avoiding an explicit evolution equation for the full spectrum, for higher spectral moments, or for auxiliary shape diagnostics such as the effective radius $r_e=\langle r^3\rangle_{\x,t}/\langle r^2\rangle_{\x,t}$.  The resulting evolution equations derived below are thus mean-field microphysical projections of the droplet spectrum, not a bin-resolved description. The corresponding coarse-grained supersaturation field $\chi$ is defined in Eq.~\eqref{Eq_def_chi_phi} of Appendix \ref{Sec_microphys_average}.

Building directly upon the foundational formalism introduced in Section \ref{Sec_Basics_SFT}, we now extend the MSR path integral to our spatially extended cloud system. Instead of tracking a single generic trajectory $q(t)$, our generating functional $\mathcal{Z}$ must sum over all mathematically possible spatiotemporal histories of the coupled droplet and vapor fields. For each observable physical field ($\phi$ and $\chi$), we introduce a corresponding conjugate response field ($\bar{\phi}$ and $\bar{\chi}$). The partition function takes the form of a functional integral over these four fields:
\be\label{Eq_Partition_Function}
\mathcal{Z} \hspace{-0.2ex}= \hspace{-0.2ex}\int_{\phi \geq 0} \hspace{-1.2ex} \mathcal{D}[\phi] \int_{i\mathbb{R}} \mathcal{D}[\bar{\phi}] \int_{\mathbb{R}} \mathcal{D}[\chi] \int_{i\mathbb{R}} \mathcal{D}[\bar{\chi}] e^{- \Action[\phi, \bar{\phi}, \chi, \bar{\chi}] }.
\ee

The integration measures $\mathcal{D}[\dots]$ represent formal, infinite-dimensional flat measures over the space of field configurations. The domains of these integrations strictly reflect the physics of the coarse-graining procedure and the mathematical rules established in Section \ref{Sec_Basics_SFT}. The supersaturation vapor field explores the entire real line ($\chi \in \mathbb{R}$), representing both sub-saturated and super-saturated macroscopic states. Conversely, the conjugate response fields $\bar{\phi}$ and $\bar{\chi}$ act as the dynamic Lagrange multipliers. Because they originate from the Fourier representation of a functional Dirac delta enforcing the equations of motion, they are integrated exclusively along the imaginary axis ($\bar{\phi}, \bar{\chi} \in i\mathbb{R}$). 

A fundamental physical requirement of this specific MSR construction is the domain of the droplet field measure, $\mathcal{D}[\phi]$. Because the physical field represents a squared macroscopic radius, the integration is formally restricted to non-negative field configurations ($\phi \ge 0$). This constraint is critical to ensure that the microphysical interaction terms involving $\sqrt{\phi}$ remain well-defined and strictly real across the spatiotemporal domain.  The singular K\"ohler drift supplies the mechanism that preserves positivity. {\it Supplementary Note 3} proves that a strictly positive droplet field cannot reach zero in finite time along a bath history with the stated local square-integrability. {\it Supplementary Note 4} establishes bath and droplet regularity estimates for the regularized coupled system and identifies the additional uniform bounds needed to remove the regularization.

As outlined in Section \ref{Sec_Basics_SFT}, the exponential weight $\exp(-\Action)$ assigns a statistical probability to each history. {The MSR action, with response fields integrated on an imaginary contour, is not a positive probability cost to be minimized. Its stationary equations identify deterministic response-free histories; probabilities and most-probable noisy paths require the complete functional integral or an appropriate large-deviation reduction.} Within the functional $\Action$, terms linear in the response fields ($\bar{\phi}, \bar{\chi}$) rigidly enforce these underlying deterministic equations of motion. Conversely, quadratic (or higher) terms in the response fields encode the variance of the stochastic noise, formally defining how far the system is allowed to turbulently fluctuate around its mean trajectory.

The total bare action is then constructed as $\Action = \Action_\phi + \Action_\chi + \Action_{int}$, as described below.
\subsubsection{The Isolated Slow Droplet Field Action ($\Action_\phi$)}\label{Sec_Slow_Action}
This term governs the deterministic local relaxation of the slow droplet field before it is coupled to the fast supersaturation bath. To construct it, we map the deterministic microphysics of the isolated parcel model into the continuous mean-square-radius field representation.

\paragraph*{\small The Isolated Deterministic Dynamics.}
In our parcel model  (Eq.~\eqref{Eq_Gbs}), the term $2DS$ represents growth driven by the surrounding vapor, whereas $-2Df(R)$ represents the intrinsic K\"ohler contribution associated with curvature and solute effects; see Section \ref{Sec_background_partI}.
In our field theory, the vapor-driven growth term, $2D\chi$, is reserved for the interaction action $\Action_{\rm int}$, leaving only the internal K\"ohler contribution for the isolated slow action; see Section \ref{Sec_S_interaction}. 
The slow droplet field has been defined in Eq.~\eqref{Eq_Def_phi} as microphysical average providing the mean-square radius per coarse-grained cloud element.

To deal with the coarse-grained supersaturation field $\chi$, we adopt a local-bath approximation; see \textit{Supplementary Note 5, Step 3}. This approximation states that the droplet-sampled mean supersaturation $\langle s\rangle_{\x,t}$, defined in Eq.~\eqref{Eq_def_chi_phi}, is, to leading order, unbiased relative to the Eulerian filtered supersaturation bath. Thus,
\be\label{Eq_local_bath_approx}
\chi(\x,t) = \langle s\rangle_{\x,t} \simeq \overline{s}(\x,t),
\ee
where the overbar denotes the Eulerian spatial filter at the same coarse-graining scale. This identification is made only for the supersaturation bath sampled by the droplets; it does not replace the microphysical average over droplet-size observables such as $\langle R\rangle_{\x,t}$.
With this convention, the microscopic supersaturation experienced by the droplets is decomposed into a resolved macroscopic bath and zero-mean sub-filter fluctuations, i.e.~$
s = \chi + s'$,  with $\langle s'\rangle_{\x,t} = 0.$
Averaging the microscopic parcel growth law $\dot{R} = 2D(s-f(R))$ thus yields at the macroscopic mean-field level
\be
\frac{D^{\bar\ub}\phi}{Dt}= \partial_t\phi +\bar\ub\cdot\nabla\phi  \simeq 2D\Big(\chi-\langle f(R)\rangle_{\x,t}\Big).
\ee
The exact averaged K\"ohler term depends on the full local droplet spectrum. For the reduced single-order-parameter theory adopted here, we systematically project this dependence onto functions of the resolved field (see Appendix \ref{App_Kohler_projection}):
\be
\langle f(R)\rangle_{\x,t} \simeq f_{\rm eff}(\phi) \simeq f(\phi)=A\phi^{-1/2}-B\phi^{-3/2}.
\ee
This is not the naive identity $\langle f(R)\rangle=f(\langle R\rangle)$. Rather, $f(\phi)$ is the leading-order projected K\"ohler response obtained by expanding the exact spectral average around the resolved mean-square-radius field (Appendix \ref{App_Kohler_projection}). {This projection requires small relative spectral-width corrections and controlled small-radius tails. Concentration in a micron-scale range alone does not establish those conditions; Appendix~\ref{App_Kohler_projection} identifies the moments that must remain subleading.}
 
We work in a Lagrangian, co-moving cloud-frame description, following the coarse-grained element along the resolved cloud-scale flow. In this frame, used throughout the action, the material derivative $D^{\bar\ub}/Dt$ reduces simply to $\partial_t$. 

The only additional term retained in the isolated slow sector is a linear residence-time loss $-\mu\phi$. This term represents unresolved removal of the area-like droplet field from the represented layer due to fallout, detrainment, dilution, or finite residence time of liquid-bearing parcels; see  {\it Supplementary Note 5}. {For a normalized mean such as $\phi$, uniform removal of droplets would change number concentration rather than mean size. The term $-\mu\phi$ is therefore an effective size-moment relaxation representing selective loss or replacement by smaller particles, not a consequence of uniform fallout alone. It is not an evaporation parameter and does not correct the microscopic condensational growth law.} Evaporation and condensation remain encoded in the combined thermodynamic drift $2D\big(\chi-f(\phi)\big)$.

The isolated deterministic slow droplet dynamics is therefore
\be\label{Eq_Isolated_Phi}
\partial_t \phi = -\mu\phi - 2D f(\phi).
\ee

\paragraph*{\small The MSR Constraint and Bare Action.}
To embed this isolated trajectory into the MSR path integral, we introduce the conjugate response field $\bar{\phi}$ through a Fourier representation of the Dirac delta functional. Its role is to enforce Eq.~\eqref{Eq_Isolated_Phi} as a dynamical constraint at each point of the coarse-grained layer and at each time.

Since the isolated droplet equation is deterministic, there is no additional noise average in this sector. The response field simply multiplies the rearranged equation of motion,
$\partial_t\phi+\mu\phi+2D f(\phi)=0,$  which gives the isolated slow droplet action
\be\label{Eq_S_phi}
\Action_\phi = \int \d t \int \d \x \left[ \bar{\phi}\left(\partial_t+\mu\right)\phi + 2D\bar{\phi} f(\phi) \right].
\ee
The first term enforces the local time evolution and the leading residence-time loss of the mean-square-radius field. The second term encodes the mean-field projection of the K\"ohler curve in the slow order parameter. The vapor-driven contribution $2D\chi$ is not included here; it enters separately through the interaction action and is responsible for coupling the slow droplet field to the fast supersaturation bath; see Section \ref{Sec_S_interaction} below.

\subsubsection{The Isolated Fast  Bath Action ($\Action_\chi$)}\label{Sec_FastBath_Action}

Unlike the slow droplet field, the background supersaturation behaves as a continuous dissipative field strongly shaped by turbulent transport and environmental exchange. In this subsection we construct the isolated fast-bath dynamics, namely the dynamics of $\chi$ before coupling it to the droplet field through the condensation sink. The droplet-induced term $-\alpha_N\sqrt{\phi}\big(\chi-f(\phi)\big)$ is therefore reserved for the interaction action $\Action_{\rm int}$.

\paragraph*{\small The Stochastic Langevin Equation for $\chi$.}  We first describe the dynamics of the isolated (coarse-grained) supersaturation field $\chi$. As detailed in {\it Supplementary Note 5}, we start from the microscopic transport equation for the local supersaturation field, $s$, which includes both the advective transport term $\ub \cdot \nabla s$ and the conservative transport  term $- s\frac{1}{\bar\rho}\partial_{x_i}\big(\bar\rho u_i\big)$ \citep[Section 5.2.7]{khain_pinsky_2018}. To avoid double counting the advection during spatial coarse-graining, we rewrite this balance in its conservative form. 
We then decompose the fields into resolved and unresolved parts, $s=\chi+s'$ and $u_i=\bar{u}_i+u_i'$, and adopt the  local-bath approximation defined in Eq.~\eqref{Eq_local_bath_approx}. 

Closing the unresolved turbulent scalar flux via a standard gradient approximation, $-\overline{u_i's'} \simeq K_\chi\partial_{x_i}\chi$, yields an effective spatial redistribution term $D_\chi\nabla^2\chi$. Constructed under a local homogeneity assumption, this term neatly combines molecular vapor diffusion with the turbulent eddy diffusivity generated by unresolved transport; see  {\it Supplementary Note 5, Step 4}. It remains to close the local exchange tendency. Moving to a Lagrangian, co-moving cloud-frame description, the resolved macroscopic advection is absorbed entirely into the material derivative $D^{\bar\ub} \chi/Dt$, which we denote simply by $\partial_t\chi$. The surviving density-weighted dilution, together with unresolved entrainment and mixing with the dry environment, constitutes a net local exchange rate. We decompose this rate into a mean component and a rapidly fluctuating component, modeled as $-\lambda\chi + \sigma\chi\eta(\x,t)$, where $\eta$ is a zero-mean Gaussian {noise, white in time and correlated in space over the Kolmogorov scale (Eq.~\eqref{Eq_noise_covariance} below)}. {This is the field counterpart of the environmental-exchange closure in Section~\ref{Sec_stochastic}: fluctuations in the relaxation rate multiply $\chi$. Under an anelastic constraint the density-weighted divergence vanishes, so unresolved entrainment and exchange, rather than that divergence alone, motivate the local rate. A nonzero environmental reference is absorbed into the mean source at this order; its fluctuating contribution is not resolved.} 

Thus, the uncoupled resulting stochastic Langevin equation for the supersaturation field $\chi$ takes the form of a reaction--diffusion equation driven by multiplicative noise: 
\be\label{Eq_Langevin_Chi} 
\partial_t \chi = D_\chi \nabla^2 \chi - \lambda \chi + \gamma + \sigma \chi {\circ}\eta,
\ee where $\gamma$
corresponds to the macroscopic adiabatic moisture forcing (due to large-scale updraft). {The noise is Gaussian, white in time, and correlated in space over the Kolmogorov scale $\eta_K$,
\be\label{Eq_noise_covariance}
\langle\eta(\x,t)\eta(\x',t')\rangle=\rho_{\eta_K}(\x-\x')\,\delta(t-t'),
\ee
Here $\rho_{\eta_K}(\x-\x')$ specifies the spatial covariance between the noise at two locations. It is smooth, symmetric and positive semidefinite, with correlations concentrated over separations of order $\eta_K$. We normalize its integral to unity, $\int\rho_{\eta_K}(\mathbf y)\,\d\mathbf y=1$, so its value at the origin determines the effective correlation volume $V_c=1/\rho_{\eta_K}(0)$ (Appendix~\ref{App_Jacobian_Regularization}). The symbol $\circ$ denotes the Stratonovich interpretation \citep{gardiner2009stochastic}. The finite correlation length regularizes the spatial forcing at the Kolmogorov scale, which provides the physical small-scale cutoff of the bath description (Appendix~\ref{Sec_Turb_Cascade}). On scales much larger than $\eta_K$, the covariance kernel can be approximated by a Dirac distribution. We use this local approximation below while retaining the finite value $\rho_{\eta_K}(0)$ wherever it enters the stochastic drift correction; Sec.~\ref{Sec_Jacobian_regularization}.

\paragraph*{\small  The MSR Constraint.}
To embed this stochastic PDE into a path integral, the MSR formalism introduces the response field $\bar{\chi}$ via a Fourier representation of the Dirac delta functional  \cite{MartinSiggiaRose, tauber2007field}. The sole mathematical purpose of $\bar{\chi}$ is to act as a continuous Lagrange multiplier, strictly enforcing Eq.~\eqref{Eq_Langevin_Chi} as a dynamical constraint across all space and time. To capture the macroscopic effect of the turbulence, the partition function must average this exponential weight over all spatiotemporal realizations of the underlying Gaussian white noise $\eta(\x,t)$.

\paragraph*{\small Fluctuation Averaging and Gaussian Integration.}
To average over the stochastic fluctuations, we must explicitly evaluate the expectation value of the noise-dependent exponential. Because $\eta(\x,t)$ is a standard spatiotemporal Gaussian white noise, its functional probability measure is defined by the bare Gaussian weight:
\be
d\mathcal{P}[\eta] = \mathcal{D}[\eta] \exp\left( -\frac{1}{2} \int \hspace{-.7ex}\d t \int \hspace{-.7ex}\d \x \, \eta^2 \right),
\ee
where the constant of normalization is conventionally absorbed into the formal flat measure $\mathcal{D}[\eta]$. The expectation value $\langle \dots \rangle_\eta$ translates directly into a functional integral over this probability measure. Combining the arguments of the exponentials yields the path integral formulation:
\begin{widetext}
\be\label{Eq_Noise_Expectation}
\Big\langle \exp\left( \int \hspace{-.7ex}\d t \int \hspace{-.7ex}\d \x \, \bar{\chi} \sigma \chi \eta \right) \Big\rangle_{\eta} = \int \mathcal{D}[\eta] \exp\left( \int \hspace{-.7ex}\d t \int \hspace{-.7ex}\d \x \left[ -\frac{1}{2} \eta^2 + \bar{\chi} \sigma \chi \eta \right] \right).
\ee
\end{widetext}
This formal integration instructs us to sum over every conceivable microscopic realization of the turbulent noise landscape. We evaluate it by completing the square in the exponent. Algebraically, the integrand can be rewritten as:
\be
-\frac{1}{2} \eta^2 + \bar{\chi} \sigma \chi \eta = -\frac{1}{2} \big( \eta - \sigma \bar{\chi} \chi \big)^2 + \frac{\sigma^2}{2} \bar{\chi}^2 \chi^2.
\ee
Substituting this  identity back into Eq.~\eqref{Eq_Noise_Expectation} allows us to pull the $\eta$-independent variance term outside of the functional integral, isolating a shifted Gaussian integral over the noise field:
\begin{widetext}
\be\label{Eq_key_eta_averaging}
\Big\langle \dots \Big\rangle_{\eta} = \exp\left( \frac{\sigma^2}{2} \int \hspace{-.7ex}\d t \int \hspace{-.7ex}\d \x \, \bar{\chi}^2 \chi^2 \right) \underbrace{ \int \mathcal{D}[\eta] \exp\left( -\frac{1}{2} \int \hspace{-.7ex}\d t \int \hspace{-.7ex}\d \x \, \big( \eta - \sigma \bar{\chi} \chi \big)^2 \right) }_{= 1},
\ee
\end{widetext}
because a uniform, field-dependent shift in the integration variable ($\eta \to \eta - \sigma \bar{\chi} \chi$) leaves the infinite-dimensional flat measure $\mathcal{D}[\eta]$ invariant, this residual Gaussian path integral perfectly normalizes to unity.

\paragraph*{\small  Emergence of the Bare Action.}
Executing this normalized integration leaves only the exponential of the residual variance. Because the statistical weight of the path integral is defined as $e^{-\Action}$, this noise integration exponentiates the variance and contributes $-\frac{\sigma^2}{2} \bar{\chi}^2 \chi^2$ to the bare action. This non-Hermitian quadratic term is the mathematical engine of the model; it explicitly captures how the intensity of turbulent fluctuations ($\sigma^2$) deforms the energetic landscape of the bath.

\paragraph*{\small The Complete Fast Bath Action.}
Combining this emergent fluctuation term with the linear deterministic terms enforcing the mean-field drift (diffusion, scavenging, and forcing), we obtain the complete bare action for the fast thermodynamic bath:
\bea\label{Eq_Action_Chi}
\Action_\chi = \int \d t \int \d \x \bigg[ &\bar{\chi} \left( \partial_t - D_\chi \nabla^2 + \lambda \right) \chi \\
&- \gamma \bar{\chi} - \frac{\sigma^2}{2} \bar{\chi}^2 \chi^2 \bigg].
\eea
{This expression collects the deterministic constraint and the noise-average contribution; it is not yet a complete midpoint-discretized action. In Section~\ref{Sec_Jacobian_regularization} we include the conversion drift and use the equivalent It\^o-discretized action for all subsequent integrations. This accounts for the Stratonovich prescription once, without retaining a second midpoint correction.} {The variance term is written here in its local approximation on scales larger than $\eta_K$. With the covariance in Eq.~\eqref{Eq_noise_covariance}, its spatially resolved form is $-\frac{\sigma^2}{2}\int\d t\int\d\x\int\d\x'\,\bar\chi\chi(\x)\rho_{\eta_K}(\x-\x')\bar\chi\chi(\x')$. We next complete the coupled droplet--supersaturation action before applying the stochastic conversion and eliminating the bath.}

\subsubsection{The Interaction Action ($\Action_{int}$) and the Mean-Field Droplet--Supersaturation System}\label{Sec_S_interaction}

To complete the  functional representation of the cloud layer, we must couple the two fields, capturing the non-equilibrium phase transitions (condensation and evaporation) that drive the system. We achieve this through the following final steps.

\paragraph*{\small The Deterministic Cross-Coupling.}
Returning to the 0D parcel model (Eq.~\eqref{Eq_growth_SR_lambda_stoch}), we isolate the explicit cross-coupling terms that were partitioned out of the previous isolated actions. The droplet growth is driven by the ambient vapor acting as a thermodynamic source ($+2D S$ in Eq.~\eqref{Eq_Gbs}). Conversely, the background vapor field experiences reciprocal depletion (or enhancement) due to condensation onto (or evaporation from) the active droplet population ($-\alpha_N \sqrt{R}\big(S - f(R)\big)$ in Eq.~\eqref{Eq_Gas}).

\paragraph*{\small  The Interaction Action.}
Transitioning to the macroscopic fields, these interaction terms are paired with their respective conjugate response fields ($\bar{\phi}$ and $\bar{\chi}$). Because the response fields enforce $\partial_t - \text{RHS} = 0$, the signs of these coupling terms are inverted when placed into the action. Summing these constraints yields the interaction action:
\be\label{Eq_Sint}
\hspace{-.4ex}\Action_{int} = \int \hspace{-.7ex}\d t \int  \hspace{-.7ex}\d \x \left[ -2D \bar{\phi} \chi + \alpha_N \bar{\chi} \sqrt{\phi} \big(\chi - f(\phi)\big) \right].
\ee
Physically, the first term ($-2D \bar{\phi} \chi$) dictates how the ambient vapor drives the macroscopic expansion of the droplet field. The second term, $\alpha_N \bar{\chi} \sqrt{\phi} \big(\chi - f(\phi)\big)$, encapsulates the collective thermodynamic feedback: the  macroscopic depletion shadow cast by the local droplet population onto the vapor bath.

\paragraph*{\small Deriving the Macroscopic Field Equations.}
To verify the integrity of our entire MSR construction, we derive the macroscopic spatial equations by taking the tree-level (classical) limit of the total action, $\Action = \Action_\phi + \Action_\chi + \Action_{int}$. This is achieved by applying the functional analogue of the principle of least action---taking the variational derivatives of $\Action$ with respect to the conjugate response fields $\bar{\phi}$ and $\bar{\chi}$. 

The terms linear in the response fields yield the deterministic drift, while the quadratic variance term in $\Action_\chi$ formally maps back to the stochastic noise source $\xi(\x,t) = \sigma \chi (\x,t) \eta(\x,t)$:
\bea\label{Eq_Mother_Langevin_Eq}
\frac{\delta \Action}{\delta \bar{\phi}(\x,t)} &= \left(\partial_t + \mu\right) \phi + 2D f(\phi) - 2D \chi = 0, \\
\frac{\delta \Action}{\delta \bar{\chi}(\x,t)} &= \left(\partial_t - D_\chi \nabla^2 + \lambda\right) \chi  \\
&\hspace{1cm} - \gamma + \alpha_N \sqrt{\phi}\big(\chi - f(\phi)\big)  \\
& = \sigma \chi (\x,t) \eta(\x,t).
\eea
By rearranging these variational derivatives, we recover the degenerate system of reaction-diffusion Langevin equations that  preserves the nonlinear K\"ohler microphysics of Eq.~\eqref{Eq_growth_SR_lambda_stoch} at its core, while elevating them to a continuous spatial topology:
\bea\label{Eq_coupled_RD}
\partial_t \phi(\x,t) &= -\mu \phi + 2D\big(\chi - f(\phi)\big) \\
\partial_t \chi(\x,t) &= D_\chi \nabla^2 \chi + \gamma - \lambda \chi \\
&\hspace{1cm}- \alpha_N \sqrt{\phi}\big(\chi - f(\phi)\big) + \sigma \chi {\circ}\eta(\x,t),
\eea
where $\eta(\x,t)$ is {the Gaussian noise of Eq.~\eqref{Eq_noise_covariance}, white in time and correlated over $\eta_K$ in space}; see {\it Supplementary Note 5} for a direct derivation from standard fluid-dynamical description of turbulent warm-cloud microphysics. Table \ref{tab:model_parameters} provides a summary of the units and physical meaning of the variables and parameters. 

This makes explicit what is being modeled: the theory evolves the cloud-scale fields $\chi(\x,t)$ and $\phi(\x,t)$, while the detailed droplet spectrum inside each local patch is projected onto the mean-square-radius field $\phi$. Higher spectral-shape diagnostics, such as the effective radius $r_e$, are not resolved explicitly in this reduced theory; their leading influence is instead absorbed into the mean-field  projection underlying  Eq.~\eqref{Eq_coupled_RD}.

\begin{table*}[ht]
\centering
\caption{
State variables and parameters of the degenerate droplet--supersaturation reaction--diffusion system \eqref{Eq_coupled_RD}. Diffusion acts only on the fast supersaturation bath $\chi$, while the slow droplet field $\phi$ evolves locally through condensation, K\"ohler feedback, and macroscopic loss.
}
\label{tab:model_parameters}
\begin{tabular}{@{}llcp{10cm}@{}}
\toprule
\textbf{Quantity} & \textbf{Symbol} & \textbf{Units} & \textbf{Physical Meaning} \\
\midrule
Mean-square-radius field
&
$\phi(\x,t)$
&
$[\mathrm{m}^2]$
&
Slow coarse-grained droplet-population field, $\phi=\langle R\rangle_{\x,t}=\langle r^2\rangle_{\x,t}$.
\\
Supersaturation bath
&
$\chi(\x,t)$
&
$[-]$
&
Fast resolved, coarse-grained supersaturation bath sampled by the droplets.
\\
\addlinespace
Diffusional growth coefficient
&
$D$
&
$[\mathrm{m}^2\,\mathrm{s}^{-1}]$
&
Microphysical condensational growth coefficient.
\\
Supersaturation diffusivity
&
$D_\chi$
&
$[\mathrm{m}^2\,\mathrm{s}^{-1}]$
&
Effective macroscopic scalar diffusivity of the supersaturation bath.
\\
\addlinespace
Updraft source
&
$\gamma$
&
$[\mathrm{s}^{-1}]$
&
Supersaturation production rate by adiabatic cooling ($A_1w$).
\\
Mixing-induced removal rate
&
$\lambda$
&
$[\mathrm{s}^{-1}]$
&
Average removal rate of supersaturation via environmental mixing.
\\
Droplet-field loss rate
&
$\mu$
&
$[\mathrm{s}^{-1}]$
&
{Effective relaxation of the normalized size moment by selective loss or replacement; uniform number loss alone does not generate it.}
\\
\addlinespace
Condensational supersaturation sink
&
$\alpha_N$
&
$[\mathrm{m}^{-1}\,\mathrm{s}^{-1}]$
&
Bulk supersaturation-depletion strength ($\propto$ aerosol number $N$) of the droplet population
\\
Spatiotemporal white noise
&
$\eta(\x,t)$
&
$[\mathrm{m}^{-d/2}\,\mathrm{s}^{-1/2}]$
&
{Gaussian noise from unresolved turbulent mixing, white in time and correlated over $\eta_K$ in space.}
\\
Mixing fluctuation amplitude
&
$\sigma$
&
$[\mathrm{m}^{d/2}\,\mathrm{s}^{-1/2}]$
&
Amplitude of unresolved turbulent fluctuations in the environmental mixing rate.
\\
\bottomrule
\end{tabular}
\end{table*}
As shown by its full derivation reported in {\it Supplementary Note 5},  Eq.~\eqref{Eq_coupled_RD} may be viewed as a coarse-grained Eulerian counterpart of the turbulent-condensation framework of Siewert et al.~\citep{siewert2017statistical}, but with two important structural extensions. 
In \citep{siewert2017statistical}, the supersaturation is treated as a passive scalar advected by an incompressible turbulent velocity field and coupled to Lagrangian droplets through a particle-resolved condensation/evaporation sink. Their subsequent stochastic reduction therefore follows the joint evolution of droplet surface area and supersaturation along droplet trajectories. By contrast, the present derivation first coarse-grains the droplet population and introduces the mean-square-radius field $\phi(\x,t)=\langle R\rangle_{\x,t}$ together with the resolved supersaturation bath $\chi(\x,t)=\langle s\rangle_{\x,t}\simeq \overline{s}(\x,t)$.

Our derivation also starts from a density-weighted supersaturation balance which contains the additional expansion/dilution term $-s\bar\rho^{-1}\partial_{x_i}(\bar\rho u_i)$. Combining this term with $\ub\cdot\nabla s$ into the conservative flux $\bar\rho^{-1}\partial_{x_i}(\bar\rho u_i s)$ is what allows the coarse-graining to separate three effects: resolved advection by $\bar\ub$, turbulent scalar transport closed as an eddy diffusivity $D_\chi=\kappa_\chi+K_\chi$, and local density-weighted exchange with the environment. The final system \eqref{Eq_coupled_RD} is written in the co-moving frame advected by $\bar\ub$, so the resolved transport term is absorbed into the material derivative, while the local exchange survives as $-\lambda\chi+\sigma\chi\eta$. Thus, relative to \citep{siewert2017statistical}, Eq.~\eqref{Eq_coupled_RD} does not only add K\"ohler thermodynamics through $f(\phi)$; it also converts the Lagrangian particle--scalar description into a spatially extended mean-field theory in which turbulent transport, density-weighted dilution, entrainment, and macroscopic stochastic modulation of supersaturation appear explicitly.

A final consistency point is the positivity of the droplet field. Since the reduced system contains both $\sqrt{\phi}$ and the singular K\"ohler term $f(\phi)=A\phi^{-1/2}-B\phi^{-3/2}$, the physical phase space is $\phi>0$, even though the supersaturation field $\chi$ is allowed to fluctuate freely into negative, subsaturated regimes. This positivity is not imposed by hand. As shown in {\it Supplementary Note 3}, for strictly positive initial data, the droplet equation preserves positivity pathwise: once a realization of $\chi$ is fixed, the $\phi$-equation is local in space and contains no direct stochastic forcing. Near $\phi=0^+$, the Raoult contribution dominates, giving $-2Df(\phi)\sim 2DB\phi^{-3/2}$ and hence a singular repulsive barrier. A barrier estimate with $V=\phi^{-1/2}$ rules out finite-time contact with $\phi=0$ whenever $\chi(\x,\cdot)\in L^2_{\rm loc}$ for almost every $\x$ and realization.  Thus $\phi(\x,t)>0$ for almost every $\x$ and almost surely on each finite interval on which this bath regularity holds, provided $\phi_0>0$; the regularized estimates supporting this condition are given in {\it Supplementary Note 4}. Consequently, the interaction term $\sqrt{\phi}$ is well-defined, and the functional integration measure $\mathcal{D}[\phi]$ in the partition function (Eq.~\eqref{Eq_Partition_Function}) is justified in being supported solely on non-negative field configurations.

\subsection{Slow Effective Action via the Gurari Variational Method}\label{Sec_slow_effective_Gurari}
We now derive an effective description for the slow droplet field by integrating out the fast supersaturation bath. The construction is motivated by the variational treatment of the polaron problem, in which a slow degree of freedom is coupled to a fast fluctuating medium and becomes dressed by the response of that medium \citep{Devreese2009}. In Gurari's formulation \citep{gurari1953}, the key step is not to expand around the unperturbed fast bath, but to shift the fast modes around the conditional response generated by the slow variable. In the present setting, this means that the supersaturation bath is first displaced to a dressed background $\chi_0^*(\phi)$ induced by the droplet field (Section \ref{Sec_Gurari_shift}), and only the residual fluctuations around this background are then integrated perturbatively (Sections \ref{Sec_Integrating_out_Fast} and \ref{Sec_Markovian_Approx}).

The same structure appears here at the level of the MSR action. The slow field is the mean-square-radius field $\phi$, while the fast bath is the supersaturation field $\chi$. We therefore use a Gurari-type variational shift of the fast field inside the path integral to compute the effective influence of supersaturation fluctuations on the droplet sector. This procedure yields a reduced action $\Action_{\rm eff}[\phi,\bar\phi]$ for the slow cloud field alone. The steps leading to this action are described below.

\subsubsection{Step 1:  The Full Fast Bath Action}
We first isolate the portion of the total MSR action governing the fast supersaturation field $\chi$ and its response field $\bar{\chi}$. We treat the slow macroscopic field $\phi$ and its response $\bar{\phi}$ as a frozen background, which is the functional analogue of the adiabatic approximation. This fast action is constructed by gathering all terms from the isolated fast thermodynamic bath action $\Action_\chi$ (Eq.~\eqref{Eq_Action_Chi}, derived in Section \ref{Sec_FastBath_Action}) and the interaction action $\Action_{int}$ (Eq.~\eqref{Eq_Sint}) that explicitly contain $\chi$ or $\bar{\chi}$. The partition function for this fast sub-system is:
\be
\mathcal{Z}_{fast}[\phi, \bar{\phi}] = \int \mathcal{D}[\chi] \mathcal{D}[\bar{\chi}] \exp\left( - \Action_{fast}[\chi, \bar{\chi}, \phi, \bar{\phi}] \right),
\ee
where the fast action takes the form:
\bea\label{Eq_S_fast}
\Action_{fast} &= \int \d t \int \d \x \bigg[ \bar{\chi} \big( \partial_t - D_\chi \nabla^2 + \Gamma(\phi) \big) \chi  \\
&- \bar{\chi} \big( \gamma + \alpha_N \sqrt{\phi} f(\phi) \big) - \underbrace{\frac{\sigma^2}{2} \bar{\chi}^2 \chi^2}_{\text{Quartic Variance}} - 2D \bar{\phi} \chi \bigg].
\eea
Notice that by grouping the background scavenging term ($\lambda \bar{\chi} \chi$) from $\Action_\chi$ with the linear condensation sink ($\alpha_N \bar{\chi} \sqrt{\phi} \chi$) from $\Action_{int}$, we have formally defined:
\be\label{Eq_Gamma_phi}
\Gamma(\phi) = \lambda + \alpha_N \sqrt{\phi},
\ee 
which represents the total, state-dependent phase-relaxation rate of the supersaturation field.

An immediate inspection of Eq.~\eqref{Eq_S_fast} reveals a critical mathematical bottleneck. It is essential here to distinguish between the statistics of the environmental noise and the statistics of the resulting field theory. While the underlying turbulent noise $\eta(\x,t)$ is strictly Gaussian---allowing us to  integrate it out in the previous section---its \textit{multiplicative} coupling to the vapor field ($\sigma \chi \eta$) dictates that its residual variance is $-\frac{\sigma^2}{2} \bar{\chi}^2 \chi^2$. Because this term is quartic (degree four) in the integration fields, it represents a nonlinear field-field interaction. This quartic term renders the subsequent functional integration over $\mathcal{D}[\chi]$ and $\mathcal{D}[\bar{\chi}]$ strictly non-Gaussian and analytically intractable. 
The latter aspect is dealt  with in Section \ref{Sec_Gurari_shift} below. We explain next how turbulent fluctuations  in the supersaturation field renormalize the fast action. 

\subsubsection{{Step 2: Stochastic Convention and Turbulent Dressing}}\label{Sec_Jacobian_regularization}

{The physical bath equation is specified in the Stratonovich convention, as the white-time limit of the modeled finite-correlation mixing fluctuations. Before eliminating the bath, we express this same process in It\^o form. With the spatial covariance of Eq.~\eqref{Eq_noise_covariance}, the conversion adds the term 
\be
c_K\chi,\textrm{ with } c_K=\tfrac12\sigma^2\rho_{\eta_K}(0)=\frac{\sigma^2}{2V_c},
\ee
to its drift. Appendix~\ref{App_Jacobian_Regularization} relates this correction to the response-field representation and introduces the forcing calibration $c_K=\sigma_S^2/\tau_K$. The dimensionless parameter $\sigma_S^2$ specifies the integrated strength of the unresolved mixing fluctuations; it is not automatically the stationary variance of $\chi$.

The corresponding action contribution is
\be\label{Eq_deltaS_sigmaS_tauK}
\Delta\Action_{\rm Jac}=-\int\d t\,\d\x\,\frac{\sigma_S^2}{\tau_K}\bar\chi\chi.
\ee
The correction reduces the relaxation coefficient in the It\^o drift,
\be\label{Eq_Gamma_star}
\Gamma^*(\phi)=\lambda+\alpha_N\sqrt\phi-\frac{\sigma_S^2}{\tau_K}.
\ee
All subsequent response-field integrations use the It\^o discretization. No additional midpoint Jacobian is to be added to the resulting dressed action,
\bea\label{Eq_S_fast_dressed}
\Action_{fast}^{\rm dressed}=\int\d t\,\d\x\,\bigg[&\bar\chi(\partial_t-D_\chi\nabla^2+\Gamma^*)\chi\\
&-\bar\chi\big(\gamma+\alpha_N\sqrt\phi f(\phi)\big)\\
&-\frac{\sigma^2}{2}\bar\chi^2\chi^2-2D\bar\phi\chi\bigg].
\eea
Here the local variance vertex is the large-scale form of the spatially regularized covariance. Turbulent dressing denotes the modification of the drift relative to the noise-free model at fixed bare exchange rate. Converting between Stratonovich and It\^o descriptions does not itself change the physical process; eliminating the bath is the separate step that produces memory and induced droplet-field noise.}

The dressing Eq.~\eqref{Eq_Gamma_star} carries an interesting physical implication when put into perspective with the bifurcation analysis of our parcel model (Section \ref{Sec_saddle_node_bif}). 
In the macroscopic field theory, the negative Stratonovich drift ($-\sigma_S^2/\tau_K$) appearing in  Eq.~\eqref{Eq_Gamma_star}  acts as a continuous outward thermodynamic pressure, directly counteracting the background mixing sink $\lambda$. By effectively lowering the total phase-relaxation rate, the multiplicative character of the turbulent noise dynamically shifts the macroscopic steady states of the field. As our parcel model's phase portraits revealed (Figure \ref{Fig_PhasePortrait_dep}), strong environmental mixing drags the slow manifold downward in the phase portrait, pushing the steady states deeper toward the sub-saturated regime. By actively fighting this downward drag, the turbulent pressure pulls the steady states back up towards the K\"ohler peak. As shown below, this feature effectively shrinks the ``forbidden zone'' and dictates that a  higher physical mixing rate is required to force the saddle-node  threshold. 

\subsubsection{Step 3: The Gurari Variational Shift and the Dressed Slow Manifold}\label{Sec_Gurari_shift}

Although the turbulent dressing successfully regularizes the measure and modifies the linear drift, the dressed fast action still retains its analytically intractable, non-Gaussian quartic structure ($-\frac{\sigma^2}{2} \bar{\chi}^2 \chi^2$). {To organize the approximation around the conditional bath response, we follow Gurari's variational insight} \cite{gurari1953} by executing a non-linear shift of the field origin. 

{Following Gurari, we treat the origin of the fast field as a variational object. We write $\chi=\chi_s+\delta\chi$, where the displacement $\chi_s(\x,t)$ is a trial field, a functional of the slow droplet field, to be determined. In Gurari's treatment of the polaron, each lattice mode is displaced by an amplitude fixed by the stationarity of the energy in the trial state, $\delta\langle\psi|H|\psi\rangle=0$ [Eq.~(2.12) of Ref.~\citep{gurari1953}]; as he shows in his Section~3, this condition amounts to completing the square in the linear electron--phonon coupling, that is, to displacing each oscillator to the minimum of its shifted potential. Feynman's path-integral treatment rests on the same operation: shifting the trajectory to the extremum of a quadratic exponent leaves the remainder only quadratically and decouples it from the source [Eqs.~(24)--(28) of Ref.~\citep{feynman1955slow}]. The MSR counterpart is the stationarity of the dressed fast action with respect to the response field at the displaced configuration,
\be\label{Eq_Gurari_stationarity_cond}
\frac{\delta \Action_{fast}^{\rm dressed}}{\delta\bar\chi(\x,t)}\bigg|_{\chi=\chi_s,\;\bar\chi=0}=0,
\ee
which removes from the action every term linear in $\bar\chi$ that does not involve the residual $\delta\chi$. Using Eq.~\eqref{Eq_S_fast_dressed}, this condition reads
\bea\label{Eq_Gurari_stationarity}
&\Big(\partial_t-D_\chi\nabla^2+\Gamma^*(\phi)\Big)\chi_s=J(\phi),\\
&J(\phi)=\gamma+\alpha_N\sqrt{\phi}f(\phi),
\eea
so that the optimal displacement is the retarded, spatially screened response of the dressed bath to the droplet source. For a frozen relaxation rate $\Gamma_0^*$, its Fourier representation is
\be\label{Eq_Gurari_displacement}
\tilde\chi_s(k,\nu)=\frac{\tilde J(k,\nu)}{\Gamma_0^*+D_\chi k^2-i\nu}.
\ee
This is the structure of Gurari's optimal amplitudes [Eq.~(2.20) of Ref.~\citep{gurari1953}], which in units $\hbar^2/2m=1$ are proportional to the coupling divided by $u^2+w^2-2(k-\gamma)wx$. The phonon frequency, $u^2=2m\omega/\hbar$, corresponds to the relaxation rate $\Gamma_0^*$; the recoil energy $w^2$ of the electron to the diffusive cost $D_\chi k^2$ of the supersaturation response; and the motion-dependent term $-2(k-\gamma)wx$, from which the polaron effective mass follows, to the lag $-i\nu$ produced by the evolution of the droplet field. Accordingly, the polaron radius $u^{-1}=(\hbar/2m\omega)^{1/2}$ corresponds to the screening length $L_{\rm scrn}=(D_\chi/\Gamma^*)^{1/2}$ of Section~\ref{Sec_Adabatic_Slaving}, and the effective mass to the lag-induced renormalization derived in Section~\ref{Sec_Effective_Mass}. One difference should be noted: in Gurari's and Feynman's Hamiltonian settings, the variational energy is an upper bound on the ground-state energy, which Feynman obtains from Jensen's inequality for a positive path weight [his Eq.~(10)]. The MSR weight is not positive and the action has no extremal property, so the displacement is selected by stationarity alone.

In the remainder of this section we work in the zero-diffusion, adiabatic limit of Eq.~\eqref{Eq_Gurari_stationarity}, $D_\chi\nabla^2\chi_s\to0$ and $\partial_t\chi_s\to0$; the screening and the lag that it discards are restored in Section~\ref{Sec_Dynamic_Inertia}. In this limit the displacement is the local balance in which the moisture source equals the dressed relaxation sink, which we denote by $\chi_0^*(\phi)$:}
\be
0 = \underbrace{\gamma + \alpha_N \sqrt{\phi} f(\phi)}_{\text{Source Force}} - \underbrace{\Gamma^*(\phi) \chi_0^*}_{\text{Dressed Relaxation Sink}}.
\ee
Solving this balance and using Eq.~\eqref{Eq_Gamma_star} yields this dressed equilibrium:
\be\label{Eq_chi_star}
\boxed{\chi_0^*(\phi) = \frac{\gamma + \alpha_N \sqrt{\phi} f(\phi)}{\Gamma(\phi) - \frac{\sigma_S^2}{\tau_K}}.}
\ee

To understand the profound physical implication of this variational shift, we map $\chi_0^*(\phi)$ directly to the deterministic topological features established in our 0D parcel analysis (Sections \ref{Sec_steady_states}, \ref{Sec_saddle_node_bif}, and \ref{Sec_expo_attraction}). Algebraically, $\chi_0^*(\phi)$ serves as the  macroscopic field-theoretic analogue of the  slow manifold $g(R)$ derived in Eq.~\eqref{Eq_balance_S} in the 0D case, that we will also refer to  as the {\it (turbulently) dressed slow manifold}. {Just as deterministic supersaturation rapidly tracks $g(R)$ with the lag quantified in Eq.~\eqref{Eq_attraction}}, the Gurari shift formalizes this timescale separation at the field level by strictly slaving the fast bath's mean-field configuration to the dressed slow manifold $\chi_0^*(\phi)$.

Furthermore, this equivalence allows us to formally trace the fate of the critical saddle-node bifurcation under the influence of turbulence. Assuming a spatially homogeneous steady state ($\partial_t \phi = 0$ with baseline decay $\mu=0$), the macroscopic droplet field stabilizes precisely when this dressed manifold intersects the K\"ohler curve: $\chi_0^*(\phi) = f(\phi)$. Substituting Eq.~\eqref{Eq_chi_star} and rearranging the terms yields:
\be
\gamma = f(\phi) \left( \lambda - \frac{\sigma_S^2}{\tau_K} \right).
\ee
Defining the turbulently dressed mixing rate as $\lambda^* \equiv \lambda - \sigma_S^2/\tau_K$, this balance perfectly recovers the structure of the cubic polynomial $P_{\lambda^*}(\sqrt{\phi}) = 0$ (Eq.~\eqref{Eq_cubic}) that governed the deterministic steady states. Crucially, because the bifurcation is now governed by $\lambda^*$, the critical mixing threshold for the saddle-node transition (Eq.~\eqref{Eq_discriminant}) is dynamically shifted:
\be\label{Eq_lambda_critical_dressed}
\lambda^\ast =\lambda_c \implies  \lambda_c^{eff}=\lambda_c + \frac{\sigma_S^2}{\tau_K}.
\ee
This suggests a physical mechanism in which  turbulence actively delays the onset of the twilight regime. For the stochastic 0D parcel model (Eq.~\eqref{Eq_growth_SR_lambda_stoch}), we have shown that as soon as the physical mixing rate exceeds $\lambda_c$, runaway droplet activation can be suppressed, leaving room for hovering twilight clouds to emerge (Section \ref{Sec_reducedSDE_main}). However, by contributing a continuous outward pressure via the Stratonovich drift, the spatio-temporal multiplicative noise effectively counteracts this background mixing. It pushes the saddle-node bifurcation to higher mixing rates, thereby shrinking the ``forbidden zone.'' Specifically, for a fixed $\lambda > \lambda_c$, the interval of homogeneous field amplitudes where $\chi_0^*(\phi) < f(\phi)$ is strictly narrower than the corresponding interval $g(R) < f(R)$. Ultimately, because the twilight cloud only exists within this saddle-node regime, turbulence dictates that the environment must reach a harsher mixing rate ($\lambda > \lambda_c^{eff}$) to suppress activation and force the system into the hovering state.

Having established the topological significance of this dressed manifold, we now shift our integration variables to center around {the optimal displacement: $\chi = \chi_s + \delta\chi$, with $\chi_s$ solving Eq.~\eqref{Eq_Gurari_stationarity} and reducing to $\chi_0^*(\phi)$ in the zero-diffusion, adiabatic limit}. {The shift absorbs the conditional mean response before the fluctuation expansion, but does not by itself control non-Gaussian residuals at arbitrary turbulent intensity. We retain the quadratic action of the residual bath and discard its higher vertices. This is a Gaussian residual-fluctuation closure, distinct from a global small-$\sigma$ expansion, whose quantitative validity requires the omitted cumulants or vertices to be small. In particular, the quadratic bath retains the additive forcing obtained by evaluating the multiplicative amplitude on the shifted background; it need not reproduce the full multiplicative bath's variance at strong forcing (Appendix~\ref{Sec_Emergent_Noise_Constraint}).}

Starting from the dressed fast action $\Action_{fast}^{\rm dressed}$ in Eq.~\eqref{Eq_S_fast_dressed}, we now expand the supersaturation bath around its shifted background, $\chi = {\chi_s} + \delta\chi$. The purpose of the Gurari shift is precisely to absorb the leading bath response into {$\chi_s$} before performing any fluctuation expansion. {By the stationarity condition \eqref{Eq_Gurari_stationarity_cond}, the shifted action contains no term linear in $\bar\chi$ alone.}
Thus the remaining approximation is not a small-$\sigma^2$ expansion of the original turbulent bath, but a Gaussian truncation of the residual fields $(\delta\chi,\bar\chi)$ around the dressed background. In the MSR formulation of Langevin dynamics with multiplicative noise, a state-dependent noise amplitude generates a nonlinear response-field vertex quadratic in the response field and quadratic in the noise amplitude \citep{Aron_al2010,aron2016dynamical}. Here, since the fast-bath noise amplitude is $\Sigma(\chi)=\sigma\chi$ and $\chi={\chi_s}+\delta\chi$, this vertex becomes
$-\frac{\sigma^2}{2}\bar\chi^2({\chi_s}+\delta\chi)^2$.

Keeping only the quadratic part of the residual fast theory retains the dominant variance term
$-\frac{\sigma^2}{2}{\chi_s^2}\bar\chi^2$,
while discarding the cubic and quartic residual vertices proportional to $\bar\chi^2\delta\chi$ and $\bar\chi^2(\delta\chi)^2$. {At the order retained below, this variance vertex is evaluated on the local component of the displacement, $\chi_s^2\simeq(\chi_0^*)^2$; the screening and lag corrections to $\chi_s$ would contribute to the noise only at higher order.}
This is the analogue, in the present stochastic field theory, of Gurari's {restriction of each lattice mode to at most one quantum, whose corrections vanish in the large-volume limit [Eqs.~(2.14) and (3.13)--(3.14) of Ref.~\citep{gurari1953}]}: the leading dressed bath response is kept nonperturbatively, while residual non-Gaussian fluctuation vertices are neglected at closure level.

Using the cascade scaling relation $\sigma^2 = 2\sigma_S^2V_c/\tau_K$, with $V_c$ the Kolmogorov volume defined in Appendix~\ref{App_Jacobian_Regularization}, the dressed fast action reduces to the Gaussian conditional action
\bea
\label{Eq_S_quad}
\Action_{quad} &= \int \d t \int \d \x \bigg[ \bar\chi \big( \partial_t - D_\chi\nabla^2 + \Gamma^*(\phi) \big) \delta\chi  \\
& - \, \frac{\sigma_S^2V_c}{\tau_K} \big(\chi_0^*(\phi)\big)^2 \bar\chi^2 - 2D\bar\phi \big( {\chi_s}+\delta\chi \big) \bigg].
\eea
This action is quadratic in the fast residual fields and can therefore be integrated exactly. The coefficient of $\bar\chi^2$ defines the induced covariance of the effective fast-bath fluctuations transmitted to the slow droplet field. Its scaling is physically transparent: it is proportional to the {dimensionless mixing-fluctuation strength} $\sigma_S^2$, distributed over the Kolmogorov volume $V_c$, injected over the Kolmogorov timescale $\tau_K$, and weighted by the squared dressed background $(\chi_0^*)^2$ around which the residual bath fluctuations are expanded.

\subsubsection{Step 4: Integrating out the Fast Fluctuation Field and Turbulence-Induced Noise Correlation}
\label{Sec_Integrating_out_Fast}
Starting from the quadratic fast action $\Action_{quad}$ in Eq.~\eqref{Eq_S_quad}, we now integrate out the residual supersaturation fluctuation field $\delta\chi$. The action separates naturally into three pieces: a dressed mean-field action ($\Action_{MF}$) proportional to $\chi_0^*(\phi)$, a linear fluctuation sector ($\Action_{lin}$) coupling $\delta\chi$ to the response fields, and a quadratic response-field variance term ($\Action_{var}$) inherited from the multiplicative turbulent bath. The linear sector is governed by the dressed fast operator
\bes
\mathcal L = \partial_t - D_\chi\nabla^2 + \Gamma^*(\phi).
\ees
Specifically,
\be
\Action_{lin} = \int \d t\int\d\x \, \left[ \bar\chi \mathcal L\delta\chi - 2D\bar\phi\delta\chi \right].
\ee
After integrating by parts in spacetime, this term may be written as
\be
\Action_{lin} = \int \d t\int\d\x \, \delta\chi \left[ \mathcal L^T\bar\chi - 2D\bar\phi \right],
\ee
where $\mathcal L^T$ is the adjoint operator. 
The role of $\Action_{lin}$ is therefore to eliminate the residual fast fluctuation $\delta\chi$ and impose the adjoint response constraint
\be\label{Eq_LT}
\mathcal L^T\bar\chi = 2D\bar\phi .
\ee
This is why the variance term, originally written in terms of the fast response field $\bar\chi$ in $\Action_{quad}$, becomes a quadratic functional of the slow response field $\bar\phi$: once the constraint Eq.~\eqref{Eq_LT} is imposed, $\bar\chi$ is the fast-bath response induced by $\bar\phi$. Substituting this constrained solution into $\Action_{var}$ and exchanging the order of integration yields a quadratic nonlocal contraction in the slow response field, with kernel $\mathcal C$ derived  in Appendix~\ref{App_Effective_Action}. 

Here we recall the main ingredients from this Appendix, and explain how it enables us to derive an effective slow acion. 
The logic goes as follows. The variance sector of $\Action_{quad}$ in Eq.~\eqref{Eq_S_quad} is
\be
\Action_{var} = -{\frac12} \int \d s\int\d\mathbf y \, V\big(\phi(\mathbf y,s)\big) \bar\chi(\mathbf y,s)^2,
\ee
where $V(\phi) = {2}\big(\chi_0^*(\phi)\big)^2 \sigma_S^2V_c/\tau_K.$
Solving Eq.~\eqref{Eq_LT}, gives
\be\label{Eq_constrained_sol}
\bar\chi(\mathbf y,s) = 2D \int \d t\int\d\x \, \mathcal G(\mathbf y,s;\x,t) \bar\phi(\x,t),
\ee
where $\mathcal G$ denote the Green's function of the adjoint problem $\mathcal L^T \mathcal G(\mathbf y,s;\x,t) = \delta(\mathbf y-\x)\delta(s-t).$
We now substitute into $\Action_{var} $ the constrained solution $\bar\chi$  slaved to the slow droplet response field $\bar\phi$ (Eq.~\eqref{Eq_constrained_sol}). 
Since $\bar\chi$ is itself an integral over $\bar\phi$, squaring it generates two independent  spacetime points, $(\x,t)$ and $(\x',t')$. Exchanging the order of integration yields a quadratic nonlocal contraction (Appendix \ref{App_Effective_Action}):
\bes
\Action_{var}\propto -\frac{1}{2} \int \d t \d\x \int \d t'\d\x'  \bar\phi(\x,t) \mathcal C(\x,t;\x',t') \bar\phi(\x',t').
\ees
with kernel:
\begin{widetext}
\be
\label{Eq_Corr_SpaceTime}
\mathcal C(\x,t;\x',t') =  \int \d s\int\d\mathbf y \, V\big(\phi(\mathbf y,s)\big) \mathcal G(\mathbf y,s;\x,t) \mathcal G(\mathbf y,s;\x',t').
\ee
\end{widetext}
{Here $V$ is the full white-noise covariance intensity of the Gaussian residual bath; the action coefficient is $-V/2$. Thus $\mathcal C$ is its propagated covariance, and $(2D)^2\mathcal C$ is the covariance transmitted to the droplet field. This convention is used throughout Eqs.~\eqref{Eq_Corr_OU_App}--\eqref{Eq_Sigma_phi} and Appendix~\ref{App_Effective_Action}.} 

Reuniting this stochastic contraction with the bare droplet action $\Action_\phi$ (Eq.~\eqref{Eq_S_phi}) and the dressed mean-field coupling ($\Action_{MF} = \int \d t \int \d \x \left[ -2D \bar{\phi} {\chi_s} \right]$) generated by {the optimal displacement $\chi_s$},
gives the effective slow action
\begin{widetext}
\be
\label{Eq_Seff}
\Action_{eff}[\phi,\bar\phi] = \int \d t\int\d\x \, \bigg[ \bar\phi(\partial_t+\mu)\phi - 2D\bar\phi \big( {\chi_s}-f(\phi) \big) \bigg] - \frac12 \int \d t\d\x \int\d t'\d\x' \, \big(2D\bar\phi(\x,t)\big) \mathcal C(\x,t;\x',t') \big(2D\bar\phi(\x',t')\big).
\ee
\end{widetext}
The first integral gives the deterministic dressed drift of the slow droplet field. It combines the bare droplet dynamics with the mean supersaturation response of the eliminated bath, grouping the K\"ohler curve $f(\phi)$ and the {optimal displacement $\chi_s$, the retarded and screened response of Eq.~\eqref{Eq_Gurari_stationarity}, into the restoring force $\chi_s-f(\phi)$. In the zero-diffusion, adiabatic limit, $\chi_s$ reduces to the dressed slow manifold $\chi_0^*(\phi)$ and the restoring force to $\chi_0^*(\phi)-f(\phi)$.} The second integral is the stochastic remnant of the fast bath. It says that the original linear coupling $-2D\bar\phi\delta\chi$ has contracted with the turbulent supersaturation variance through two adjoint bath propagators. Physically, the eliminated supersaturation fluctuations become an induced colored noise acting on $\phi$, with covariance $\mathcal C$ determined by the dressed bath response and the local intensity $V(\phi)$.

\subsubsection{Step 5: Markovian Approximation in the Zero-Diffusion Bath Limit}\label{Sec_Markovian_Approx}
To explicitly evaluate the correlation integral (Eq.~\eqref{Eq_Corr_SpaceTime}), we mathematically isolate the fast local relaxation by momentarily neglecting the spatial diffusion of the bath ($D_\chi \nabla^2 \to 0$). In this strictly local limit, the linear operator governing the bath simplifies to $\mathcal{L} \approx \partial_t + \Gamma^*(\phi)$. Consequently, the spatiotemporal Green's function defined by $\mathcal{L} \mathcal{G} = \delta(\x-\mathbf{y})\delta(t-s)$ collapses into a spatial Dirac delta multiplied by a pure temporal exponential decay:
\be\label{Eq_G_zero_diff}
\mathcal{G}(\mathbf{y},s; \x,t) \approx \delta(\mathbf{y}-\x) \Theta(t-s) e^{-\Gamma^*(\phi)(t-s)}.
\ee

This explicit functional form clarifies precisely under which conditions the dressed variance $V\big(\phi(\mathbf{y},s)\big)$ can be treated as a constant and factored out of the integral. While the integration variable $s$ in Eq.~\eqref{Eq_Corr_SpaceTime} spans the entire past history of the system, the key control parameter is the relaxation rate $\Gamma^*$. When the latter is large, the exponential kernel $e^{-\Gamma^*(t-s)}$ decays to zero almost instantaneously, acting as a strict temporal filter. In this case, the correlation integral receives non-negligible contributions only from a microscopic ``memory window'' where $t - s \lesssim 1/\Gamma^*$.

Because the droplet radius evolves on a slow, polynomial diffusion-limited timescale (Eq.~\eqref{Eq_Law_RS}), the macroscopic droplet field can be, in a first approximation, effectively frozen over this short temporal window. Furthermore, the spatial Dirac delta strictly localizes the interaction to $\mathbf{y} = \x$. Thus, under this approximation, $V\big(\phi(\mathbf{y},s)\big) \approx V\big(\phi(\x,t)\big)$ and can be factored entirely out of the convolution. Evaluating the remaining temporal integral for the correlation yields the  covariance of an Ornstein-Uhlenbeck (OU) process:
\be\label{Eq_Corr_OU_App}
\mathcal{C}(t, t') \approx V\big(\phi(\x,t)\big) \frac{e^{-\Gamma^* |t-t'|}}{2 \Gamma^*(\phi)}.
\ee
Because the slow macroscopic dynamics cannot resolve the microscopic memory timescale $1/\Gamma^*$, this exponentially decaying colored noise acts mathematically as a $\delta$-correlated stochastic forcing. We coarse-grain this memory into an effective Markovian white noise {whose intensity is the time integral of the covariance, that is, its zero-frequency spectral density: $\int\d\tau\,V e^{-\Gamma^*|\tau|}/(2\Gamma^*)=V/(\Gamma^*)^2$. Matching the equal-time variance instead would not preserve the low-frequency content of the forcing, which is what the slow droplet field integrates.}

Thus, the {intensity} of the induced macroscopic noise simplifies  to:
\be\label{Eq_Sigma_phi}
\boxed{{\Sigma_{\phi}(\phi) \approx (2D)^2 \int \d\tau\,\mathcal{C}(\tau) = \frac{{8} D^2 \sigma_S^2 V_c (\chi_0^*(\phi))^2}{ \tau_K \big(\Gamma^*(\phi)\big)^2}},}
\ee
with $\Gamma^*(\phi)$ given by Eq.~\eqref{Eq_Gamma_star} and $\chi_0^*(\phi)$ given by Eq.~\eqref{Eq_chi_star}.

However, we must strictly delineate the physical regime where this approximation remains valid. Because the dressed relaxation rate is 
$\Gamma^*(\phi) = \underbrace{\lambda + \alpha_N \sqrt{\phi}}_{\Gamma(\phi)} - \frac{\sigma_S^2}{\tau_K}$, 
increasing the turbulent intensity ($\sigma_S^2$) actively slows down the supersaturation relaxation. Under sufficiently strong turbulence, this critical slowing down expands the memory window, preventing the timescale separation and the Markovian approximation (Eq.~\eqref{Eq_Corr_OU_App}) from holding, provided the physical constraint Eq.~\eqref{Eq_cond_validity_noise} is met (see Appendix \ref{Sec_Emergent_Noise_Constraint}). In Section \ref{Sec_Dynamic_Inertia} below, we analyze the  consequences of a Markovian approximation breakdown. For the moment, we remain within this approximation to derive the closed Langevin equation for the droplet field evolution in the zero-diffusion limit of the supersaturation bath. 

\subsubsection{Step 6: The  Markovian Closure  of Droplet Field Dynamics}\label{Sec_Markov_closure}

To extract the macroscopic Langevin equation from the effective action (Eq.~\eqref{Eq_Seff}), we apply the functional analogue of the principle of least action by varying $\Action_{eff}$ with respect to the response field $\bar{\phi}(\x,t)$. However, before we can execute this variational principle, we must address the  non-local stochastic contraction term sitting at the end of the action:
\begin{widetext}
\be\label{Eq_Nonlocal_Term_Step5}
\Action_{stoch} = - \frac{1}{2} \int \d t \int \d \x \int \d t' \int \d \x' \, \bar{\phi}(\x,t) \Big[ 4D^2 \, \mathcal{C}(\x,t; \x',t') \Big] \bar{\phi}(\x',t').
\ee
\end{widetext}
In this  non-local representation, the response field $\bar{\phi}$ at coordinate ($\x,t$) is dynamically coupled to its historical and spatial neighbors at ($\x',t'$). If we were to decouple this interaction directly, it would birth a colored noise with complex spatiotemporal memory. 

To uncover the zeroth-order local dynamics, we go back to the local Markovian limit used at the end of Section \ref{Sec_Integrating_out_Fast}. In this limit, the underlying fast bath fluctuations encoded in $\mathcal{C}$ are assumed to be infinitely fast and infinitely localized compared to the slow, macroscopic droplet field, leading the Ornstein-Uhlenbeck temporal kernel in Eq.~\eqref{Eq_Corr_OU_App} to collapse into a temporal delta function, while the spatial operator remains localized to a point. Mathematically, we approximate the exact correlation operator by a state-dependent spatiotemporal Dirac delta function:
\be\label{Eq_C_OU_App}
4D^2 \, \mathcal{C}(\x,t; \x',t') \approx \Sigma_\phi\big(\phi(\x,t)\big) \, \delta(\x-\x')\delta(t-t'),
\ee
where $\Sigma_\phi(\phi)$ (Eq.~\eqref{Eq_Sigma_phi}) represents the {time-integrated intensity} of those local fluctuations. Substituting this delta-function approximation back into the  non-local double integral (Eq.~\eqref{Eq_Nonlocal_Term_Step5}), the spatiotemporal deltas instantly collapse the integration over $\x'$ and $t'$. Evaluating the fields at equal coordinates maps $\bar{\phi}(\x',t') \to \bar{\phi}(\x,t)$, forcing the double integral to collapse into a clean, single local variance term:
\be\label{Eq_Seff_var_collapsed}
\Action_{stoch} \approx - \frac{1}{2} \int \d t \int \d \x \, \bar{\phi}(\x,t)^2 \Sigma_{\phi}\big(\phi(\x,t)\big).
\ee
In the MSR formalism, an effective action containing this quadratic term in the response field ($-\frac{1}{2} \bar{\phi}^2 \Sigma_{\phi}$) cannot be deterministically set to zero via simple variation. Instead, this quadratic term formally encodes the variance of a Gaussian probability distribution over field histories. To recover the explicit stochastic trajectory of the droplets, we decouple this self-interaction by applying a Hubbard-Stratonovich transformation \cite{stratonovich1957method,hubbard1959calculation,byczuk2023generalized}. This transformation is mathematically devised to eliminate quadratic interaction terms from the partition function, while preserving the overall statistical weight of the system by trading those deterministic interactions for a linear coupling to a fluctuating auxiliary field. 

At its most fundamental scalar level, this method relies on the Gaussian integral identity \cite{byczuk2023generalized}:
\bea\label{Eq_HS_Scalar}
\exp\left( \frac{1}{2} a x^2 \right) &= \frac{1}{\sqrt{2\pi}} \int_{-\infty}^{\infty} \d y \, \exp\left( -\frac{1}{2} y^2 + x \sqrt{a} y \right)\\
& \equiv \big\langle \exp\big( x \sqrt{a} y \big) \big\rangle_y.
\eea
Here, the deterministic quadratic nonlinearity in $x$ is elegantly eliminated. In its place, $x$ appears purely linearly, coupled to a new dummy integration variable $y$, which is statistically weighted by a standard normal distribution ($\langle \dots \rangle_y$). 

By extending this algebraic identity from scalar variables to infinite-dimensional continuous fields, the integral over the stochastic contraction ($\bar{\phi} \mathcal{C} \bar{\phi}$) is mathematically responsible for birthing the macroscopic white noise $\eta_\phi$ in the final Langevin equation. Because we collapsed the spatiotemporal memory of $\mathcal{C}$ into the local scalar function of $\phi$, $\Sigma_\phi(\phi)$ in Eq.~\eqref{Eq_Seff_var_collapsed}, the transformation births an auxiliary, zero-mean spatiotemporal Gaussian noise field $\eta_{\phi}(\x,t)$ that is strictly delta-correlated (white) in both space and time, {with unit covariance intensity.} The Hubbard-Stratonovich transformation  precisely maps the quadratic response term back into a linear coupling with this white noise field according to:
\begin{widetext}
\be
\exp\left( \frac{1}{2} \int \d t \int \d \x \, \bar{\phi}(\x,t)^2 \Sigma_{\phi}(\phi) \right) = \Big\langle \exp\left( \int \d t \int \d \x \, \bar{\phi}(\x,t) \sqrt{\Sigma_{\phi}(\phi)} \eta_{\phi}(\x,t) \right) \Big\rangle_{\eta_{\phi}}.
\ee
\end{widetext}

Substituting this mathematical identity back into the path integral replaces the $\bar{\phi}^2$ variance term with a new auxiliary stochastic source term ($\bar{\phi} \sqrt{\Sigma_{\phi}(\phi)} \eta_{\phi}$) that is strictly linear in the response field $\bar{\phi}$. For any specific spatiotemporal realization of the auxiliary noise $\eta_{\phi}$, the effective action $\Action_{eff}$(Eq.~\eqref{Eq_Seff}) then takes the form:
\bea\label{Eq_Seff_final}
\Action_{eff}[\phi, &\bar{\phi},  \eta_{\phi}] = \int \d t \int \d \x \, \bar{\phi} \bigg[ \left( \partial_t + \mu \right) \phi  \\
&- 2D \Big( \chi_0^*(\phi) - f(\phi) \Big) - \sqrt{\Sigma_{\phi}(\phi)} \eta_{\phi}(\x,t) \bigg],
\eea
with the macroscopic noise variance $\Sigma_{\phi}(\phi)$ given by Eq.~\eqref{Eq_Sigma_phi}, evaluated strictly under the physical validity condition established in Eq.~\eqref{Eq_cond_validity_noise} in Appendix \ref{Sec_Emergent_Noise_Constraint}. 

Because the path integral over the response field $\mathcal{D}[\bar{\phi}]$ is performed along the imaginary axis, exponentiating this strictly linear action generates a functional Dirac delta. This  integration enforces the constraint that the term inside the brackets must vanish everywhere, yielding the {\it stochastic Markovian closure for coarse-grained droplet population}:
\bea\label{Eq_Macro_Langevin}
\partial_t \phi(\x,t) = -\mu \phi + 2D &\Big[ \chi_0^*(\phi) - f(\phi) \Big] \\
&+ \sqrt{\Sigma_{\phi}(\phi)} \eta_{\phi}(\x,t),
\eea
where $\chi_0^*(\phi)$ and $\Sigma_{\phi}(\phi)$ are given by Eq.~\eqref{Eq_chi_star} and Eq.~\eqref{Eq_Sigma_phi}, respectively. {We interpret this local equation in the It\^o discretization inherited from the retained effective action, with the spatial cutoff or a coarse-grained grid understood. The Hubbard--Stratonovich identity fixes the covariance, not a stochastic convention by itself. Additional homogenization-induced drift terms beyond the retained closure are not determined by matching the covariance alone.} 

\subsection{Turbulent Dressing and Dragged Supersaturation Envelope}\label{Sec_dragged_env}
Equation~\eqref{Eq_Macro_Langevin} provides the first mean-field closure for the droplet field, obtained at leading order under the Markovian and zero-diffusion bath approximations. Although local in this limit, the equation already carries nontrivial physical content. It describes the effective thermodynamics of the coarse-grained field $\phi$ after the nonlocal spatiotemporal Green's functions of the supersaturation bath have been collapsed to an instantaneous local response. Its deterministic part is therefore not the bare parcel drift with $\chi$ replaced by an ordinary mean value in the $\phi$-equation of the droplet--supersaturation  system (Eq.~\eqref{Eq_coupled_RD}). Rather, the fast supersaturation bath has been variationally shifted, as in Section~\ref{Sec_Gurari_shift}, producing the dressed slow manifold $\chi_0^*(\phi)$ given by Eq.~\eqref{Eq_chi_star}.

This gives the cloud analogue of Gurari's ``polarization envelope'' \cite{gurari1953}. In the polaron problem, the electron lowers its energy by dragging a deformation of the surrounding medium. Here, the slow object is not an individual microscopic droplet, but the local mean-square-radius field $\phi(\x,t)=\langle R\rangle_{\x,t}$, i.e.~the coarse-grained droplet population inside a cloud element. This local droplet population drags a statistically localized supersaturation envelope generated by the turbulent vapor bath. {The crucial contribution is the relaxation-rate correction $-\sigma_S^2/\tau_K$ appearing in $\chi_0^*(\phi)$. It follows from converting the Kolmogorov-scale multiplicative bath noise from Stratonovich to It\^o form (Section~\ref{Sec_Jacobian_regularization}), rather than from a phenomenological variance correction. An equivalent midpoint action retains the corresponding measure terms.}

Within Eq.~\eqref{Eq_chi_star}, {this conversion term} acts against the bare local phase-relaxation sink $\Gamma(\phi)$. It therefore weakens the effective drying/relaxation experienced by the local droplet population and sustains a turbulence-fed supersaturation response around the coarse-grained cloud element. This is the dragged envelope: not a microscopic vapor shell around each droplet, but a collective statistical vapor buffer attached to the slow field $\phi$. As anticipated in Section~\ref{Sec_Gurari_shift}, the effect is topological at the mean-field level: it shifts the saddle-node threshold controlling the transition into the hovering ``twilight'' regime. In the isolated parcel theory, runaway activation is halted once the mixing rate reaches the critical value $\lambda_c$. In the dressed theory, the turbulent envelope counteracts this drying, so that the effective threshold is displaced to $\lambda_c^{\rm eff}=\lambda_c+\sigma_S^2/\tau_K$; see Eq.~\eqref{Eq_lambda_critical_dressed}. Thus a stronger mixing environment is required to suppress activation of the local droplet population: turbulence shrinks the forbidden zone and delays the onset of the twilight hovering state.

The stochastic term in Eq.~\eqref{Eq_Macro_Langevin} has a complementary origin. {The drift already includes the Stratonovich-to-It\^o correction, whereas} the macroscopic multiplicative noise $\sqrt{\Sigma_\phi(\phi)}\eta_\phi$ is generated by functionally integrating over the fast supersaturation fluctuations around the dressed state. As carried out in Section~\ref{Sec_Integrating_out_Fast}, the contraction of the bath fluctuations produces the quadratic response contribution $\bar\phi\mathcal C\bar\phi$ in Eq.~\eqref{Eq_Seff}. In the Markovian limit this contraction yields the variance $\Sigma_\phi(\phi)$ in Eq.~\eqref{Eq_Sigma_phi}. The noise in Eq.~\eqref{Eq_Macro_Langevin} is therefore not postulated externally: it is the residual stochastic imprint of the turbulently dressed supersaturation landscape after the fast bath has been eliminated.

\section{The Non-Markovian Cloud Droplet Field Equation and Dynamic Inertia}\label{Sec_Dynamic_Inertia}

In the previous section, we derived the macroscopic dynamics in the strict local and Markovian limit ($D_\chi \nabla^2 \to 0$, $e^{-\Gamma^* |t-t'|} \to \delta(t-t')$). While this zeroth-order reduction beautifully isolates the turbulent dressing of the K\"ohler curve, it mathematically treats the cloud field as a collection of uncoupled, zero-dimensional stochastic parcels. The  effective action (Eq.~\eqref{Eq_Seff}), however, encodes a physically richer reality: droplets communicate across space via vapor diffusion, and the supersaturation bath retains a finite temporal memory. 

To systematically unpack these non-local effects, we proceed in two stages. First, we relax only the spatial isolation ($D_\chi \nabla^2 \neq 0$) while maintaining the instantaneous temporal slaving ($\partial_t \chi \approx 0$). This isolates the purely spatial architecture of droplet competition. Subsequently, in Section \ref{Sec_Low_Pass_Truncation}, we {write this screened response in local, pseudo-parabolic form, and in Section~\ref{Sec_Effective_Mass} we restore the finite temporal memory of the bath at first order in the ratio of slow to fast rates, which renormalizes the inertia of the droplet field. Both steps are expansions of the optimal Gurari displacement, Eq.~\eqref{Eq_Gurari_displacement}: the first keeps its diffusive cost $D_\chi k^2$, the second its lag $-i\nu$.}

\subsection{Nonlinear Adiabatic Slaving and the 2D Bessel Kernel}\label{Sec_Adabatic_Slaving}

To understand the spatial structures generated by our MSR field theory, let us examine the classical (tree-level) equation for the fast supersaturation bath derived in Eq.~\eqref{Eq_Mother_Langevin_Eq}. Ignoring the white noise momentarily to isolate the deterministic spatial operator, we group the terms linear in $\chi$. Defining the local relaxation rate $\Gamma(\phi) = \lambda + \alpha_N \sqrt{\phi}$ and a local thermodynamic source term $J(\phi) = \gamma + \alpha_N \sqrt{\phi} f(\phi)$, the deterministic bath evolution simplifies directly to:
\be\label{Eq_Bath}
\partial_t \chi = D_\chi \nabla^2 \chi - \Gamma(\phi) \chi + J(\phi).
\ee

Because the supersaturation field ($\chi$) diffuses and relaxes orders of magnitude faster than the macroscopic mean-square radius field ($\phi$) evolves, we apply the adiabatic slaving principle by setting $\partial_t \chi \approx 0$:
\bes
\big(\Gamma(\phi) - D_\chi \nabla^2\big) \chi \approx J(\phi).
\ees
Inverting this differential operator yields the ``slaved'' quasi-equilibrium supersaturation field:
\be
\chi_0(\phi) \approx \big(\Gamma(\phi) - D_\chi \nabla^2\big)^{-1} J(\phi).
\ee
Substituting the slaved bath back into the slow droplet equation, $\partial_t\phi = -\mu\phi + 2D\big(\chi - f(\phi)\big)$, produces the nonlocal macroscopic closure 
\be\label{Eq_Nonlocal_Slaved}
\partial_t \phi = -\mu \phi  + 2D \Big[ \mathcal{M}_{\phi} \big(\gamma + \alpha_N\sqrt{\phi}f(\phi)\big) - f(\phi) \Big],
\ee
where $\mathcal{M}_{\phi}$ denotes the inverse elliptic operator 
\be\label{Def_M_phi}
\mathcal{M}_{\phi} = \big(\Gamma(\phi) - D_\chi\nabla^2\big)^{-1}.
\ee 
Once the turbulent dressing of Section~\ref{Sec_Jacobian_regularization} is included, $\Gamma$ is replaced by $\Gamma^*$ of Eq.~\eqref{Eq_Gamma_star}; we denote the dressed operator by $\mathcal L_\phi=\Gamma^*(\phi)-D_\chi\nabla^2$, so that $\mathcal L_\phi^{-1}$ is the dressed counterpart of $\mathcal M_\phi$ (Section~\ref{Sec_Spatial_Markovian}).
Compared to the stochastic Markovian closure (Eq.~\eqref{Eq_Macro_Langevin}), this operator is the key new feature which replaces the purely local Markovian response in the drift part of Eq.~\eqref{Eq_Macro_Langevin} by a spatially mediated supersaturation response. To understand its meaning, freeze the slowly varying coefficient $\Gamma(\phi)$ over a local cloud element, and let us look at the Green's function $G$ of $\mathcal{M}_{\phi}$. 

In two dimensions, recall that the fundamental solution to the modified Helmholtz equation $(1 - L^2 \nabla^2)u = 0$ is the screened Bessel function of the second kind, $K_0(\rho/L)$ \cite{abramowitz1966handbook}, where $\rho = \|\x\|$ is the macroscopic radial distance. Then, 
\bea\label{Eq_Bessel_Kernel}
G(\rho) &= \frac{1}{2\pi D_\chi} K_0\left(\frac{\rho}{L_{\rm scrn}}\right), \\
L_{\rm scrn} &= \sqrt{\frac{D_\chi}{\Gamma(\phi)}}. 
\eea

Thus, under the local frozen-coefficient approximation, applying $\mathcal{M}_{\phi}$ is equivalent to convolving the source $J(\phi)$ with a screened Bessel kernel. The function $K_0$ has a weak logarithmic singularity at the origin and decays exponentially at large distance, $K_0(z) \sim \sqrt{\pi/(2z)} \, e^{-z}$ as $z \to \infty$ \cite{abramowitz1966handbook}. The length $L_{\rm scrn}$ is therefore the spatial range over which the fast supersaturation bath transmits the thermodynamic influence of one coarse-grained cloud element to its neighbors.

Physically, this is the point at which we enable the dragged supersaturation envelopes (Section \ref{Sec_dragged_env}) of neighboring cloud elements  to interact. In the Markovian and zero-diffusion bath limit, each coarse-grained droplet population, represented by $\phi(\x,t)$, is dressed by a statistically localized supersaturation envelope, but the response remains pointwise in space. Here the Markovian slaving assumption is still retained, $\partial_t\chi\simeq0$, but the zero-diffusion approximation is relaxed. 
The elimination of the fast-bath diffusive operator, causes then a nonlocal coupling on the droplet field level that is mediated by the  screened operator
$\mathcal M_\phi$.  Its Green's function gives the spatial range over which the vapor response attached to one cloud element contributes to the dressed response felt by neighboring elements. Thus the cloud becomes a self-screening thermodynamic medium: each local droplet population remains dressed by the turbulent supersaturation bath, but these dressed envelopes now overlap and couple nonlocally through the screened operator $\mathcal M_\phi$, with interaction range $L_{\rm scrn}=\sqrt{D_\chi/\Gamma(\phi)}$.

\subsection{Emergence of Dynamic Inertia in Droplet Field  Dynamics}\label{Sec_Low_Pass_Truncation}

The strict adiabatic limit explored in the previous section sets $\partial_t\chi\simeq0$: the supersaturation bath is assumed to adjust instantaneously to the slow droplet field. Once the zero-diffusion approximation is relaxed, this limit already reveals spatial screening through the inverse elliptic operator $\mathcal M_\phi$ (Eq.~\eqref{Def_M_phi}) and its Bessel Green's function. However, the response remains instantaneous in time. The dragged supersaturation envelope has a finite spatial range, but no intrinsic delay: the bath rearranges at once around the evolving coarse-grained droplet population.

We now relax this remaining adiabatic assumption. The vapor bath has not only a screening length $L_{\rm scrn}$, inherited from $\mathcal M_\phi$, but also a finite relaxation time of order $1/\Gamma^*$. This is the dynamical content of Gurari's polarization-envelope picture \cite{gurari1953}: the slow field does not merely sit inside a static medium, but drags a surrounding bath response that takes time to reorganize. Retaining $\partial_t\chi$ in the bath equation, equivalently the factor $-i\nu$ in Fourier space, keeps this temporal lag. {As shown below, the expansion of the resulting dynamic nonlocal response in the ratio $\delta$ of slow to fast rates has two levels. At leading order, it yields the mixed derivative term $\nabla\cdot(\epsilon\nabla\partial_t\phi)$, which is the local form of the instantaneous screened response of Section~\ref{Sec_Adabatic_Slaving} and records the spatial footprint of the halo. At first order, the lag itself appears and renormalizes the coefficient of $\partial_t\phi$ (Section~\ref{Sec_Effective_Mass}). Together they constitute the emergence of dynamic inertia: not a microscopic droplet mass effect, but the macroscopic consequence of dragging a spatially screened supersaturation envelope that adjusts with a finite delay.}

To expose this mechanism within the mean-field theory, we linearize the bath response around a representative background droplet state $\phi_0$. This linearization must be applied to the turbulently dressed bath obtained in Section~\ref{Sec_Integrating_out_Fast}, not to the bare supersaturation equation. We therefore freeze the dressed relaxation rate at the constant value
$\Gamma_0^*\equiv\Gamma^*(\phi_0)=\lambda+\alpha_N\sqrt{\phi_0}-\sigma_S^2/\tau_K$.
This fixes the screening and relaxation scale of the bath. However, the thermodynamic forcing of the bath is not frozen: we retain the full spatiotemporal dependence of
$J(\phi(\x,t))=\gamma+\alpha_N\sqrt{\phi(\x,t)}f(\phi(\x,t))$.
Thus the bath is treated as a linear dressed response medium driven by the evolving droplet field, allowing the finite lag of the supersaturation envelope to feed back into the macroscopic dynamics of $\phi$.

The dressed supersaturation bath before adiabatic elimination obeys
\be
\partial_t\chi = D_\chi\nabla^2\chi - \Gamma^*(\phi)\chi + J(\phi).
\ee
{We split the dressed relaxation term as $\Gamma^*(\phi)\chi=\Gamma_0^*\chi+\big[\Gamma^*(\phi)-\Gamma_0^*\big]\chi$ and keep the second part, which carries the dependence of the condensational sink $\alpha_N\sqrt\phi$ on the droplet field, in the source,
\be\label{Eq_effective_source}
\hat J(\phi,\chi)=J(\phi)-\big[\Gamma^*(\phi)-\Gamma_0^*\big]\chi .
\ee
This gives}
\be
\Big( \partial_t - D_\chi\nabla^2 + \Gamma_0^* \Big)\chi(\x,t) = {\hat J}.
\ee
Thus the fast bath is represented by a linear constant-coefficient response operator, whose Fourier symbol is $-i\nu + D_\chi k^2 + \Gamma_0^*$. {In the local adiabatic limit, $\hat J/\Gamma_0^*=\chi_0^*(\phi)$ exactly, since $\Gamma_0^*\chi=\hat J$ is then equivalent to $\Gamma^*(\phi)\chi=J(\phi)$. Linearized about the dressed slow manifold, $\hat J\simeq\Gamma_0^*(\chi_0^*)'(\phi_0)\,\delta\phi$, so that the bath responds to the slope of the dressed slow manifold, including the dependence of the sink on $\phi$; freezing $\Gamma^*$ inside $J$ alone would have retained only $J'(\phi_0)>0$ and lost the sign change discussed in Section~\ref{Sec_Phys_nonFickian}.}

The dynamically slaved supersaturation field in $(k, \nu)$ space is the algebraic inversion of this operator acting on the Fourier-transformed source, ${\tilde{\hat J}}(k, \nu)$:
\be\label{Eq_slaved_bath_sol}
\tilde{\chi}(k, \nu) = \frac{{\tilde{\hat J}}(k, \nu)}{\Gamma_0^* + D_\chi k^2 - i\nu}.
\ee
This is the optimal Gurari displacement, Eq.~\eqref{Eq_Gurari_displacement}, with the effective source.

In the time domain, the response in Eq.~\eqref{Eq_slaved_bath_sol} weights the past source at each spatial wavenumber $k$ by the causal kernel $e^{-(\Gamma_0^*+D_\chi k^2)s}$, $s\ge0$, together with the decaying contribution of the initial bath state. The present supersaturation therefore retains a spatially dependent memory of earlier forcing. Memory coupled to spatial transport also appears in viscoelastic flow and Coleman--Gurtin heat conduction, where integrals over past velocity or temperature gradients enter the constitutive response \cite{chekroun_glatt-holtz,Chek_al11_memo}. Here the kernel follows from the relaxation and diffusion of the supersaturation bath, providing the response whose low-frequency expansion is developed below.

To see the emergence of dynamic inertia, we substitute this dynamically slaved response back into the full macroscopic droplet equation. Recalling the $\phi$-equation in the droplet-supersaturation system Eq.~\eqref{Eq_coupled_RD} and temporarily omitting the stochastic noise for deterministic reduction, we get:
\be
\partial_t \phi = -\mu \phi - 2D f(\phi) + 2D \chi.
\ee
Taking the Fourier transform of this entire equation and substituting our slaved bath solution (Eq.~\eqref{Eq_slaved_bath_sol}), we obtain:
\be
-i\nu \tilde{\phi} = -\mu \tilde{\phi} - 2D \tilde{f} + 2D \left( \frac{{\tilde{\hat J}}}{\Gamma_0^* + D_\chi k^2 - i\nu} \right),
\ee
where $\tilde{f} = \mathcal{F}\{f(\phi)\}$. Gathering the bare droplet dynamics to the left-hand side:
\be
-i\nu \tilde{\phi} + \mu \tilde{\phi} + 2D \tilde{f} = \frac{2D {\tilde{\hat J}}}{\Gamma_0^* + D_\chi k^2 - i\nu}.
\ee

To unpack the dynamic nonlocal response, we multiply both sides by the denominator and factor out the dressed background relaxation rate $\Gamma_0^*$:
\be\label{Eq_Fourier_Full}
\left( 1 + \frac{D_\chi}{\Gamma_0^*}k^2 - \frac{i\nu}{\Gamma_0^*} \right) \Big( -i\nu\tilde{\phi} + \mu\tilde{\phi} + 2D\tilde f \Big) = \frac{2D}{\Gamma_0^*}{\tilde{\hat J}} .
\ee
To pass to the macroscopic limit, we now make explicit the separation between the fast dressed bath and the slow cloud dynamics. The bath relaxes on the time scale $\tau_0^* = (\Gamma_0^*)^{-1}$, while the resolved droplet field evolves on longer times. We therefore introduce a temporal low-pass projection $\mathcal P_\Lambda$ onto the slow band
\be
\Omega = \{\nu:\ |\nu|\le \Lambda\}, \qquad \Lambda \ll \Gamma_0^* .
\ee
The macroscopic fields $\tilde\phi$, $\tilde f$, and ${\tilde{\hat J}}$ are assumed to be supported predominantly in this band, so that they are essentially asymptotically invariant under $\mathcal P_\Lambda$.
Within this projected subspace, the temporal lag of the fast bath is controlled by the small parameter
\be\label{Eq_low_pass_band}
\left| \frac{\nu}{\Gamma_0^*} \right| \le \frac{\Lambda}{\Gamma_0^*} \equiv \delta \ll 1 .
\ee
Hence the factor $i\nu/\Gamma_0^*$ in the bath response is of order $\mathcal O(\delta)$ and may be discarded at leading order in the bath operator. This truncation does not remove the slow time derivative $-i\nu\tilde\phi$ appearing in the droplet dynamics; it only suppresses the subleading temporal lag inside the fast response denominator. By contrast, the spatial screening factor $D_\chi k^2/\Gamma_0^*$ is independent of $\nu$ and is retained. Thus the leading low-frequency approximation preserves the nonlocal spatial response of the supersaturation bath while eliminating only its higher-frequency temporal correction.
Applying $\mathcal P_\Lambda$ at order $\mathcal O(\delta^0)$ therefore reduces Eq.~\eqref{Eq_Fourier_Full} to
\be\label{Eq_low_pass_band2}
\left( 1 + \frac{D_\chi}{\Gamma_0^*}k^2 \right) \Big( -i\nu\tilde{\phi} + \mu\tilde{\phi} + 2D\tilde f \Big) = \frac{2D}{\Gamma_0^*}{\tilde{\hat J}} .
\ee
{Relaxing this low-pass truncation amounts to retaining higher orders in the small parameter $\delta$. Note that the discarded factor $-i\nu/\Gamma_0^*$ multiplies the whole bracket $-i\nu\tilde\phi+\mu\tilde\phi+2D\tilde f$, which equals $2D\tilde\chi$ and is therefore of order one in the slow rates, not only the slow derivative $-i\nu\tilde\phi$. Its first contribution is thus of order $\mathcal O(\delta)$, not $\mathcal O(\delta^2)$: in real time and at long wavelengths, it is the term $-(2D/\Gamma_0^{*2})\,\partial_t\hat J$ on the right-hand side, which records the lag of the bath behind the evolving droplet field and renormalizes the coefficient of $\partial_t\phi$. We derive it in Section~\ref{Sec_Effective_Mass}; terms such as $\partial_t^2\phi$ only appear at $\mathcal O(\delta^2)$.}

Transforming the truncated equation (Eq.~\eqref{Eq_low_pass_band2}) back to real space, using $k^2\to-\nabla^2$ and $-i\nu\to\partial_t$, we introduce the local dynamic-inertia coefficient $\epsilon(\phi_0) = D_\chi/\Gamma_0^*$.
At the frozen-coefficient level, the right-hand side is the slaved dressed bath response,
${\hat J}/\Gamma_0^*  = \chi_0^*(\phi)$, within the local linearized response approximation.

To obtain a nonlinear field equation, we then promote the background state $\phi_0$ to the full slowly varying field $\phi(\x,t)$. This gives the state-dependent coefficient
\be\label{Eq_epsilon_phi_dressed}
\epsilon(\phi) = \frac{D_\chi}{\Gamma^*(\phi)},
\ee 
and restores the nonlinear dressed manifold $\chi_0^*(\phi)$ from Eq.~\eqref{Eq_chi_star}. {At this stage we adopt a conservative nonlinear extension of the frozen-coefficient response, replacing $\epsilon(\phi)\nabla^2$ by $\nabla\cdot[\epsilon(\phi)\nabla]$. This is a closure choice, not an operator identity: the latter adds $\nabla\epsilon\cdot\nabla$ and therefore agrees with the former at linear order about a homogeneous state but changes nonlinear gradient terms. The exact instantaneous variable-coefficient elimination instead gives $(I-\epsilon(\phi) \nabla^2)[\partial_t\phi+\mu\phi+2Df(\phi)]=2D\chi_0^*(\phi)$. We use the divergence-form extension below to retain conservative spatial coupling, and interpret its nonlinear interface terms within that extension.} The resulting deterministic macroscopic droplet-field equation is a nonlinear pseudo-parabolic equation of Barenblatt--Zheltov--Kochina type \cite{barenblatt1960basic,barenblast1960fundamental}:
\be\label{Eq_pseudo_PDE1}
\Big( I - \nabla\cdot\epsilon(\phi)\nabla \Big) \Big( \partial_t\phi + \mu\phi + 2D f(\phi) \Big) = 2D\chi_0^*(\phi).
\ee
At this order, Eq.~\eqref{Eq_pseudo_PDE1} is the local, differential form of the instantaneous screened closure of Section~\ref{Sec_Adabatic_Slaving}: for frozen coefficients, applying $ \mathcal L_\phi=\Gamma_0^*\big(I-\epsilon\nabla^2\big)$ to Eq.~\eqref{Eq_Nonlocal_Slaved} and dividing by $\Gamma_0^*$ gives Eq.~\eqref{Eq_pseudo_PDE1}. Its mixed derivative therefore records the spatial footprint of the dragged halo, not yet its delay; the delay enters at $\mathcal O(\delta)$ (Section~\ref{Sec_Effective_Mass}).

The mixed spatiotemporal derivative $\nabla \cdot \epsilon(\phi) \nabla \partial_t$ is what gives this equation its pseudoparabolic character \citep{showalter1970pseudoparabolic}. Pseudoparabolic models  are uniquely suited for describing physical systems dominated by non-local interactions, where the dynamics at a given point in space are fundamentally influenced by conditions at distant points \cite{carroll1977singular}. The inclusion of a time derivative on higher-order spatial gradient terms allows the model to account for the temporal evolution of these non-local effects. These models arise across a wealth of macroscopic physics applications, including radiative transfer theory \cite{sobolev1963treatise}, heat conduction in materials with two distinct temperatures \cite{Chen_Gurtin1968theory}, and fluid seepage in fissured rock formations\cite{barenblatt1960basic}. In continuum mechanics, these equations are instrumental for analyzing non-Newtonian fluid flows, providing a framework to describe fluids whose stress-strain relationships depend heavily on memory and non-local effects. This is particularly prevalent in the modeling of viscoelasticity \cite{joseph1985hyperbolicity}, where materials exhibit a complex amalgamation of both elastic (solid-like) and viscous (liquid-like) properties \cite{Coleman_Noll1974, huilgol1968second, ting1963certain}.

 A related spatial coupling of the rate of change occurs in anelastic convection: eliminating perturbation pressure yields an elliptic equation for the vertical mass-flux tendency, so that the acceleration depends on the spatial distribution of forcing \cite{kuo2022conditions,kuo2025anelastic}. There the coupling enforces mass continuity through pressure adjustment; here it emerges from supersaturation transport and relaxation, which couple the growth tendencies of neighboring droplet populations.

{Thus, in the cloud-physics setting, the leading-order truncation that produced Eq.~\eqref{Eq_pseudo_PDE1} exposes the spatial half of the mechanism: the mixed derivative is the macroscopic signature of the footprint of the screened supersaturation response. A change of the coarse-grained droplet population at position $\x$ alters the bath, and hence the growth tendency, over the whole screening disk of radius $L_{\rm scrn}$, so that the rate of change of $\phi$ at $\x$ is tied to its rate of change over that disk. The bath response is transmitted through the screened operator inherited from $\Gamma^*(\phi)-D_\chi\nabla^2$, and the mixed derivative term records this shared adjustment between the evolving droplet field and its surrounding supersaturation envelope. Its temporal half, the delay of the envelope, is the subject of Section~\ref{Sec_Effective_Mass}.}

The central structure is the pseudoparabolic operator $\big(I-\nabla\cdot\epsilon(\phi)\nabla\big)\partial_t$. To see how it alters the evolution of sharp spatial gradients, we examine its spectral response. Linearizing Eq.~\eqref{Eq_pseudo_PDE1} around a constant background state $\phi_0$, and momentarily fixing $\epsilon(\phi_0)\simeq\epsilon$, we perturb the field as $\phi=\phi_0+\delta\phi e^{\sigma t+i\kvec\cdot\x}$, where $\sigma(k)$ is the real growth or decay rate.

This linearization separates the bare deterministic local relaxation rate, $\gamma_0 \equiv \mu+2D f'(\phi_0)$, from the dressed bath response, $\mathcal{R}_0 \equiv 2D(\chi_0^*)'(\phi_0)$. Since the screening operator becomes $1+\epsilon k^2$ in Fourier space, the dispersion relation is
\be
\sigma(k) = \frac{\mathcal{R}_0}{1+\epsilon k^2} - \gamma_0.
\ee
{Here $\mathcal R_0$ is the slope of the dressed slow manifold, including the dependence of the condensational sink on $\phi$ carried by the effective source \eqref{Eq_effective_source}. This response differs sharply from that of a gradient, or Fickian, closure of the same coupling. Expanding $\mathcal R_0/(1+\epsilon k^2)\simeq\mathcal R_0-\epsilon\mathcal R_0k^2$ shows that, at long wavelengths, the halo-mediated coupling acts as a diffusion with coefficient $\epsilon\mathcal R_0=2D\epsilon(\chi_0^*)'(\phi_0)$, positive where the dressed slow manifold increases and negative where it decreases (Section~\ref{Sec_Phys_nonFickian}). Truncated at this order, the coupling would damp fine scales at rates growing as $k^2$ where the coefficient is positive and, where it is negative, amplify them at rates growing as $k^2$ without bound, which is the ill-posed backward diffusion equation.

In the pseudo-parabolic dynamics, by contrast, the high-$k$ response saturates. As $k\to\infty$, the nonlocal thermodynamic bath response is screened out, $\mathcal{R}_0 /(1+\epsilon k^2)\to 0$, and the rate approaches the finite local value, $\sigma(k)\to-\gamma_0$. Hence arbitrarily fine spatial modes are neither erased nor amplified by an unbounded rate; they evolve at the same finite local rate, and the differential response between high-wavenumber modes disappears. The pseudo-parabolic operator thus regularizes forward--backward diffusion, as in the viscous diffusion and aggregation models studied by Novick-Cohen and Pego \citep{novick1991stable} and by Padr\'on \citep{padron2004effect}, which share this structure. The term $-\nabla\cdot\big(\epsilon(\phi)\nabla\partial_t\phi\big)$ acts as a macroscopic dashpot: it resists instantaneous spatial rearrangement by distributing the response of the droplet field over the footprint of the screened supersaturation envelope. In this sense, dynamic inertia helps preserve sharp cloud boundaries against the dry environment, replacing an unbounded, gradient-driven response by a bounded, screened thermodynamic adjustment.}

\subsection{{First-Order Bath Lag and the Effective Mass of the Cloud Polaron}}\label{Sec_Effective_Mass}
{We now retain the lag $-i\nu$ of the optimal displacement, Eq.~\eqref{Eq_Gurari_displacement}, which was discarded at $\mathcal O(\delta^0)$. In Gurari's  \citep{gurari1953} and Feynman's  \citep{feynman1955slow} treatments of the polaron, it is the corresponding motion-dependent term that produces the effective mass: the polarization cloud must be displaced together with the electron, and the energy acquires the form $-W+\hbar^2k^2/2m^*$, with $m^*/m=1+W/(6\hbar\omega)$ at weak coupling [Eqs.~(1.4) and (2.32) of Ref.~\citep{gurari1953}; Eq.~(46) of Ref.~\citep{feynman1955slow}, where $m^*/m=1+\alpha/6$]. The enhancement is the ratio of the coupling energy to the phonon energy. The droplet field is overdamped, so its inertia is the coefficient of $\partial_t\phi$ rather than of a second time derivative, and the question is how the delay of the halo renormalizes it.

The answer is exact at the linear level. Linearizing the bath and droplet equations about a homogeneous state $\big(\phi_0,\chi_0^*(\phi_0)\big)$, with the effective source $\hat J\simeq\Gamma_0^*(\chi_0^*)'(\phi_0)\,\delta\phi$, and keeping the full bath operator, a mode $e^{\sigma t+i\kvec\cdot\x}$ satisfies
\be\label{Eq_dispersion_lag}
\big(\sigma+\gamma_0\big)\Big(\sigma+\Gamma_0^*\big(1+\epsilon k^2\big)\Big)=\mathcal R_0\,\Gamma_0^*,
\ee
with $\gamma_0$ and $\mathcal R_0$ as in the dispersion relation of Section~\ref{Sec_Low_Pass_Truncation}. Equation~\eqref{Eq_dispersion_lag} has a fast branch, $\sigma\simeq-\Gamma_0^*(1+\epsilon k^2)$, which is the relaxation of the bath itself, and a slow branch. Solving for the slow branch to first order in $\delta$ gives
\be\label{Eq_slow_branch_lag}
\sigma(k)=\frac{\dfrac{\mathcal R_0}{1+\epsilon k^2}-\gamma_0}{1+\dfrac{\mathcal R_0}{\Gamma_0^*\big(1+\epsilon k^2\big)^2}}+\mathcal O(\delta^2).
\ee
The numerator is the $\mathcal O(\delta^0)$ result. The denominator is the lag correction: the slow rate is divided by the factor
\be\label{Eq_effective_mass}
\frac{m^*(k)}{m}\equiv1+\frac{\mathcal R_0}{\Gamma_0^*\big(1+\epsilon k^2\big)^2},
\ee
which is the cloud analogue of the polaron mass ratio. As for the polaron, the enhancement is the ratio of the bath-mediated coupling rate, $\mathcal R_0=2D(\chi_0^*)'(\phi_0)$, to the bath relaxation rate, $\Gamma_0^*$, and it is small, of order $\delta$, in the regime of validity of the expansion. It is largest at long wavelengths and decays as $(1+\epsilon k^2)^{-2}$ on scales shorter than the screening length, where the halo no longer follows the droplet field. Its sign is that of $\mathcal R_0$: in the haze regime, where the dressed slow manifold increases with $\phi$, the lag slows the response of the droplet field; beyond the maximum of the manifold, it accelerates it.

At the nonlinear level, the same correction follows from the adiabatic expansion of the optimal displacement in the zero-diffusion limit,
\be\label{Eq_adiabatic_expansion_lag}
\chi_s=\chi_0^*(\phi)-\frac{1}{\Gamma^*(\phi)}\,\partial_t\chi_0^*(\phi)+\mathcal O(\delta^2),
\ee
which, inserted into the droplet equation, gives the local closure
\bea\label{Eq_local_closure_lag}
&\big(1+a(\phi)\big)\partial_t\phi=-\mu\phi+2D\big(\chi_0^*(\phi)-f(\phi)\big),\\
&a(\phi)=\frac{2D(\chi_0^*)'(\phi)}{\Gamma^*(\phi)} .
\eea
Combining the lag with the screened response, to first order in $\delta$ and to leading order in $\epsilon k^2$ in the lag term, the deterministic part of the closure becomes
\be\label{Eq_pseudo_PDE_lag}
\Big( I - \nabla\cdot\epsilon(\phi)\nabla \Big) \Big( \partial_t\phi + \mu\phi + 2D f(\phi) \Big) + a(\phi)\,\partial_t\phi = 2D\chi_0^*(\phi),
\ee
whose linearization reproduces Eq.~\eqref{Eq_slow_branch_lag} with $(1+\epsilon k^2)^2$ replaced by $1+\epsilon k^2$ in the lag term. The operator acting on $\partial_t\phi$,
\be\label{Eq_dynamic_inertia_operator}
\mathcal I_\phi=\big(1+a(\phi)\big)I-\nabla\cdot\epsilon(\phi)\nabla ,
\ee
is the dynamic-inertia operator of the droplet field. Its two parts are the spatial and temporal faces of the same dressed object: $-\nabla\cdot\epsilon(\phi)\nabla$ records the footprint of the halo, of radius $L_{\rm scrn}$, the analogue of the polaron radius; $a(\phi)$ records its delay, the analogue of the polaron mass enhancement. Both descend from the single optimal displacement of Eq.~\eqref{Eq_Gurari_displacement}, through its diffusive cost and its lag, respectively.}

\subsection{Emergent Spatiotemporal Memory of the Macroscopic Noise}\label{Sec_Emerging_OU_noise}
When we constructed our MSR action and integrated out the fast supersaturation bath (Section \ref{Sec_Integrating_out_Fast}), we executed a classical field-theoretic equivalent of Feynman's first step \cite{feynman1955slow}. By projecting out the fast degrees of freedom, we generated a retarded self-interaction---the Fluctuation-Dissipation Contraction. As derived in Eq.~\eqref{Eq_Seff}, this manifests in the effective action $\Action_{eff}$ as a non-local quadratic coupling mediated by the bath's spatiotemporal correlation function $\mathcal{C}$:
\be\label{Eq_Noise_generator}
- \frac{1}{2} \int \d t \d \x \int \d t' \d \x' \, (2D\bar{\phi}) \, \mathcal{C}(\x,t; \x',t') \, (2D\bar{\phi}).
\ee
In the Markovian approximation in the zero-diffusion limit, we "washed" this term as a spacetime white noise term 
$\sqrt{\Sigma_{\phi}(\phi)} \eta_{\phi}(\x,t)$, with nonlinear state-dependence variance $\Sigma_{\phi}(\phi)$ (Eq.~\eqref{Eq_Sigma_phi}), that emerged at the field level closure for the droplet field (Eq.~\eqref{Eq_Macro_Langevin}).  

Such a strict adiabatic limit and Markovian approximation assumes the bath to have localized effects on, and to adjust  instantaneously to,  the slow droplet evolution.
However, the non-local term (Eq.~\eqref{Eq_Noise_generator})  formally encodes the finite temporal memory and spatial screening of the thermodynamic environment.

To unpack the spatial screening component of the noise correlation $\mathcal{C}$, we must evaluate the macroscopic Green's function $\mathcal{G}$, since $\mathcal{C}$ is given as the integral of $V \mathcal{G} \mathcal{G}$ (Eq.~\eqref{Eq_Corr_SpaceTime}). The  Green's function is obtained by inverting the linear operator of the bath, $\mathcal{L} = \partial_t - D_\chi \nabla^2 + \Gamma^*(\phi)$. Because this operator is strictly parabolic, it generates a classical heat kernel where spatial spreading and temporal decay are inextricably coupled.

In Fourier space, the  bath propagator is:
\be\label{Eq_G_Fourier}
\tilde{\mathcal{G}}(k,\nu) = \frac{1}{\Gamma^* + D_\chi k^2 - i\nu}.
\ee
To help uncover a spatiotemporally decoupled approximation, we multiply the numerator and denominator by $\Gamma^*$ and utilize the  algebraic identity $\Gamma^*(\Gamma^* + D_\chi k^2 - i\nu) = (\Gamma^* + D_\chi k^2)(\Gamma^* - i\nu) + i\nu D_\chi k^2$. This allows us to rewrite the  propagator by factoring out the decoupled spatial and temporal terms:
\be\label{Eq_G_R}
\tilde{\mathcal{G}}(k, \nu) = \frac{\Gamma^*}{(\Gamma^* + D_\chi k^2)(\Gamma^* - i\nu)} \Big[ 1 + \mathcal{R}(k, \nu) \Big]^{-1},
\ee
where the dimensionless spatiotemporal cross-term $\mathcal{R}(k, \nu)$ is defined as:
\be\label{Eq_R}
\mathcal{R}(k, \nu) \equiv \left( \frac{D_\chi k^2}{\Gamma^* + D_\chi k^2} \right) \left( \frac{i\nu}{\Gamma^* - i\nu} \right).
\ee
Factorizing the second term by $\Gamma^*$, we note that within the low-pass band $\delta \sim \nu/\Gamma^* \ll 1$ (as in Eq.~\eqref{Eq_low_pass_band}): 
\bes
\frac{i(\nu/\Gamma^*)}{1 - i(\nu/\Gamma^*)} = \frac{i\delta}{1 - i\delta} = \mathcal{O}(\delta),
\ees
leading to
\be\label{Eq_key_estimateR}
|\mathcal{R}(k, \nu)| \le \mathcal{O}\left(\frac{\nu}{\Gamma^*}\right) \ll 1.
\ee
This operation is consistent with the  low-pass spectral truncation ($\mathcal{P}_{\Lambda}$) used in Section \ref{Sec_Low_Pass_Truncation} to isolate the slow deterministic dynamics of the droplet field. 
Here, within this projected subspace, the spatial fraction of $\mathcal{R}$ is strictly bounded by unity for all real $k$, while the temporal fraction is governed by the small asymptotic parameter $\delta \sim \nu/\Gamma^* \ll 1$. 

Because the cross-term is strictly $\mathcal{O}(\nu/\Gamma^*)$ (Eq.~\eqref{Eq_key_estimateR}), we can expand the inverse term in Eq.~\eqref{Eq_G_R} as a convergent Taylor series:
$$[1 + \mathcal{R}(k, \nu)]^{-1} = 1 - \mathcal{R}(k, \nu) + \mathcal{O}(\mathcal{R}^2).$$ 
By truncating at the leading order ($\mathcal{O}(\delta^0)$), the cross-term is thus annihilated. The propagator drops the $\mathcal{R}$-term and reduces thus  to the factorized form:
\be\label{Eq_G_key_approx}
\tilde{\mathcal{G}}(k,\nu) \approx \frac{\Gamma^*}{(\Gamma^* + D_\chi k^2)(\Gamma^* - i\nu)}.
\ee

Going back to the denominator in Eq.~\eqref{Eq_G_Fourier}, written as
\bes
\Gamma^* + D_\chi k^2 - i\nu = \frac{(\Gamma^* + D_\chi k^2)(\Gamma^* - i\nu)}{\Gamma^*} + \frac{D_\chi}{\Gamma^*} k^2 i\nu,
\ees
we understand how the approximation of  $\tilde{\mathcal{G}}$  in Eq.~\eqref{Eq_G_key_approx} is guided by physics. Indeed, since the macroscopic droplet field evolves on a timescale vastly slower than the bath's relaxation ($|\nu| \ll \Gamma^*$), it says that within this projected low-frequency subspace, the cross-term $ k^2 (i\nu / \Gamma^*)$ is strictly a higher-order transient ($\mathcal{O}(\nu/\Gamma^*)$). Because the droplets are slow, they are completely "blind" to this ultra-fast spatiotemporal cross-term. 

Transforming the decoupled fraction in Eq.~\eqref{Eq_G_key_approx}  back to real space, the macroscopic Green's function perfectly decouples into a spatial Bessel screening operator $G^*$ and a pure temporal exponential delay:
\bes\label{Eq_G_Macroscopic_Factorized}
\mathcal{G}(\mathbf{y},s; \x,t) \approx \Gamma^*(\phi) G^*(|\x-\mathbf{y}|) \Theta(t-s) e^{-\Gamma^*(\phi)(t-s)},
\ees
where $\Theta$ denotes the Heaviside step function ensuring causality, and the spatial screening kernel is defined as $G^*(r) = (2\pi D_\chi)^{-1} K_0(r/L_{scrn}^*)$ with the dressed screening length $L_{scrn}^* = \sqrt{D_\chi/\Gamma^*(\phi)}$. 

The physical meaning of this mathematical truncation is profound. By applying $\mathcal{P}_\Lambda$, we acknowledge that the  noise correlation $\mathcal{C}$ is ultimately filtered through the response of the macroscopic droplet field. Because the {droplet field evolves} on a timescale vastly slower than the fast thermodynamic bath{, it filters the bath response}. Consequently, the droplets act as a physical low-pass filter, completely blind to the ultra-fast transient radial spreading of the vapor (represented by the truncated $\mathcal{R}$ term). From the perspective of the slow macroscopic field, the vapor shadow equilibrates to an effective "steady-state" spatial footprint  so rapidly that the droplets only perceive the fully developed Bessel profile decaying purely in time.

We can then complete the noise derivation. 
From Eq.~\eqref{Eq_G_key_approx}, we get  
\begin{widetext}
\bea
\mathcal{C}(\x,t;& \x',t') = \int \d \mathbf{y} \, G^*(|\x - \mathbf{y}|) G^*(|\x' - \mathbf{y}|)  \\
& \times \underbrace{ (\Gamma^*)^2 \int \d s \, V\big(\phi(\mathbf{y},s)\big) \Theta(t-s)\Theta(t'-s) e^{-\Gamma^*(\phi( \mathbf{y},s))(t+t'-2s)} }_{\text{Temporal coarse-graining } \mathcal{T}(t,t')}.
\eea
\end{widetext}
We focus entirely on the temporal coarse-graining term, $\mathcal{T}(t,t')$. The upper limit of the integral over the past history $s$ is strictly governed by the Heaviside step functions, which restrict the integration to $s \in (-\infty, \min(t,t')]$. Exploiting the timescale separation, we treat the local variance $V$ and the dressed relaxation rate $\Gamma^*$ as constants relative to the fast integration variable $s$. Pulling these slow variables outside the integral, we show after elementary manipulations that the temporal memory resolves  into a symmetric exponential decay:
\be\label{Eq_Temporal_OU}
\mathcal{T}(t,t') \approx V\big(\phi(\mathbf{y})\big) \frac{\Gamma^*(\phi(\mathbf{y}))}{2} e^{-\Gamma^*(\phi(\mathbf{y}))|t-t'|}.
\ee
  Substituting this  temporal kernel back into the full convolution, the fully non-Markovian spatiotemporal correlation function is:
\bea\label{Eq_Corr_NonMarkovian_Full}
\hspace{-2ex}\mathcal{C}(\x,t; \x',t') &= \int \d \mathbf{y} \, G^*(|\x - \mathbf{y}|) G^*(|\x' - \mathbf{y}|)  \\
& \times \left[ V\big(\phi(\mathbf{y})\big) \frac{\Gamma^*(\phi(\mathbf{y}))}{2} e^{-\Gamma^*(\phi(\mathbf{y}))|t-t'|} \right].
\eea
This  non-Markovian, non-local expression of the noise correlation seamlessly integrates with the structure of our pseudo-parabolic BZK formulation (Eq.~\eqref{Eq_pseudo_PDE1}) as we show next. 

\subsection{The Non-Markovian Closure of Droplet-Field Dynamics}
The macroscopic Langevin equation is now driven by a noise field whose covariance is given by Eq.~\eqref{Eq_Corr_NonMarkovian_Full}. This covariance contains two distinct structures: a screened spatial convolution through the Bessel Green's function $G^*$, and a finite temporal memory through the exponential Ornstein--Uhlenbeck (OU)  kernel. The operator $I-\nabla\cdot\epsilon(\phi)\nabla$ on the left-hand side of the pseudo-parabolic Eq.~\eqref{Eq_pseudo_PDE1}, is precisely the macroscopic remnant of the same screened bath response. Applying this operator to the explicit Langevin equation removes the explicit Bessel convolution from the stochastic forcing, while leaving the finite temporal memory intact. {Since $G^*$ is the Green's function of $\Gamma^*-D_\chi\nabla^2$, one has $\big(I-\epsilon\nabla^2\big)G^*=\delta/\Gamma^*$; each of the two Bessel kernels in Eq.~\eqref{Eq_Corr_NonMarkovian_Full} is therefore replaced by a Dirac distribution divided by $\Gamma^*$, and the localized covariance carries the factor $(\Gamma^*)^{-2}$.}

The resulting localized noise generator is therefore no longer a spatiotemporal white noise, as in the Markovian zero-diffusion closure Eq.~\eqref{Eq_Macro_Langevin}. It is spatially local, but temporally colored. {{We retain an OU representation of the finite bath memory together with the deterministic lag correction of Section~\ref{Sec_Effective_Mass}. The OU spectrum is even in frequency; its departure from its zero-frequency level begins at order $(\nu/\Gamma^*)^2$, so its color is not itself a first-order spectral correction.}} The corresponding {\it stochastic non-Markovian droplet-field closure} takes the form of a stochastic pseudo-parabolic equation:
\bea\label{Eq_PseudoParabolic_NonMarkovian}
\hspace{-1ex}\Big[ I-\nabla\cdot\epsilon(\phi)\nabla \Big] &\Big[ \partial_t\phi+\mu\phi+2D f(\phi) \Big] {+a(\phi)\,\partial_t\phi} \\
&= 2D\chi_0^*(\phi) +  \xi_{OU}(\x,t),
\eea
{with $a(\phi)$ given by Eq.~\eqref{Eq_local_closure_lag},} where $\xi_{OU}(\x,t)$ is spatially uncorrelated but temporally colored, {with the following frozen-background stationary covariance}
\begin{widetext}
\be\label{Eq_covariance_OU}
\langle \xi_{OU}(\x,t)\xi_{OU}(\x',t')\rangle = (2D)^2\delta(\x-\x') {\frac{V\big(\phi(\x)\big)}{2\Gamma^*(\phi(\x))}} e^{-\Gamma^*(\phi(\x))|t-t'|}.
\ee
\end{widetext}
 {\it Supplementary Note 6} summarizes the construction of Eq.~\eqref{Eq_PseudoParabolic_NonMarkovian}, distinguishing the stochastic conversion, Gaussian bath approximation, and spatial and temporal reductions.

This closure unifies the two consequences of eliminating the fast supersaturation bath: the left-hand side carries the dynamic inertia generated by the screened bath response{, footprint and lag,} while the right-hand side carries the finite temporal memory of the bath fluctuations. {{The deterministic lag is retained to first order in $\delta$ and leading spatial order in that correction. The factorized OU covariance is a frozen-background, low-frequency approximation, not an exact covariance for an evolving inhomogeneous field.}}

{To clarify the physical content of this emerging noise in our non-Markovian closure, we compare the equal-time variance of the localized OU generator, denoted $\Sigma_{OU}(\phi)$, with the intensity $\Sigma_\phi(\phi)$ of the local Markovian closure, Eq.~\eqref{Eq_Sigma_phi}. Setting $t=t'$ in Eq.~\eqref{Eq_covariance_OU} and using $V(\phi)={2}\sigma_S^2V_c\big(\chi_0^*(\phi)\big)^2/\tau_K$ gives
\be\label{Eq_Sigma_OU}
\Sigma_{OU}(\phi)=\frac{(2D)^2V(\phi)}{2\Gamma^*(\phi)}=\frac{{4}D^2\sigma_S^2V_c\big(\chi_0^*(\phi)\big)^2}{\tau_K\Gamma^*(\phi)},
\ee
so that
\be\label{Eq_Variance_Scaling}
\Sigma_{OU}(\phi) = \frac{\Gamma^*(\phi)}{2}\, \Sigma_\phi(\phi).
\ee
This is the expected consistency relation. An Ornstein--Uhlenbeck forcing of variance $\Sigma_{OU}$ and correlation time $1/\Gamma^*$ has zero-frequency intensity $2\Sigma_{OU}/\Gamma^*=\Sigma_\phi$: the Markovian closure \eqref{Eq_Macro_Langevin} is the white-noise limit of the non-Markovian one, and the localization by the screening operator leaves the local intensity of the forcing unchanged. The screening appears instead in the explicit form \eqref{Eq_Corr_NonMarkovian_Full}, where the forcing felt by the droplet field is correlated over the footprint $L_{\rm scrn}$ of the dragged envelope: each bath fluctuation is spread over the screening disk, so that neighboring cloud elements receive correlated kicks. In the localized form \eqref{Eq_PseudoParabolic_NonMarkovian}, the same spreading is carried by the operator on the left-hand side, which acts on the response of $\phi$ to the forcing.

The consequence is clearest near a critical relaxation threshold. As $\Gamma^*\to0$, the intensity grows as $\Sigma_\phi\propto(\chi_0^*)^2/(\Gamma^*)^2=J^2/(\Gamma^*)^4$, while the correlation time $1/\Gamma^*$ of the forcing diverges: the induced noise becomes strong and slow at once. This is a critical slowing down of the bath-induced forcing, and it is where the separation $\delta\ll1$ fails and the expansion must be abandoned (Appendix~\ref{Sec_Emergent_Noise_Constraint}).}

\subsection{A Spatially Correlated Markovian Closure}\label{Sec_Spatial_Markovian}
The local Markovian closure of Eq.~\eqref{Eq_Macro_Langevin} was obtained in a double limit: vanishing bath diffusion, $D_\chi\to0$, and negligible bath memory on the droplet timescale. Here we retain $D_\chi>0$ while making only the temporal adiabatic approximation. The bath then redistributes fluctuations spatially before relaxing, so its elimination produces screened spatial coupling and correlated noise increments. The resulting closure remains Markovian in $\phi$: its drift and noise amplitude depend on the current field, without retaining its past history. It therefore lies between the local closure of Eq.~\eqref{Eq_Macro_Langevin} and the finite-memory closure of Eq.~\eqref{Eq_PseudoParabolic_NonMarkovian}, separating the effects of spatial screening from those of temporal memory.

To isolate this regime within the retained Gaussian bath approximation, define
\bea\label{Eq_Spatial_Markovian_operator}
\mathcal L_\phi&=\Gamma^*(\phi)-D_\chi  \nabla^2,\\
J(\phi)&=\gamma+\alpha_N\sqrt\phi f(\phi),\qquad
b_b(\phi)=\sigma\chi_0^*(\phi).
\eea
 The amplitude $b_b$ is evaluated on the local adiabatic background $\chi_0^*(\phi)$, as in the quadratic action \eqref{Eq_S_quad}. Its squared value is the residual bath's white-noise intensity,
\be\label{Eq_bath_noise_intensity}
V(\phi)=b_b(\phi)^2=\sigma^2\big(\chi_0^*(\phi)\big)^2
=\frac{2\sigma_S^2V_c}{\tau_K}\big(\chi_0^*(\phi)\big)^2,
\ee
where the last equality uses Eq.~\eqref{Eq_sigma_scaling}. Thus the local quadratic action contains $-V\bar\chi^2/2$, and $V$ is the intensity propagated in Eq.~\eqref{Eq_Corr_SpaceTime}. Let $W_{\eta_K}$ be a $Q$-Wiener process \citep{lord2014introduction}, with independent increments in disjoint time intervals and spatial covariance operator
\be\label{Eq_spatial_noise_operator}
(Q_{\eta_K}v)(\x)=\int\rho_{\eta_K}(\x-\mathbf y)v(\mathbf y)\,\d\mathbf y.
\ee
Here $v$ is a test field, and $\rho_{\eta_K}$ is the unit-integral covariance kernel specified after Eq.~\eqref{Eq_noise_covariance}. The covariance of an increment $\d W_{\eta_K}$ is $Q_{\eta_K}\d t$; formally, its time derivative is the noise $\eta$ in that equation. The bath propagator is $(\mathcal L_\phi-i\nu)^{-1}$ when the slow field is frozen during relaxation. Temporal coarse-graining replaces this propagator by its zero-frequency value $\mathcal L_\phi^{-1}$ while retaining $D_\chi>0$. We assume $\Gamma^*(\phi)$ remains bounded away from zero over the states considered, so the screened inverse is defined on the periodic domain.

This gives the spatially coupled Markovian closure
\bea\label{Eq_Spatial_Markovian_closure}
\d\phi={}&\big[-\mu\phi-2Df(\phi)
+2D\mathcal L_\phi^{-1}J(\phi)\big]\d t\\
&+2D\mathcal L_\phi^{-1}\big[b_b(\phi)\,\d W_{\eta_K}\big].
\eea
The drift and noise depend on the current field, without a history integral or an auxiliary colored-noise variable. The equation is therefore Markovian in $\phi$, despite its spatial nonlocality. As in Eq.~\eqref{Eq_Macro_Langevin}, we adopt the It\^o interpretation of the effective closure. Matching the bath's zero-frequency response fixes the leading drift and covariance retained here; a separate controlled homogenization would be needed to determine any additional state-dependent drift corrections beyond this approximation.

The covariance per unit time of the induced white-noise increments is obtained by integrating the frozen bath covariance over the time difference and multiplying by $(2D)^2$, as in Eq.~\eqref{Eq_Sigma_phi}. This differs from evaluating the bath covariance at equal times in Eq.~\eqref{Eq_Corr_OU_App}, which gives $V/(2\Gamma^*)$ in the local limit; the corresponding variance transmitted to the droplet forcing is $\Sigma_{OU}$ in Eq.~\eqref{Eq_Sigma_OU}. To write the spatial counterpart, define the multiplication operator by $(M_{b_b}v)(\x)=b_b(\phi(\x))v(\x)$. The noise in Eq.~\eqref{Eq_Spatial_Markovian_closure} first weights each bath increment by this local amplitude and then spreads it through $\mathcal L_\phi^{-1}$. Its covariance operator is therefore
\be\label{Eq_Spatial_Markovian_cov}
Q_{\rm eff}(\phi)=4D^2\mathcal L_\phi^{-1}
M_{b_b}Q_{\eta_K}M_{b_b}(\mathcal L_\phi^*)^{-1}.
\ee
 Here $\mathcal L_\phi^*$ denotes the $L^2$-adjoint of $\mathcal L_\phi$. With real $\Gamma^*(\phi)$, constant $D_\chi$ and the periodic boundary conditions used here, $\mathcal L_\phi^*=\mathcal L_\phi$. Equation~\eqref{Eq_Spatial_Markovian_cov} applies the screened noise map on one side of $Q_{\eta_K}$ and its adjoint on the other, ensuring a nonnegative covariance. It retains the microscopic spatial cutoff and keeps the varying amplitude inside the screening operator, since multiplication by $b_b(\phi)$ and spatial screening generally do not commute.

Around a homogeneous background $\phi_0$, put $\Gamma_* = \Gamma^*(\phi_0)$, $V_*=V(\phi_0)$ and $\ell_{\rm scrn}^2=D_\chi/\Gamma_*$. Each Fourier mode then has noise intensity
\bea\label{Eq_Spatial_Markovian_spectrum}
\widehat Q_{\rm eff}(k)
&=\frac{4D^2V_*\,\widehat\rho_{\eta_K}(k)}
{(\Gamma_*+D_\chi|k|^2)^2}\\
&=\Sigma_{\phi,*}\,
\frac{\widehat\rho_{\eta_K}(k)}{(1+\ell_{\rm scrn}^2|k|^2)^2},
\textrm{with } \Sigma_{\phi,*}=\frac{4D^2V_*}{\Gamma_*^2}.
\eea
The local intensity $\Sigma_{\phi,*}$ agrees with Eq.~\eqref{Eq_Sigma_phi}. Screening suppresses fine-scale forcing and extends its spatial correlations over $\ell_{\rm scrn}$, in addition to the microscopic correlation scale $\eta_K$. The same squared screening factor follows by integrating the frozen-background covariance in Eq.~\eqref{Eq_Corr_NonMarkovian_Full} over the time difference.

The connection with the pseudo-parabolic description is immediate. Multiplying Eq.~\eqref{Eq_Spatial_Markovian_closure} by $\mathcal L_\phi$ and dividing pointwise by $\Gamma^*(\phi)$ yields
\bea\label{Eq_Spatial_Markovian_pseudo}
&\big[I-\epsilon(\phi) \nabla^2\big]
\big[\d\phi+\big(\mu\phi+2Df(\phi)\big)\d t\big]\\
&\qquad=2D\chi_0^*(\phi)\d t
+\frac{2D b_b(\phi)}{\Gamma^*(\phi)}\,\d W_{\eta_K},
\eea
where $\epsilon=D_\chi/\Gamma^*$. This is the instantaneous screened, non-divergence form discussed after Eq.~\eqref{Eq_low_pass_band2}; replacing $\epsilon\nabla^2$ by $\nabla\cdot(\epsilon\nabla)$ remains a separate nonlinear extension. The mixed derivative represents spatially distributed adjustment and does not, by itself, imply temporal memory. Here the bath-lag correction $a(\phi)$ is omitted at leading adiabatic order, and the forcing is white in time.

The three closures can be distinguished by two dimensionless ratios,
\be\label{Eq_closure_scale_ratios}
 r_k=\frac{D_\chi|k|^2}{\Gamma_*}
 =\ell_{\rm scrn}^2|k|^2,
 \qquad \delta_\nu=\frac{|\nu|}{\Gamma_*}.
\ee
The first compares spatial redistribution with local bath relaxation; the second compares the resolved temporal frequency with that relaxation rate. The exact frozen-mode denominator is $\Gamma_*[1+r_k-i\nu/\Gamma_*]$, so its temporal adiabatic parameter is $\delta_\nu/(1+r_k)$. Small $D_\chi$ is therefore meaningful only relative to the wavenumbers retained: the same diffusivity may be negligible at large scales and appreciable at fine scales.

In the \emph{local Markovian regime}, both $r_k\ll1$ and $\delta_\nu\ll1$ over the resolved band, and both corrections are discarded. The propagator reduces to $\Gamma_*^{-1}$, recovering Eq.~\eqref{Eq_Macro_Langevin} and its local noise intensity, with the microscopic spatial cutoff understood. Setting $D_\chi=0$ is the exact spatially local limit; it does not remove temporal memory unless the temporal approximation is also made.

In the \emph{spatially correlated Markovian regime}, $\delta_\nu/(1+r_k)\ll1$ allows temporal memory to be discarded, while the spatial factor $1+r_k$ is retained. Equation~\eqref{Eq_Spatial_Markovian_closure} then includes weak spatial corrections when $0<r_k\ll1$ and appreciable screening when $r_k$ is of order one. In particular, for small $r_k$ the noise spectrum is $\widehat Q_{\rm eff}=\Sigma_{\phi,*}\widehat\rho_{\eta_K}[1-2r_k+\mathcal O(r_k^2)]$. We retain the unexpanded denominator in Eq.~\eqref{Eq_Spatial_Markovian_spectrum}, since a truncated spectrum need not remain nonnegative outside its long-wavelength range.

In the \emph{finite-memory pseudo-parabolic regime}, spatial screening is retained and the leading temporal corrections are also resolved: the deterministic response includes the bath-lag term $a(\phi)\partial_t\phi$, while the forcing retains its OU correlation time, as in Eq.~\eqref{Eq_PseudoParabolic_NonMarkovian}. The low-frequency derivation still requires $\delta_\nu\ll1$; it retains finite-memory effects within that approximation rather than establishing a closure for arbitrarily slow bath relaxation. If $\delta_\nu$ becomes order one, the expansion must be replaced by the full bath dynamics or its unexpanded memory kernel. Thus these are levels of resolution of the same response, not three phases separated by $D_\chi$ alone. The intermediate closure isolates the case in which temporal memory is unresolved but neighboring droplet populations still share spatially distributed forcing.

\subsection{Physical Implications}\label{Sec_Phys_nonFickian}
Expanding Eq.~\eqref{Eq_PseudoParabolic_NonMarkovian} makes the full spatiotemporal coupling explicit:
\bea\label{Eq_PseudoParabolic_Final_Expanded}
{\big(1+a(\phi)\big)}\partial_t \phi &- \nabla \cdot \big( \epsilon(\phi) \nabla \partial_t \phi \big) \\
= &-\mu \phi +  2D \Big[ \chi_0^*(\phi) - f(\phi) \Big] + \mu \nabla \cdot \big( \epsilon(\phi) \nabla \phi \big)\\
&+ 2D \nabla \cdot \big( \epsilon(\phi) \nabla f(\phi) \big) +\xi_{OU}(\x,t), \eea
where $\chi_0^*(\phi)$ is  given by Eq.~\eqref{Eq_chi_star},  $f(\phi) = A/\sqrt{\phi} - B/\phi^{3/2}$, $\epsilon(\phi) = D_\chi / \Gamma^*(\phi)$, {$a(\phi)$ by Eq.~\eqref{Eq_local_closure_lag},} and $\Gamma^*(\phi)$ is given by Eq.~\eqref{Eq_Gamma_star}.

{The term $-2D f(\phi)$ represents the mean-field projection of the K\"ohler condensation/evaporation drift (Appendix \ref{App_Kohler_projection}). The two divergence terms on the right-hand side, $\mu\nabla\cdot\big(\epsilon\nabla\phi\big)$ and $2D\nabla\cdot\big(\epsilon\nabla f(\phi)\big)$, should not be read as independent transport mechanisms. They arise from distributing the screening operator over the local drift, and at long wavelengths they are compensated by the mixed derivative on the left-hand side. Indeed, to first order in $\epsilon$, the time derivative in $\nabla\cdot(\epsilon\nabla\partial_t\phi)$ may be replaced by its leading-order value, $-\mu\phi+2D(\chi_0^*-f)$, and the $\mu$- and $f$-terms then cancel, leaving
\bea\label{Eq_longwave_closure}
\big(1+a(\phi)\big)\partial_t\phi=&-\mu\phi+2D\big(\chi_0^*(\phi)-f(\phi)\big)\\
&+2D\nabla\cdot\big(\epsilon(\phi)\nabla\chi_0^*(\phi)\big)+\hat\xi(\x,t),
\eea
up to terms of order $\epsilon^2\nabla^4$ and $\epsilon\delta$, where $\hat\xi=\big(I+\nabla\cdot\epsilon\nabla\big)\xi_{OU}$ is the forcing at the same order. {Within this conservative nonlinear extension, the long-wave spatial coupling is therefore the screened diffusion of the dressed slow manifold,} $2D\nabla\cdot\big(\epsilon\nabla\chi_0^*\big)$, not of the K\"ohler curve.

Because local droplet populations share a common supersaturation bath, a dense coarse-grained cluster of droplets experiencing joint growth does not evolve as an isolated local variable. It drags a supersaturation halo, and the spatial overlap of these halos couples neighboring cloud elements through the screened response. Expanding the coupling in Eq.~\eqref{Eq_longwave_closure} ({\it Supplementary Note 7}) shows that its Laplacian part is controlled by the state-dependent coefficient
\be\label{Eq_Deff}
\mathcal D_{\rm eff}(\phi)=2D\,\epsilon(\phi)\,(\chi_0^*)'(\phi),
\ee
which, at $\phi=\phi_0$, is the long-wave coefficient $\epsilon\mathcal R_0$ of the dispersion relation of Section~\ref{Sec_Low_Pass_Truncation}. Since $\epsilon(\phi)>0$, the direction of the effective transport is set by the slope of the dressed slow manifold,
\be
\chi_0^*(\phi)=\frac{\gamma+\alpha_NA-\alpha_NB/\phi}{\lambda^*+\alpha_N\sqrt\phi},
\qquad
\lambda^*=\lambda-\frac{\sigma_S^2}{\tau_K}.
\ee
This function increases with $\phi$ at small $\phi$, through the solute term, and decreases at large $\phi$, through the growing condensational sink. Its maximum $\phi_m$ solves
\be\label{Eq_phi_m}
(\gamma+\alpha_NA)\,\phi_m-3\alpha_NB-2B\lambda^*\phi_m^{-1/2}=0 .
\ee
At the saddle-node threshold, $\lambda^*=\lambda_c$, Eq.~\eqref{Eq_phi_m} gives exactly $\phi_m=3B/A$, the K\"ohler critical value: the dressed slow manifold is then tangent to the K\"ohler curve at its peak. For $\lambda^*>\lambda_c$, $\phi_m$ moves to larger values but remains inside the forbidden zone, between the K\"ohler critical radius and the edge state. For the parameters of Table~\ref{Table_RegimeA} with $\lambda^*=1.65\lambda_c$ and $2.1\lambda_c$, $\sqrt{\phi_m}$ exceeds the K\"ohler critical radius by 10--16\% for $N=50\,{\rm cm}^{-3}$ and by 2--4\% for $N=500\,{\rm cm}^{-3}$, while the edge state lies at more than twice that radius.

Thus, for $\phi<\phi_m$, which includes the haze regime, $\mathcal D_{\rm eff}(\phi)>0$ and the screened supersaturation coupling behaves as a standard diffusion: spatial variations in the unactivated aerosol population are smoothed through the surrounding supersaturation halo. For $\phi>\phi_m$, just beyond the K\"ohler critical radius, $\mathcal D_{\rm eff}<0$, and the supersaturation-mediated coupling becomes \textit{anti-diffusive}: the system enters a non-Fickian regime. The physical reason is that a locally larger droplet population lowers its own dressed supersaturation, so that the screened bath carries supersaturation from neighboring regions into it: the activated core draws on the supersaturation of its surroundings, and its contrast with them grows. In a gradient closure, this backward diffusion would be ill posed; in the pseudo-parabolic closure, the screening operator bounds the growth of fine scales (Section~\ref{Sec_Low_Pass_Truncation}), as in the forward--backward pseudo-parabolic equations of Refs.~\citep{novick1991stable,padron2004effect}, where it produces stable steep patterns rather than blow-up.}
 In non-equilibrium physics, such sign reversal is a standard mechanism for phase separation, boundary sharpening, and macroscopic pattern formation \cite{Cross_Hohenberg93}. Related non-Fickian transport mechanisms appear in reversed Ostwald processes in active phase-separating fluids \cite{fausti2024statistical} and in spontaneous layering of stratified turbulent flows \cite{middleton2025reconciling}. In the present cloud-physics setting, this emergent anti-diffusion means that activated local droplet populations effectively hoard the surrounding supersaturation bath, creating a non-equilibrium thermodynamic sink. Through their dragged supersaturation halos, they reinforce spatial contrasts in the activation field, sharpening the transition between cloudy and dry regions and helping stabilize distinct cloud edges.

Alongside this thermodynamically driven sharpening, the expansion of $2D\nabla\cdot\big(\epsilon\nabla\chi_0^*\big)$ also generates a nonlinear squared-gradient contribution, $2D\big[\epsilon'(\chi_0^*)'+\epsilon(\chi_0^*)''\big]|\nabla\phi|^2$. For an interface-height variable, a term of this form generates Kardar--Parisi--Zhang (KPZ) growth \cite{KPZ86,TGM97,corwin2012kardar}; {here it acts on an order parameter within the adopted nonlinear closure. It can affect interface propagation, but neither KPZ scaling nor a specific propagation asymmetry follows from its presence alone. Thus the pseudo-parabolic closure contains two complementary edge mechanisms: dynamic inertia, which bounds the response of fine scales through the footprint and delay of the screened halo, and halo-mediated anti-diffusion, which can sharpen cloud boundaries where the dressed slow manifold decreases ($\phi>\phi_m$). See {\it Supplementary Note 7} for the explicit expansion.}

Figure~\ref{Fig_MSR_schematic} summarizes how these ingredients follow from the elimination of the bath: the dressed slow manifold sets the local thermodynamic target, the screened supersaturation halo couples neighboring cloud elements, and the delayed adjustment of the halo renormalizes the inertia of the droplet field.

\begin{figure*}[t]
    \centering
   \includegraphics[width=.98\textwidth]{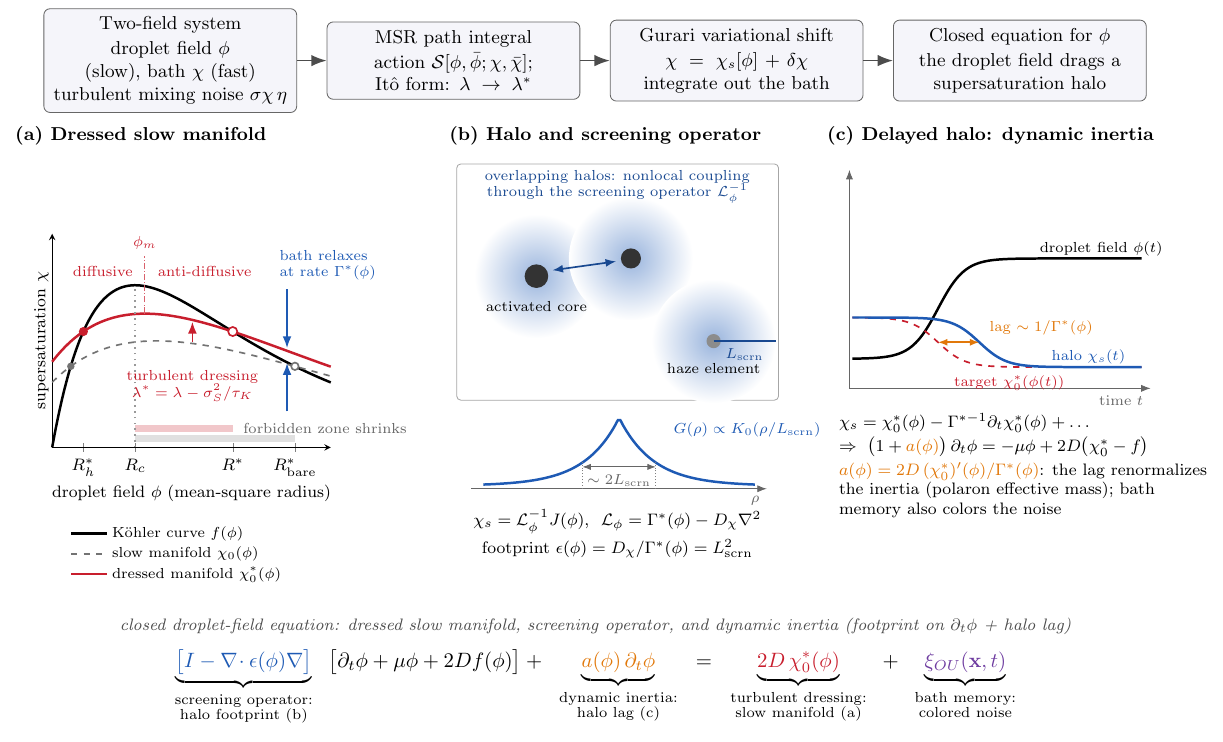}
    \caption{{\bf What the elimination of the supersaturation bath reveals.} Top: reduction of the droplet--supersaturation system (Eq.~\eqref{Eq_coupled_RD}) through the MSR action and the Gurari variational shift (Section~\ref{Sec_Field_Theory}). (a) Turbulent dressing lifts the slow manifold $\chi_0(\phi)$ to $\chi_0^*(\phi)$ (Eq.~\eqref{Eq_chi_star}) moves the edge state $R^*$ toward critical (squared) radius $R_c$, and shrinks the forbidden zone; the slope of $\chi_0^*$ changes sign at $\phi_m$ (Eq.~\eqref{Eq_phi_m}) separating diffusive from anti-diffusive coupling; see Eq.~\eqref{Eq_Deff}. Curves are drawn for illustrative parameters. (b) Each coarse-grained droplet population drags a supersaturation halo, the screened response $\mathcal L_\phi^{-1}J$ with Bessel kernel (Eq.~\eqref{Eq_Bessel_Kernel}) and footprint $\epsilon=L_{\rm scrn}^2$ (Eq.~\eqref{Eq_epsilon_phi_dressed}) while the overlapping halos couple neighboring cloud elements. (c) The halo relaxes at rate $\Gamma^*$ and lags its target, renormalizing the inertia through $a(\phi)$; see Eq.~\eqref{Eq_local_closure_lag}. Bottom: the resulting closure, Eq.~\eqref{Eq_PseudoParabolic_NonMarkovian}. Noise-driven hovering (Section~\ref{Sec_Lamperti}) is not drawn here: in (a), it would appear as stochastic excursions of $\phi$ along and transverse to the dressed manifold (as in Figs.~\ref{Fig_2Dcase2} and \ref{Fig_2Dcase3}) from the haze state toward the weakly repelling edge state $R^*$, lingering in the forbidden zone before returning to haze or activating.  {\it Supplementary Note 6} summarizes the construction of the closed droplet-field equation (Eq.~\eqref{Eq_PseudoParabolic_NonMarkovian}).}
    \label{Fig_MSR_schematic}
\end{figure*}

 The spatially correlated Markovian closure (Eq.~\eqref{Eq_Spatial_Markovian_closure}) is recovered by neglecting the bath-lag correction $a(\phi)\partial_t\phi$ and replacing the OU forcing by white noise at fixed time-integrated covariance, while retaining $D_\chi>0$. Spatial screening then remains: inversion of the screening operator spreads the white-time forcing across neighboring cloud elements, giving the covariance in Eq.~\eqref{Eq_Spatial_Markovian_cov}. At the level of the unexpanded bath response, this is precisely the zero-frequency reduction leading to Eq.~\eqref{Eq_Spatial_Markovian_closure}. For the divergence-form pseudo-parabolic equation above, the same limit gives its conservative nonlinear counterpart; the two agree on a homogeneous background, but differ in their treatment of spatially varying coefficients, as explained after Eq.~\eqref{Eq_Spatial_Markovian_pseudo}. Recovering the local closure \eqref{Eq_Macro_Langevin} additionally requires $\epsilon\to0$. Conversely, taking $\epsilon\to0$ alone removes screening but retains the lag and colored forcing. At finite screening length, neighboring coarse-grained populations respond through the shared supersaturation bath. The frozen-coefficient calculation fixes the linear coupling and its bounded high-wavenumber response; the divergence-form extension specifies the nonlinear coupling used to investigate cloud interfaces. Persistence and sharpening are mechanisms suggested by this structure, to be tested in solutions of the spatial model.

\section{Closure Results}\label{Sec_Closure_Results}

\subsection{Cloud-layer Organization From the Edge State}

We test whether the bath elimination derived above preserves both the evolving radius distribution and the spatial organization of the droplet field. We first compare the spatially correlated Markovian closure, Eq.~\eqref{Eq_Spatial_Markovian_closure}, with the droplet--supersaturation system \eqref{Eq_coupled_RD} from which it descends. The two-field system resolves the supersaturation bath and serves as the reference; the closure evolves only $\phi$, retaining the bath's screened, zero-frequency response in its drift and forcing. Both are integrated in It\^o form on a doubly periodic  domain of $128h\times128h$, that is, $128\times128$ cells of spacing $h$, with time step $0.01$\,s, for $N=500\,{\rm cm^{-3}}$, $\lambda^*=1.65\lambda_c$, $\sigma=0.1$, $\mu=0$, and the remaining parameters of Table~\ref{Table_RegimeA}.  For these numerical comparisons, we prescribe a smooth, positive-semidefinite spatial covariance $\rho_\eta$,  the Gaussian $\rho_\eta(r)=\exp(-r^2/\ell_{\rm noise}^2)$, with unit point variance $\rho_\eta(0)=1$  (a normalization by the point value, rather than to unit integral as for $\rho_{\eta_K}$) and correlation length $\ell_{\rm noise}=4h$, and denote the associated  $Q$-Wiener process by $W_\eta$.
  The covariance $\rho_\eta$ replaces $\rho_{\eta_K}$ in Eq.~\eqref{Eq_noise_covariance}, so that the bath noise satisfies $\langle\eta(\x,t)\eta(\x',t')\rangle=\rho_\eta(\x-\x')\,\delta(t-t')$; its correlation length $\ell_{\rm noise}$ is a numerical choice and is not identified with the physical Kolmogorov scale $\eta_K$.

 The bath-elimination formulas retain their form under this substitution, provided the covariance normalization is accounted for consistently in the noise amplitude and the Stratonovich-to-It\^o correction. The same covariance drives the parent system and the closure, and all spatial lengths are reported in units of the grid spacing $h$. The bath diffusivity, $D_\chi/h^2=1.326\,{\rm s^{-1}}$, sets the haze screening length $\ell_{\rm scrn}=\big(D_\chi/\Gamma^*(R_h)\big)^{1/2}$ to $4h$. The two lengths coincide in these runs but describe different things: the covariance supplied to the bath and the spatial response generated by bath diffusion.

The halo of the theory therefore extends over a few cloud elements, which is the situation in which neighboring droplet populations share their supersaturation. Both models start from the uniform edge state, $\phi=R^*$, with the bath of the two-field system on its slow manifold, and both are driven by the same realization of $W_\eta$. Sharing the forcing permits a direct comparison of the responses of the two models. Since $\phi$ is the local mean-square radius, $r=\sqrt\phi$ is the local root-mean-square radius; its distribution across grid cells describes the coarse-grained field, rather than the full distribution of individual droplet radii. We call a cell haze when $r=\sqrt\phi<r_c$, hovering when $r_c\le r\le\sqrt2\,r^*$, and activated when $r>\sqrt2\,r^*$, where $r_c$ is the K\"ohler critical radius and $r^*=\sqrt{R^*}$.

The reference field passes through three stages (Fig.~\ref{Fig_closure_maps}, top). For about ten minutes it hovers: nearly nine cells in ten remain in the hovering band, while haze opens in compact holes a few cells across. The holes then grow and connect into a labyrinth, and activated cloudlets appear on its ridges; by $t=20$\,min the hovering band has shrunk to the rims of these cloudlets. At $t=40$\,min the layer {consists predominantly of} compact activated cloudlets embedded in haze, separated from it by a boundary one or two cells wide. The closure reproduces each stage (Fig.~\ref{Fig_closure_maps}, bottom). The fractions of haze, hovering and activated cells {differ by at most about two percentage points at the four displayed times}{.} {The maps also show close correspondence in the haze holes, the labyrinthine organization and the placement of emerging cloudlets.} Under a common noise realization corresponding structures remain recognizable up to $t=20$\,min, after the first cloudlets have formed; the reported distributional and pattern diagnostics remain closely matched thereafter.

\begin{figure*}[t]
    \centering
   \includegraphics[width=.98\textwidth]{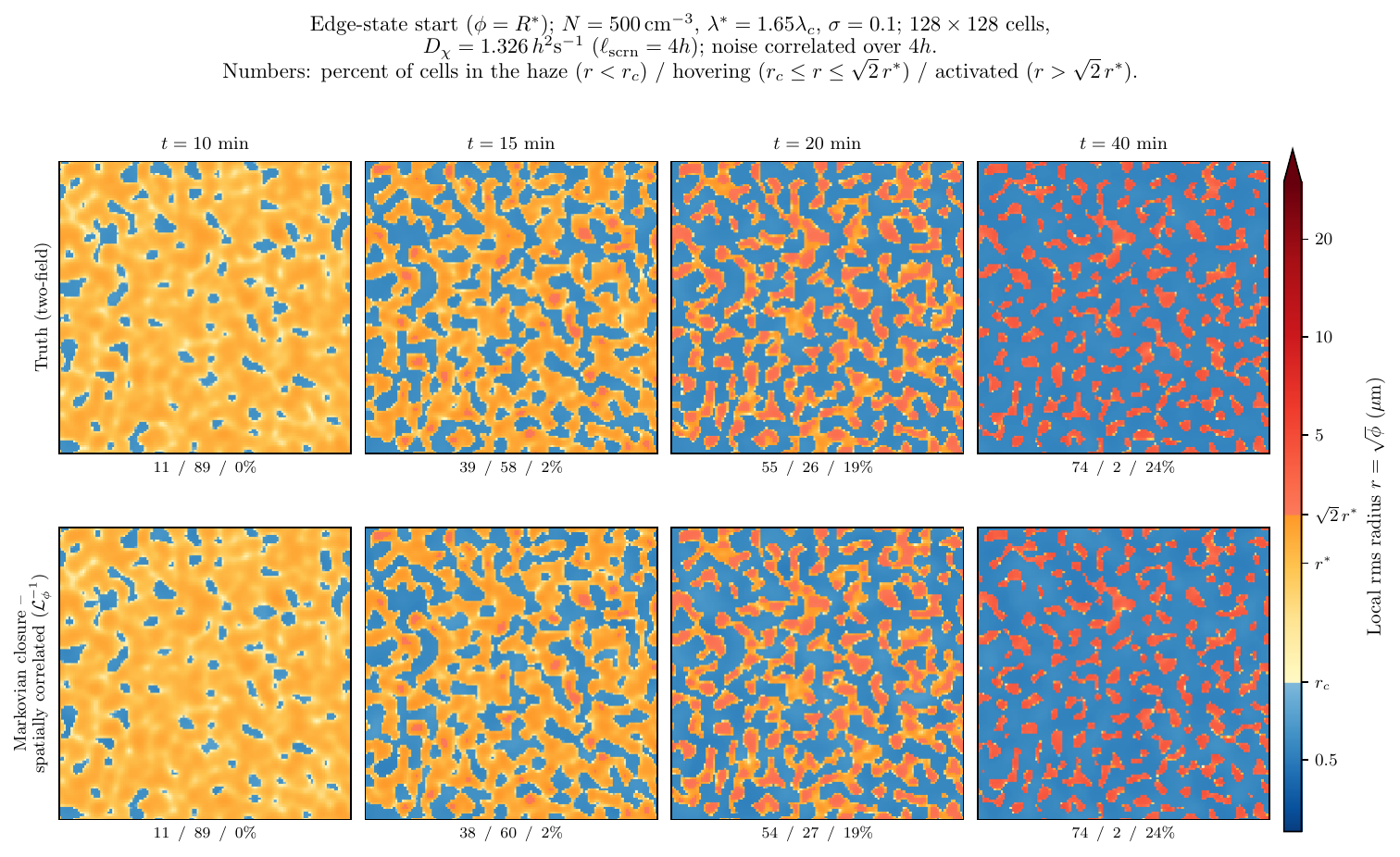}
    \caption{{\bf Cloud-layer organization from the edge state.} {Local root-mean-square-radius field $r=\sqrt\phi$} of the droplet--supersaturation system, Eq.~\eqref{Eq_coupled_RD} (top), and of the spatially correlated Markovian closure, Eq.~\eqref{Eq_Spatial_Markovian_closure} (bottom), at $t=10$, $15$, $20$ and $40$\,min, for $N=500\,{\rm cm^{-3}}$, $\lambda^*=1.65\lambda_c$ and $\sigma=0.1$. Both models start from the uniform edge state and are driven by the same noise realization. The prescribed bath-noise correlation length $\ell_{\rm noise}$ and the haze screening length $\ell_{\rm scrn}$ are both $4h$, with $D_\chi/h^2=1.326\,{\rm s^{-1}}$ on a $128h\times128h$ domain; the closure retains both the supplied noise covariance and the screened response. The numbers below each panel give the percentages of haze, hovering and activated cells.}
    \label{Fig_closure_maps}
\end{figure*}

The spatial distributions of root-mean-square radius in Fig.~\ref{Fig_closure_diagnostics} quantify this agreement. At the four displayed times the distribution is bimodal, with a haze peak near $0.55\,\mu$m and a second mode that moves from the edge state, near $2\,\mu$m, to about $4\,\mu$m as cloudlets grow. The closure places both modes, their relative weights and the plateau between them where the reference places them. Its Kullback--Leibler divergence from the reference remains below $0.07$ up to $t=20$\,min and reaches $0.16$ at $t=40$\,min, where the remaining discrepancy sits in the sparsely populated band just above $r_c$. The pattern statistics in the bottom row of Fig.~\ref{Fig_closure_diagnostics} follow the reference over the whole fifty minutes.  The number of hovering patches measures the fragmentation of the hovering set rather than its extent. At the edge-state start the hovering cells form a single domain-filling patch, and the count remains equal to one while haze opens isolated holes in it. It rises after the first quarter of an hour, when the holes merge and the hovering region breaks into the rims of the cloudlets, peaks near $24$\,min, and falls again as these rims are absorbed into the cloudlets or return to haze. The amount of hovering is read instead from the class percentages in Fig.~\ref{Fig_closure_maps}, which fall from $89\%$ at $t=10$\,min to $58\%$, $26\%$ and $2\%$ at $t=15$, $20$ and $40$\,min.
The number of cloudlets rises rapidly after the first activation, at $11.5$\,min in the reference and $12$\,min in the closure, reaches about $170$, then decreases before increasing again. The closure reproduces the initial rise and subsequent nonmonotonic evolution, although its intermediate dip is shallower than in the reference. The mean cloudlet area peaks near $28$ cells and then relaxes as cloudlets become compact.

The structure of the closure provides a physical interpretation of these agreements (Table~\ref{Tab_closure_terms}); the comparison tests their combined action rather than isolating the contribution of each term. The dressed slow manifold $\chi_0^*(\phi)$ of Eq.~\eqref{Eq_chi_star} fixes the haze state, the edge state and the shallow landscape between them, with the dressed removal rate $\lambda^*$ of Eq.~\eqref{Eq_Gamma_star}. This landscape organizes the initial hovering state and the return toward haze. The duration of hovering and the subsequent growth of the activated-radius mode depend on the combined drift, forcing and spatial coupling.

\begin{figure*}[t]
    \centering
   \includegraphics[width=.98\textwidth]{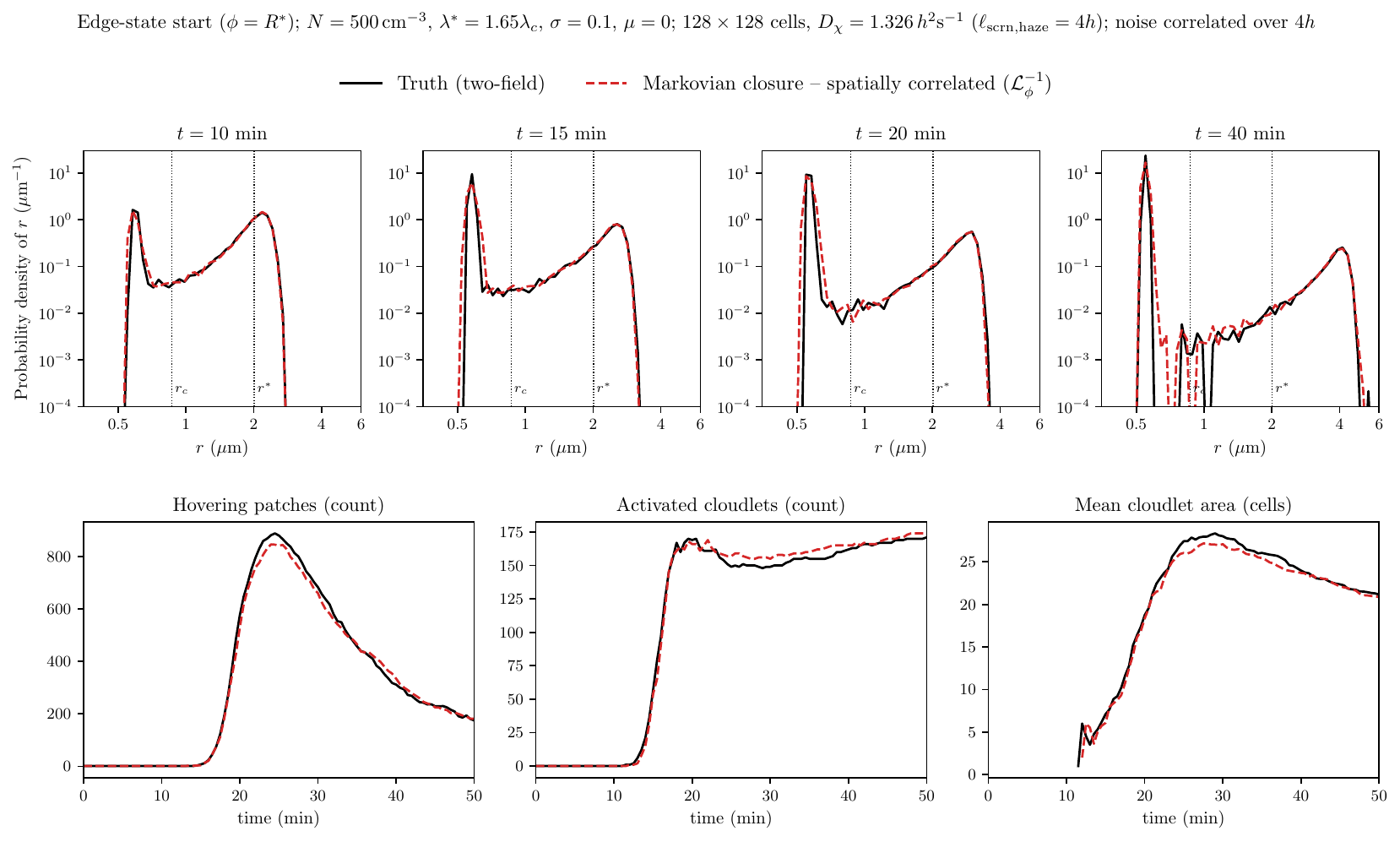}
    \caption{{{\bf {Radius distributions and pattern statistics}} for the simulations of Fig.~\ref{Fig_closure_maps}. Top: {spatial probability density of the local root-mean-square radius $r=\sqrt\phi$} over the $128^2$ cells at $t=10$, $15$, $20$ and $40$\,min (60 logarithmic bins); the dotted lines mark $r_c$ and $r^*$. Bottom: number of hovering patches, number of activated cloudlets and mean cloudlet area, in grid cells, as functions of time. Patches and cloudlets are the connected components, with periodic four-neighbor connectivity, of the hovering band $r_c\le r\le\sqrt2\,r^*$ and of the activated set $r>\sqrt2\,r^*$. Black: droplet--supersaturation system; red dashed: spatially correlated Markovian closure.}}
    \label{Fig_closure_diagnostics}
\end{figure*}
\begin{table*}[t]
\centering
\caption{{Terms of the spatially correlated Markovian closure, Eq.~\eqref{Eq_Spatial_Markovian_closure}, their physical content, and {associated diagnostics} in Figs.~\ref{Fig_closure_maps} and \ref{Fig_closure_diagnostics}.}}
\label{Tab_closure_terms}
\begin{tabular}{@{}p{4.2cm}p{6.0cm}p{6.6cm}@{}}
\toprule
\raggedright {Term} & \raggedright {Physical content} & \raggedright {{Related diagnostics}} \tabularnewline
\midrule
\raggedright {Dressed slow manifold $\chi_0^*(\phi)$, Eqs.~\eqref{Eq_Gamma_star} and \eqref{Eq_chi_star}} & \raggedright {Turbulently dressed activation landscape: haze state, edge state $R^*$, and the shallow region between them} & \raggedright {{Initial hovering near $R^*$ and emergence of a haze mode; activation timing and subsequent growth also depend on forcing and coupling}} \tabularnewline[3pt]
\raggedright {Screened drift $2D\,\mathcal L_\phi^{-1}J(\phi)$} & \raggedright {{Decaying, state-dependent halo coupling neighboring elements; long-wave coupling changes sign at $\phi_m$, Eqs.~\eqref{Eq_Deff} and \eqref{Eq_phi_m}}} & \raggedright {Compact haze holes merging into connected domains; compact cloudlets with sharp, persistent boundaries; cloudlet count and mean area} \tabularnewline[3pt]
\raggedright {Screened forcing  $2D\,\mathcal L_\phi^{-1}[b_b\,\d W_\eta]$, Eq.~\eqref{Eq_Spatial_Markovian_spectrum}} & \raggedright {Supplied covariance at scale  $\ell_{\rm noise}$ modified by screening at $\ell_{\rm scrn}$; fine-scale forcing suppressed} & \raggedright {Collective departures from the edge state; hovering and haze nucleate as patches; history of hovering patches} \tabularnewline[3pt]
\raggedright {Zero-frequency intensity $\Sigma_\phi$, Eq.~\eqref{Eq_Sigma_phi}} & \raggedright {Amplitude of the bath-induced forcing; Gaussian residual bath, accurate when $b_K^2\ll2\Gamma^*$} & \raggedright {{Radius-distribution widths and population fractions, jointly determined with drift and spatial coupling}} \tabularnewline
\bottomrule
\end{tabular}
\end{table*}

The screened response $\mathcal L_\phi^{-1}$ supplies the halo. In the drift, each cloud element responds through a surrounding supersaturation halo with a decaying, rather than sharply bounded, spatial influence. The Bessel-kernel picture of Section~\ref{Sec_Adabatic_Slaving} describes a homogeneous background; in the evolving field, the response also depends on the spatially varying $\Gamma^*(\phi)$. The long-wave expansion identifies a change in the sign of the effective coupling at $\phi_m$, Eq.~\eqref{Eq_phi_m}, where the dressed manifold is maximal. Below $\phi_m$, in the haze, the effective coefficient $\mathcal D_{\rm eff}$ of Eq.~\eqref{Eq_Deff} is positive, consistent with smoothing of haze-side variations. The compact holes and connected haze domains are compatible with this mechanism, acting together with the spatially correlated forcing.  Above $\phi_m$ it is negative, and the activated core draws supersaturation from its surroundings. {This contrast-amplifying contribution provides a mechanism consistent with the compact cloudlets and narrow boundaries seen in Fig.~\ref{Fig_closure_maps}.} Since $\sqrt{\phi_m}$ lies only a few percent above $r_c$ at $N=500\,{\rm cm^{-3}}$, {both signs of the long-wave coupling occur within the simulated field. The sign change is set by $\phi_m$, not by the operational threshold $r=\sqrt2\,r^*$ used to count activated cloudlets.}

The forcing is screened by the same operator. Equation~\eqref{Eq_Spatial_Markovian_spectrum} separates the prescribed bath spectrum $\widehat\rho_\eta(k)$, with correlation length $\ell_{\rm noise}$, from the additional screening factor $(1+\ell_{\rm scrn}^2|k|^2)^{-2}$. Both contribute to the spatial coherence of the increments in this experiment. The closure therefore retains the supplied bath correlations and their modification by the bath response; the appearance of patches cannot be attributed to screening alone. The zero-frequency intensity $\Sigma_\phi$ of Eq.~\eqref{Eq_Sigma_phi} fixes the leading forcing amplitude within the Gaussian residual approximation. With the unit-point-variance normalization used here, $b_K^2=\sigma^2$, so $b_K^2/(2\Gamma^*)$ is about $0.06$ at haze and $0.02$ at the edge state. These small ratios support the frozen-background Gaussian residual approximation of Appendix~\ref{Sec_Emergent_Noise_Constraint}. Temporal adiabaticity is a separate condition, governed by $|\nu|/(\Gamma_*+D_\chi|k|^2)$ in Eq.~\eqref{Eq_closure_scale_ratios}; the numerical comparison tests its practical adequacy for the reported observables.

The most visible early discrepancy is a slight widening of the haze peak toward small radii at $t=15$--$20$\,min, where the smaller $\Gamma^*$ makes bath elimination less accurate; at $t=40$\,min differences remain in the sparsely populated intermediate-radius band. Nevertheless, eliminating $\chi$ preserves the main evolution of both the radius distribution and the spatial pattern, including cloudlet formation and the subsequent changes in their number and area. These results support the instantaneous screened closure in this regime. They do not separately test the finite-memory correction or the nonlinear divergence-form extension discussed in Section~\ref{Sec_Phys_nonFickian}. The central result is that retaining the bath's state-dependent spatial response in both drift and forcing allows a single-field Markovian equation to reproduce the organization generated by the coupled droplet--supersaturation dynamics.

\subsection{Recurrent Hovering Near the Fold: The Role of Bath Memory}
\label{Sec_nearfold_memory}

The preceding comparison follows cloud organization from an initially
hovering field. We now ask whether the closures also reproduce hovering
patches that emerge from haze and subsequently dissolve. To favor these
excursions without requiring strong, non-Gaussian bath fluctuations, we
move closer to the saddle-node threshold, setting
$\lambda^*=1.25\lambda_c$ at $N=500\,{\rm cm}^{-3}$.
This lowers the deterministic barrier between the haze state and the
activation edge while allowing a moderate noise amplitude,
$\sigma=0.13$. At the haze equilibrium, the local diagnostic
$\sigma^2/[2\Gamma^*(R_h)]\simeq0.10$ supports the Gaussian
residual approximation. Temporal elimination is a separate issue:
the bath still filters fluctuations on timescales relevant to
haze--hovering transitions.

To isolate this temporal filtering, we retain the deterministic drift
and spatial operator of Eq.~\eqref{Eq_Spatial_Markovian_closure},
\be\label{Eq_nearfold_drift}
B[\phi]=-\mu\phi-2Df(\phi)
       +2D\mathcal L_\phi^{-1}J(\phi),
\ee
and restore the finite relaxation of the residual bath through an
auxiliary field $z$:
\bea\label{Eq_zbath_closure}
\d z&=-\mathcal L_\phi z\,\d t
      +b_b(\phi)\,\d W_\eta,\\
\partial_t\phi&=B[\phi]+2Dz.
\eea
This auxiliary-field construction fits within stochastic parameterization approaches that represent unresolved memory through additional dynamical variables \cite{MSM2015,SantosGutierrez2021,LucariniChekroun2023}. A related enlargement of the dynamical state underlies the history-space formulations of viscoelastic flow and Coleman--Gurtin heat conduction, where an additional variable carries the past evolution needed to determine the present response \cite{chekroun_glatt-holtz,Chek_al11_memo}.

Here $\mathcal L_\phi$ and $b_b$ are defined in
Eq.~\eqref{Eq_Spatial_Markovian_operator}, and $W_\eta$ has the
prescribed numerical covariance $\rho_\eta$ introduced above.
The instantaneous mean bath remains
$\mathcal L_\phi^{-1}J(\phi)$, whereas $z$ retains the temporal
correlation of its Gaussian residual fluctuations. We call
Eq.~\eqref{Eq_zbath_closure} the \emph{non-Markovian closure}
in this comparison: the augmented system $(\phi,z)$ is Markovian,
but eliminating $z$ introduces memory into the evolution of $\phi$ \cite{MSM2015}.
It is the colored-noise version tested here, rather than the full
pseudo-parabolic closure
\eqref{Eq_PseudoParabolic_NonMarkovian}. Testing the full pseudo-parabolic closure would change the spatial operator and introduce the deterministic bath-lag correction $a(\phi)\partial_t\phi$ alongside colored forcing. Equation~\eqref{Eq_zbath_closure} instead isolates fluctuation memory while retaining the instantaneous screened mean response.

The connection to the Markovian closure is explicit on a frozen
homogeneous background $\phi_0$. Writing
$\Gamma_*=\Gamma^*(\phi_0)$ and $b_{b,*}=b_b(\phi_0)$, the
power spectral density of the forcing $2Dz$has a Lorentzian frequency dependence \cite{Chekroun_al_RP2} at each wavenumber $k$:
\be\label{Eq_zbath_spectrum}
\widehat C_{2Dz}(k,\omega)
=
\frac{4D^2b_{b,*}^{\,2}\widehat\rho_\eta(k)}
     {(\Gamma_*+D_\chi|k|^2)^2+\omega^2}.
\ee
Its zero-frequency value is precisely the noise intensity in
Eq.~\eqref{Eq_Spatial_Markovian_spectrum}.  On this frozen background, the construction therefore
preserves the time-integrated forcing covariance while filtering rapid
fluctuations through the bath's own mode-dependent relaxation rates.
No additional correlation time is fitted.

We compare the parent two-field system, the Markovian closure and
Eq.~\eqref{Eq_zbath_closure} on a periodic $64h\times64h$ domain,
with $\mu=0$, $D_\chi/h^2=1.326\,{\rm s}^{-1}$,
$\ell_{\rm noise}=4h$ and time step $0.01$\,s.
All three start from uniform haze, $\phi=R_h$, and evolve for
20 minutes under the same realization of the supplied noise.
The parent bath and the residual field $z$ are prepared with
$\phi$ held at $R_h$ before release. The comparison uses one
matched realization and no recalibration.

\begin{figure*}[t]
    \centering
   \includegraphics[width=.98\textwidth]{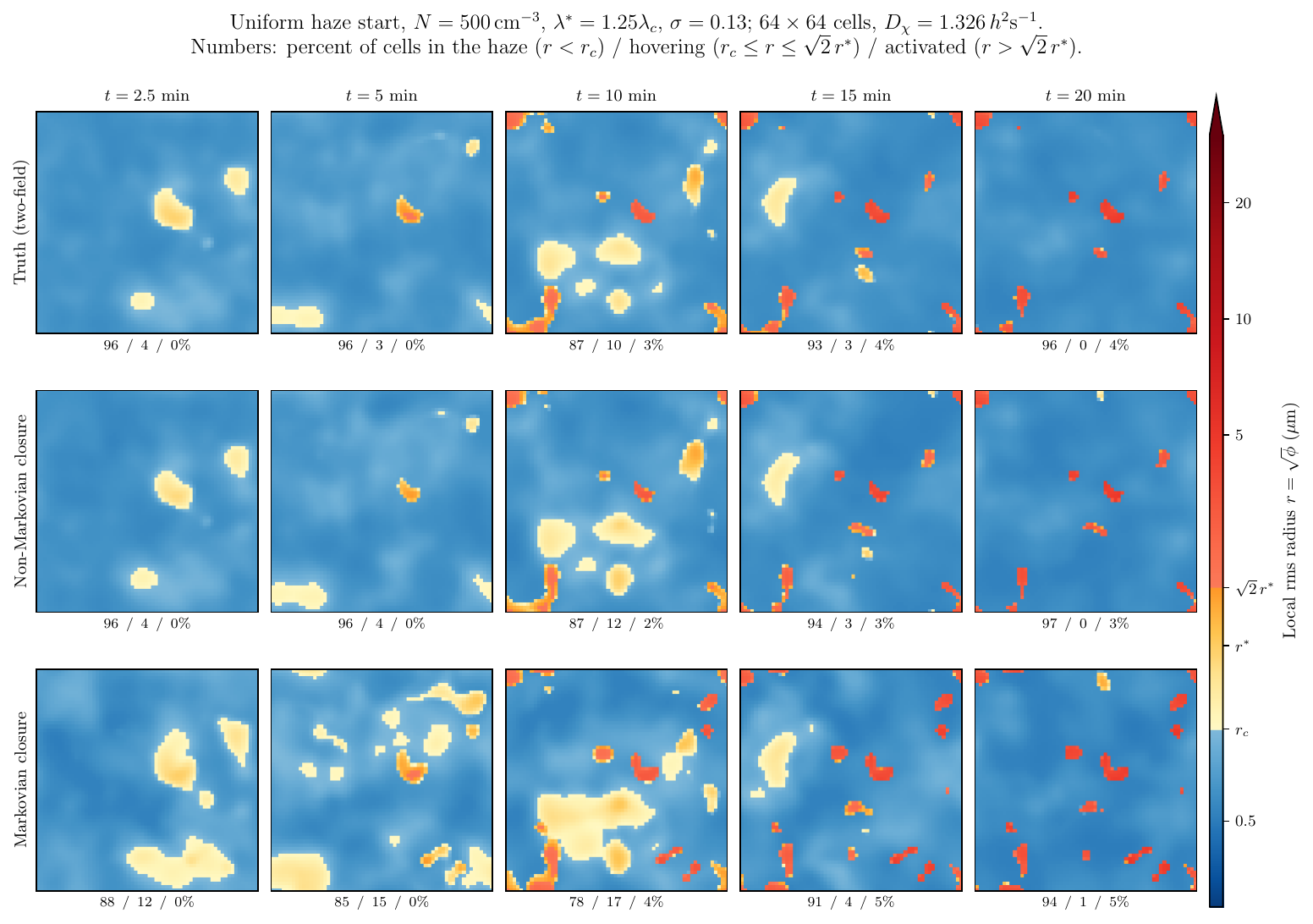}
    \caption{{\bf Recurrent hovering near the fold bifurcation.} Local root-mean-square-radius field $r=\sqrt\phi$ of the droplet--supersaturation system, Eq.~\eqref{Eq_coupled_RD} (top), the non-Markovian colored-noise closure with residual bath $z$ Eq.~\eqref{Eq_zbath_closure} (middle), and the spatially correlated Markovian closure, Eq.~\eqref{Eq_Spatial_Markovian_closure} (bottom), at $t=2.5$, $5$, $10$, $15$ and $20$\,min, for $N=500\,{\rm cm^{-3}}$, $\lambda^*=1.25\lambda_c$ and $\sigma=0.13$ on a $64h\times64h$ domain, with $D_\chi/h^2=1.326\,{\rm s^{-1}}$ and $\ell_{\rm noise}=4h$. All three models start from uniform haze, $\phi=R_h$, and are driven by the same noise realization. The numbers below each panel give the percentages of haze, hovering and activated cells.}
    \label{Fig_nearfold_maps}
\end{figure*}

Hovering patches emerge repeatedly from haze, grow and disappear,
while a small fraction develops into persistent activated structures
(Fig.~\ref{Fig_nearfold_maps}).
The Markovian closure captures most of their spatial placement but at the expense of over predicting hovering regions that are too  extensive.
Retaining bath fluctuation memory brings their extent, location and evolution
substantially closer to the parent system. 
At 10 minutes, the
hovering fraction is $12\%$ in the non-Markovian closure,
compared with $10\%$ in the parent and $17\%$ in the Markovian
closure.
Throughout its evolution, the system remains predominantly haze: this regime
supports recurrent local hovering excursions rather than conversion of
the whole domain into cloud.

 Figure~\ref{Fig_nearfold_diagnostics} separates the frequency,
duration and spatial extent of these excursions. The non-Markovian
closure follows the time-evolution of the hovering fraction much more closely
(panel a), particularly during the first ten minutes when the
Markovian closure overstates its peaks. To distinguish resolved
spells from threshold grazing, we require hovering spells and
intervening haze intervals to last at least 5\,s.
The entry rate is normalized by the time spent in haze, so that
different available haze populations do not bias the comparison.
Across the four five-minute windows, the Markovian closure
overestimates this rate by factors of $1.5$--$3.4$; retaining
fluctuation memory reduces the factors to $1.3$--$1.7$
(Fig.~\ref{Fig_nearfold_diagnostics}b). 

The non-Markovian correction largely preserves the residence-time distributions
conditioned on return to haze or progression to activation
(panel c), while suppressing the anomalous large hovering patches
generated by instantaneous forcing in the Markovian closure (panel d).
Repeated excursions also occur at individual locations: when both
the hovering spell and the subsequent haze interval must last at
least 30\,s, $6\%$ of cells complete two or more cycles over the
20-minute run in the parent system, compared with $14\%$ in the
Markovian closure and $9\%$ when bath fluctuation memory is retained.

These results identify finite bath fluctuation memory as an
important ingredient in capturing recurrent hovering by a closure. The instantaneous
screened closure already retains much of the spatial response
and the duration of resolved spells; restoring temporal filtering
substantially improves how often those spells begin and how far
their patches extend. The remaining excess in entry rate shows
that this correction is not complete, but the agreement is
improved without changing the deterministic drift or fitting a
new timescale. The experiment thus tests the physical role of
colored bath forcing separately from the deterministic lag
and nonlinear spatial extension of the full pseudo-parabolic
theory. Numerical tests of the full, non-Markovian pseudo-parabolic closure (Eq.~\eqref{Eq_PseudoParabolic_NonMarkovian}), including regimes in which mean-response lag becomes important, are reserved for a separate study. 

\begin{figure*}[t]
    \centering
   \includegraphics[width=.90\textwidth]{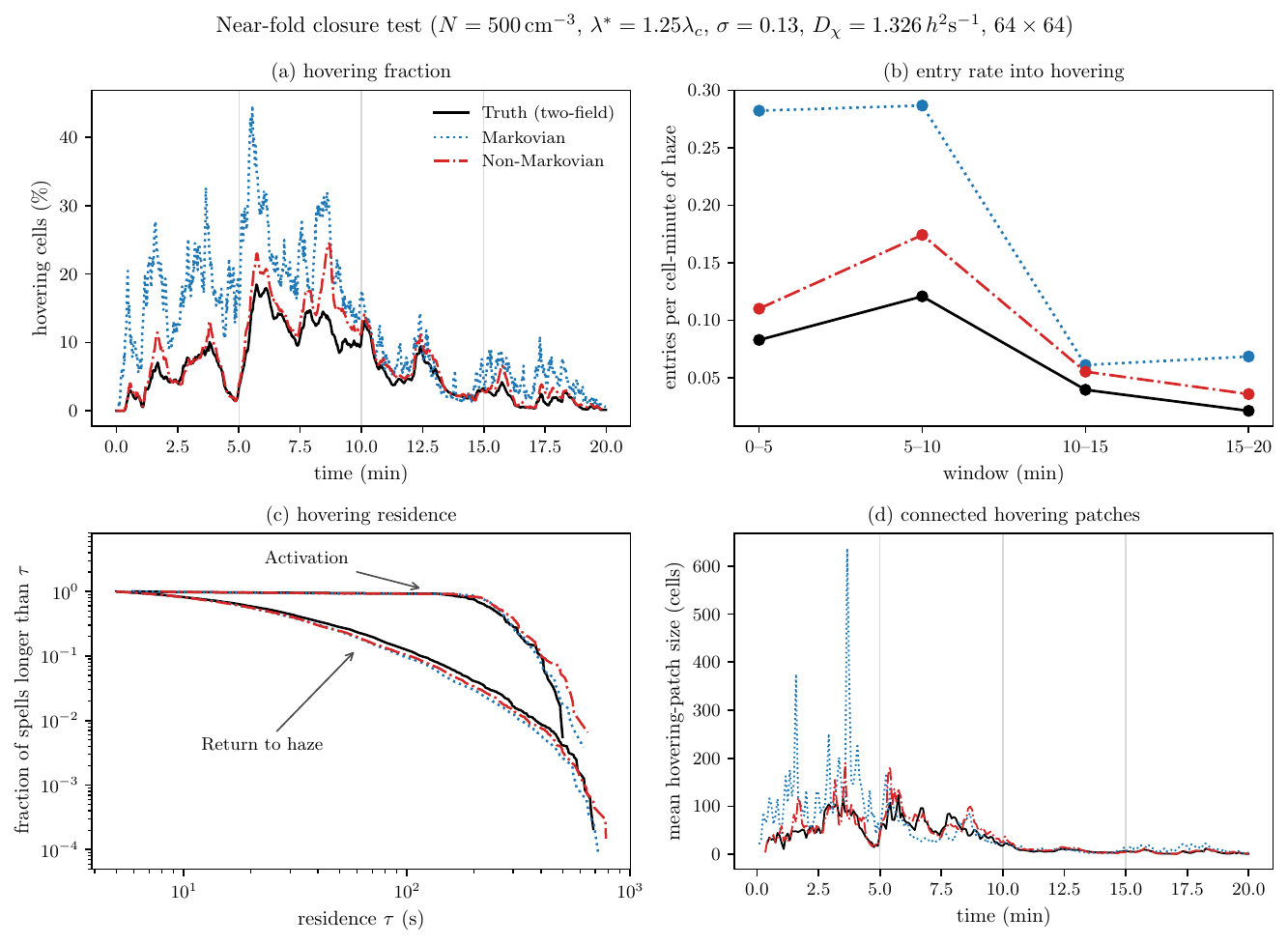}
    \caption{ {\bf Frequency, duration and extent of hovering excursions} for the simulations of Fig.~\ref{Fig_nearfold_maps}. (a) Percentage of hovering cells. (b) Rate of entry into resolved hovering spells per cell-minute spent in haze, in four five-minute windows. (c) Residence-time survival curves, giving the fraction of resolved hovering spells lasting longer than $\tau$, separately for spells ending in haze and in activation. (d) Mean size, in grid cells, of the connected hovering patches, with periodic four-neighbor connectivity. Resolved hovering spells and the intervening haze intervals last at least 5\,s; the improvement persists when this minimum duration is varied from 2 to 10\,s. Black: droplet--supersaturation system; blue dotted: Markovian closure; red dash-dotted: non-Markovian colored-noise closure, Eq.~\eqref{Eq_zbath_closure}.}
    \label{Fig_nearfold_diagnostics}
\end{figure*}

\section{Discussion}
\label{Sec_Discussion}

The formation and persistence of twilight clouds---optically thin transitional wisps between humidified aerosol and fully activated cloud---pose a genuinely multi-scale problem in atmospheric physics. Their constituent wet aerosol and small droplets sit near an activation boundary, where small changes in mixing, supersaturation, or aerosol loading can reorganize the entire cloud state. In this study, we moved beyond a purely deterministic K\"ohler picture by constructing a theoretical framework that connects the local nonlinear thermodynamics of droplet activation to the macroscopic field dynamics of an extended cloud layer.

The starting point was the timescale separation already present in the classical droplet-growth equations. {The fast supersaturation variable tracks a quasi-steady balance set by the slower droplet-growth coordinate. This curve provides a leading approximation to the slow dynamics, rather than an invariant manifold; the finite tracking lag is quantified in Section~\ref{Sec_expo_attraction}. The resulting slow geometry organizes the parcel dynamics and determines how activation proceeds.} Introducing an advective supersaturation removal rate $\lambda$ drives the reduced system through a saddle-node bifurcation, replacing the classical K\"ohler critical radius by a dynamically shifted activation threshold $R^*$.

This saddle-node topology exposes a forbidden zone within the classical K\"ohler picture: a regime where deterministic thermodynamics would predict immediate return to haze, but where the slow-fast system supports long-lived transitional dynamics. When turbulent fluctuations are introduced into the mixing rate, the Lamperti transformation of the reduced stochastic system  reveals a flattened effective potential near the edge state (Section~\ref{Sec_Lamperti}). In the supercritical mixing regime $\lambda>\lambda_c$, droplets do not simply activate or evaporate. Instead, they hover near the activation boundary, spending extended periods in a suspended intermediate state. This provides a microscopic dynamical mechanism for the persistence of twilight clouds and for the typical $1$--$2\mu{\rm m}$ droplet-size distributions observed in weak-updraft boundary layers \citep{hirsch2014transition}; see Figures~\ref{Fig_Hovering_N=500_sig0dot3}, \ref{Fig_Hovering_N=500_sig0dot35}, and \ref{Fig_DSD_comp}.

The 0D parcel theory, however, cannot by itself describe a physical cloud layer. It contains no spatial thermodynamic communication between neighboring cloud elements and cannot explain how hovering states organize across space. To address this, we lifted the slow-manifold dynamics into a non-equilibrium statistical field theory using the MSR path-integral formalism (Section~\ref{Sec_Field_Theory}). The slow field is the mean-square-radius field $\phi(\x,t)=\langle R\rangle_{\x,t}$, representing a local coarse-grained droplet population, while the fast field $\chi(\x,t)$ is the resolved supersaturation bath sampled by that population.

A central theoretical result is the correspondence between this cloud-field problem and the polaron problem in many-body physics. In the Fr\"ohlich--Gurari picture, a slow electron becomes dressed by a polarization envelope of the surrounding fast medium, modifying its effective dynamics \citep{feynman1955slow,gurari1953}. In the cloud problem, the slow object is not a single droplet, but the local droplet population represented by $\phi$. This population becomes dressed by a statistically localized supersaturation halo generated by the fast turbulent bath $\chi$. The resulting cloud polaron is therefore collective: a coarse-grained droplet field dragging a supersaturation envelope, not an individual droplet carrying a microscopic vapor shell.

This correspondence becomes mathematically explicit through the Gurari variational shift used to integrate out the fast supersaturation bath (Section \ref{Sec_slow_effective_Gurari}). {The shift is fixed by a stationarity condition, the MSR counterpart of Gurari's variational condition and of Feynman's extremal shift of the path, and the optimal displacement has the same denominator as Gurari's optimal phonon amplitudes: bath relaxation rate, diffusive cost, and lag, in place of phonon frequency, recoil energy, and electron motion. Its local, adiabatic component} produces the dressed slow manifold $\chi_0^*(\phi)$ and shows that turbulence modifies the effective K\"ohler thermodynamics {within the stated microphysical projection and bath closure.} {In particular, the relaxation-rate correction $-\sigma_S^2/\tau_K$ follows from the Stratonovich-to-It\^o conversion of the spatially regularized bath equation, equivalently accounted for by the measure terms of a consistent midpoint formulation.} Its effect is to reduce the effective supersaturation relaxation rate and oppose the drying action of the background mixing rate $\lambda$. As a result, the turbulent dressing shifts the saddle-node threshold to $\lambda_c^{\rm eff}=\lambda_c+\sigma_S^2/\tau_K$, shrinking the forbidden zone and requiring a stronger mixing environment to suppress activation than would be predicted by the deterministic parcel theory. {The parcel model, read in the same Stratonovich interpretation, shows the same shift, $\lambda_c+\sigma^2/2$, and in the regimes where hovering occurs the shift is comparable to, or larger than, $\lambda_c$ itself (Section~\ref{Sec_stochastic}).}

The second major consequence appears when the zero-diffusion and instantaneous-bath limits are relaxed (Section \ref{Sec_Dynamic_Inertia}). In the local Markovian closure, the dragged supersaturation halo is already present, but it is effectively collapsed to a point: neighboring halos do not interact (Section \ref{Sec_dragged_env}). Once the finite spatial response of the supersaturation bath is retained, the inverse screened operator $\mathcal M_\phi=(\Gamma(\phi)-D_\chi\nabla^2)^{-1}$ couples neighboring cloud elements through a Bessel-type Green's function. The screening length $L_{\rm scrn}=\sqrt{D_\chi/\Gamma^*(\phi)}$ determines the range over which the dressed supersaturation halo attached to one coarse-grained droplet population overlaps with those of its neighbors (Section \ref{Sec_Adabatic_Slaving}).

{Written in local form, this screened response becomes the pseudo-parabolic operator $-\nabla\cdot[\epsilon(\phi)\nabla\partial_t\phi]$, with $\epsilon(\phi)=D_\chi/\Gamma^*(\phi)=L_{\rm scrn}^2$, which records the footprint of the dragged halo (Section \ref{Sec_Low_Pass_Truncation}). Retaining the finite relaxation time of the bath at first order then shows that the halo also lags behind the droplet field, and this lag renormalizes the coefficient of $\partial_t\phi$ by the factor $1+a(\phi)$, with $a=2D(\chi_0^*)'/\Gamma^*$ (Section~\ref{Sec_Effective_Mass}). The analogy with the polaron is then complete but collective: the footprint plays the role of the polaron radius, and the lag correction that of the mass enhancement, which, as for the polaron at weak coupling, is the ratio of the bath-mediated coupling rate to the bath relaxation rate. Together they form the dynamic-inertia operator, which bounds the response of fine spatial scales instead of letting a gradient-driven coupling damp or amplify them without bound.

The same pseudo-parabolic closure also reveals a second, more topological, spatial mechanism (Section \ref{Sec_Phys_nonFickian}). At long wavelengths, the halo-mediated coupling is the screened diffusion of the dressed slow manifold, $2D\nabla\cdot[\epsilon(\phi)\nabla\chi_0^*(\phi)]$, with coefficient $\mathcal D_{\rm eff}(\phi)=2D\epsilon(\phi)(\chi_0^*)'(\phi)$. Since $\epsilon(\phi)>0$, the sign of the effective transport is set by the slope of the dressed slow manifold. Below its maximum $\phi_m$, which includes the haze regime, the coupling acts diffusively and smooths unactivated aerosol fluctuations. Beyond $\phi_m$, which coincides with the K\"ohler critical value at the saddle-node threshold and lies just above it in the hovering regimes, the same coupling becomes anti-diffusive. Activated local droplet populations then draw supersaturation from their surroundings, creating a non-equilibrium thermodynamic sink, while {the frozen-coefficient pseudo-parabolic response bounds its high-wavenumber growth; nonlinear well-posedness requires additional analysis.}}

At the stochastic level, integrating out the fast bath does not merely add scalar white noise to the droplet equation. It generates a non-Markovian, spatially screened noise structure (Section \ref{Sec_Emerging_OU_noise}). {The same screened bath response that produces dynamic inertia also shapes the macroscopic fluctuations: the forcing felt by the droplet field is correlated over the footprint of the dragged halo and in time over the bath relaxation time, and its Markovian limit has the same zero-frequency intensity. Thus the supersaturation bath acts not only as a source of turbulent fluctuations, but also as a thermodynamic buffer that distributes those fluctuations over a finite response envelope before they drive the macroscopic droplet field. Near a critical relaxation threshold, this forcing becomes both stronger and slower, a critical slowing down that marks the limit of the closure.}

{Figures~\ref{Fig_closure_maps} and \ref{Fig_closure_diagnostics} demonstrate that this spatial response survives the elimination of the bath in a quantitatively useful form. In the tested regime, the single-field Markovian closure follows the two-field system from an initially uniform edge state through the opening and connection of haze regions to a field of activated cloudlets. It reproduces the evolving distribution of local root-mean-square radius, the rise and decline of hovering-patch counts, and the growth and subsequent evolution of cloudlet area. The cloudlet count follows the same nonmonotonic course, with a shallower intermediate dip in the closure. The agreement therefore extends beyond local hovering statistics to the changing geometry of the field.}

This comparison tests the joint action of the screened drift and forcing in Eq.~\eqref{Eq_Spatial_Markovian_closure}. The bath forcing already has spatial correlations, and the eliminated bath modifies them through the same response operator that mediates the deterministic coupling. Preserving this operator allows the closure to retain spatial organization without evolving supersaturation explicitly. The results support the instantaneous screened reduction over the simulated interval; deterministic bath-lag corrections and the separate nonlinear divergence-form extension are not tested by this comparison. They thus establish a practical link between the analytical bath elimination and the evolving cloud structures it is intended to describe.

 The near-fold comparison (Section~\ref{Sec_nearfold_memory}) separately tests fluctuation memory. Starting from haze, all three models generate recurrent hovering, but the instantaneous closure overestimates its frequency and patch extent. Restoring the residual bath's relaxation, while retaining the same deterministic drift and frozen-background time-integrated forcing covariance, reduces the excess entry-rate factors from $1.5$--$3.4$ to $1.3$--$1.7$ across the four time windows. Conditional residence-time distributions remain comparatively similar. Thus bath fluctuation memory chiefly improves how often resolved hovering spells begin and how far their patches extend in this experiment. This improvement uses the bath's own relaxation rates, with no fitted correlation time or recalibration.

These results clarify the coupled role of aerosol loading and turbulent mixing.  Laboratory experiments show that increasing droplet concentration accelerates supersaturation relaxation and reduces fluctuation-driven broadening of the droplet-size distribution \citep{chandrakar2016}. These chamber results also informed the Brownian-model description of coexisting haze and cloud droplets developed in \citep{Santos_al24_SciAdv}. Here, increasing the aerosol number concentration $N$ strengthens the slow-fast separation and deepens the nonlinear structure that makes hovering possible. The fate of marginal cloud droplets is therefore not controlled solely by local K\"ohler thermodynamics, nor solely by stochastic mixing, but by their coupled effect: aerosol microphysics shapes the slow manifold, while turbulent supersaturation fluctuations dress it, shift its bifurcations, and mediate its spatial organization.

The theory developed here therefore closes the activation problem dynamically near the K\"ohler threshold. At parcel level, supersaturation behaves as a fast thermodynamic bath slaved to a slow droplet coordinate, producing saddle-node activation, a forbidden zone, and stochastic hovering. At field level, integrating out that bath dresses the coarse-grained droplet population, producing supersaturation halos, retarded self-interaction, colored fluctuations, dynamic inertia, and {anti-diffusion controlled by the dressed slow manifold}.

Ultimately, these results show that the interaction between microphysical thermodynamics and stochastic turbulent mixing is not a secondary perturbation to cloud evolution. It is a primary organizing mechanism. Turbulent dressing reshapes the activation threshold, nonlocal supersaturation halos couple neighboring cloud elements, dynamic inertia {spreads and} delays their response, and anti-diffusive {halo-mediated} transport sharpens activated cloud boundaries. Together, these mechanisms explain how a microscopic principle---supersaturation relaxation toward the droplet-growth slow manifold---is lifted into macroscopic spatiotemporal operators governing the persistence, morphology, and edge structure of twilight clouds.

By casting warm-cloud microphysics as a statistical field theory, this work offers a complementary route to the traditional treatment of subgrid cloud processes. Instead of representing the droplet spectrum through computationally intensive bin microphysics and then closing turbulence--activation feedback empirically, the MSR construction derives an interpretable reduced field theory in which the relevant closures appear as explicit operators: turbulent dressing, screened supersaturation halos, colored noise, dynamic inertia, and K\"ohler-mediated anti-diffusion. The resulting non-Markovian pseudo-parabolic stochastic PDE, Eq.~\eqref{Eq_PseudoParabolic_Final_Expanded}, is therefore not only a reduced model, but also an analytical object on which cloud regimes, bifurcations, and stability boundaries can be studied directly. A natural next step is to carry out such a bifurcation and regime analysis of the full spatiotemporal closure. 

Equally important, the MSR framework is intrinsically designed to compute response functions. It therefore provides an analytic route to perturbation response and uncertainty quantification in cloud physics, questions that are often accessible only through large ensembles of costly numerical simulations. In this sense, the field-theoretic formulation does not merely enable the derivation of macroscopic spatiotemporal models consistent with known microphysical mechanisms. It opens a systematic response theory for how twilight clouds react to fluctuations in aerosol loading, mixing, updraft strength, and supersaturation forcing.

Beyond cloud microphysics, the construction developed here suggests a general route to effective theories of slow fields coupled to relaxing, diffusive environments. In degenerate reaction--diffusion systems, a slow field without intrinsic diffusion can acquire spatial interactions through the environmental field that it modifies. The MSR formulation and the Gurari--Feynman reduction give this mechanism a collective-polaron interpretation: the slow field becomes dressed by an environmental halo, whose spatial extent mediates screened coupling and whose delayed adjustment generates memory and dynamic inertia, while the same response filters environmental fluctuations. This structure connects naturally to vegetation--water feedbacks, where slowly spreading biomass reorganizes a mobile water resource \citep{gilad2007ecosystem,meron2016pattern}, and to chemical and biological systems whose local dynamics communicate through a diffusing mediator \citep{tanaka2003nonlocal}. 

Plasma physics provides a classic precedent. A test charge carries a Debye shielding cloud, and the dressed-test-particle picture of Rostoker and Rosenbluth builds plasma kinetics on such dressed particles \citep{rostoker1960test,rostoker1961fluctuations}. In collisional, diffusive regimes, the shielding obeys a screened relaxation operator of the same form as $\mathcal L_\phi$, with the Debye length set by the diffusivity and the charge-relaxation time; at finite speed, the cloud lags behind the charge, producing wakes and polarization drag \citep{peter1991energy}. In dense plasmas, eliminating the fast electrons likewise leaves the ions with screened interactions, electron--ion friction, and correlated noise \citep{dai2010unified,mabey2017strong}, and in complex plasmas it reduces the dust dynamics to screened, wake-mediated interactions \citep{morfill2009complex}. The MSR generating functional is itself a standard tool in the statistical theory of plasma turbulence \citep{krommes2002fundamental}.

A complementary extension concerns bacterial colonies with density- and nutrient-dependent mobility, as in the model of Kawasaki et al.~\citep{kawasaki1997patterns}, where nutrient availability controls both population growth and a nonlinear diffusivity that vanishes at zero bacterial density or nutrient concentration. Such systems motivate extending the present environmental-dressing approach to resource-dependent transport, in which eliminating the resource field would couple growth and mobility through its spatial response and temporal memory. It also suggests extensions to transport in heterogeneous media, where distributions of environmental relaxation times generate richer memory kernels \citep{dentz2003transport}. The cloud problem thus provides a concrete realization of a broader organizing principle: spatial organization and temporal response emerging together from the dynamics of the environment surrounding the slow degrees of freedom.

\begin{acknowledgements}
The authors are grateful for inspiring discussions with Yi-Hung Kuo, Honghu Liu, David Neelin,  and Manuel Santos-Guti\'errez. 
This research was supported by the Institute for Environmental Sustainability (IES) at the Weizmann Institute of Science.
This research was also supported by the European Research Council (ERC) under the European Union's Horizon 2020 research and innovation program (grant agreement no.~810370), the Knell Family Institute for Artificial Intelligence, Weizmann Institute,  by the Planning and Budgeting Committee (VATAT) of the Council for Higher Education in Israel, as part of a grant for the climate research center titled "Integrating Climate Dynamics, Clouds, and Extreme Events through Teleconnections in Climate Networks," and by the National Science Foundation Grant DMS-2407484. 
\end{acknowledgements}

\subsection*{Data Availability Statement}
The simulation data and codes supporting the numerical results are available from the corresponding author upon reasonable request.

\begin{appendices}
{\small

\addtocontents{toc}{\protect\vspace{-7pt}}
\section{Mean-Field Projection of the K\"ohler Curve}
\label{App_Kohler_projection}
We clarify here the closure step by which the microscopic K\"ohler curve $f(R)$ is represented in the macroscopic equation for the Mean-Square Radius Field $\phi$. Let $h_{\x,t}(R)$ denote the normalized local distribution of squared droplet radii,
\be
\int_0^\infty h_{\x,t}(R)\,\d R=1,
\qquad
\phi(\x,t)=\int_0^\infty R\,h_{\x,t}(R)\,\d R .
\ee
The exact local average of the K\"ohler contribution is
\be
\langle f(R)\rangle_{\x,t} = A\left\langle R^{-1/2}\right\rangle_{\x,t} - B\left\langle R^{-3/2}\right\rangle_{\x,t}.
\label{Eq_exact_Kohler_average}
\ee
Thus, in general,
\be
\langle f(R)\rangle_{\x,t} \neq f\big(\langle R\rangle_{\x,t}\big).
\ee
The thermodynamic closure used in the Main Text is therefore not an exact identity, but a projection of the underlying spectral dependence in Eq.~\eqref{Eq_exact_Kohler_average} onto the single resolved order parameter $\phi$.

A clean way to visualize the structure of this projection is to write the local spectrum in the scaled form
\be
h_{\x,t}(R) = \frac{1}{\phi(\x,t)} H_{\x,t}\left(\frac{R}{\phi(\x,t)}\right),
\qquad
y=\frac{R}{\phi(\x,t)} ,
\ee
where the shape function $H_{\x,t}$ is normalized and has unit first moment:
\be
\int_0^\infty H_{\x,t}(y)\,\d y=1,
\qquad
\int_0^\infty y\,H_{\x,t}(y)\,\d y=1 .
\ee
The negative moments then scale directly with $\phi$,
\beas
\left\langle R^{-p}\right\rangle_{\x,t} &=& \phi(\x,t)^{-p} C_p(\x,t),\\
C_p(\x,t)&=&\int_0^\infty y^{-p}H_{\x,t}(y)\,\d y .
\eeas
Consequently,
\be
\langle f(R)\rangle_{\x,t}
=
A\,C_{1/2}(\x,t)\phi^{-1/2}
-
B\,C_{3/2}(\x,t)\phi^{-3/2}.
\label{Eq_Kohler_projected_shape}
\ee
Since $H_{\x,t}$ is a nonnegative normalized distribution on $y>0$, the factors $C_{1/2}$ and $C_{3/2}$ are positive whenever the corresponding negative moments are finite. Thus the projected average preserves the Kelvin--Raoult sign structure of the microscopic K\"ohler curve. However, unless the spectral shape is specified, $C_{1/2}$ and $C_{3/2}$ remain unresolved spectral quantities. The one-field closures derived in the Main Text (Eq.~\eqref{Eq_Macro_Langevin} and Eq.~\eqref{Eq_PseudoParabolic_NonMarkovian}) therefore does not attempt to evolve these factors. Instead, it keeps the leading contribution in an expansion around the resolved Mean-Square Radius Field.

To see this explicitly, let us write
\bes
R=\phi(1+\delta), \mbox{ with } \langle\delta\rangle_{\x,t}=0.
\ees
Then
\bea
\left\langle R^{-1/2}\right\rangle_{\x,t}
&=&
\phi^{-1/2}
\left[
1+\frac{3}{8}\langle\delta^2\rangle
-\frac{5}{16}\langle\delta^3\rangle
+\cdots
\right],
\\
\left\langle R^{-3/2}\right\rangle_{\x,t}
&=&
\phi^{-3/2}
\left[
1+\frac{15}{8}\langle\delta^2\rangle
-\frac{35}{16}\langle\delta^3\rangle
+\cdots
\right].
\eea
Therefore from Eq.~\eqref{Eq_exact_Kohler_average}
\beas
\langle f(R)\rangle_{\x,t}
&= A\phi^{-1/2}- B\phi^{-3/2}
\\
& +\left[
\frac{3A}{8}\phi^{-1/2}- \frac{15B}{8}\phi^{-3/2}
\right]\langle\delta^2\rangle
\\
& - \left[ \frac{5A}{16}\phi^{-1/2} - \frac{35B}{16}\phi^{-3/2} \right]\langle\delta^3\rangle
+\cdots.
\eeas

The single-order-parameter closure consists in retaining the leading term,
\be
f_{\rm eff}(\phi) \simeq f(\phi) = A\phi^{-1/2} - B\phi^{-3/2},
\ee
and neglecting the explicit spectral-shape corrections proportional to $\langle\delta^2\rangle$, $\langle\delta^3\rangle$, and higher moments. Thus $f(\phi)$ is not the exact identity $\langle f(R)\rangle=f(\langle R\rangle)$; it is the leading projection of the K\"ohler response onto the resolved Mean-Square Radius Field.

Notably, for a given relative spectral width, the solute/Raoult contribution is five times more sensitive to variance than the Kelvin contribution, since the corresponding coefficients are $15/8$ and $3/8$, respectively. Physically, this reflects the stronger weighting of the small-radius tail by the more singular moment $\langle R^{-3/2}\rangle$.

{The closure retains the singular barrier and the Kelvin--Raoult sign structure, but its quantitative accuracy requires small relative-width corrections and finite, controlled negative moments. The parcel histograms illustrate possible radius ranges; they are not by themselves measurements of the simultaneous local distribution $h_{\x,t}$. A test of the closure should therefore evaluate $C_{1/2}$ and $C_{3/2}$, or the relative-width corrections above, in a resolved droplet population. Broader spectra would require additional spectral variables or a parameterization of these shape factors.}

\addtocontents{toc}{\protect\vspace{-7pt}}
\section{Microphysical Averages}\label{Sec_microphys_average}
The local microphysical average is taken over the droplet ensemble contained in the coarse-grained element. Let
$\mathcal{H}_{\x,t}(R,s)$  denote the normalized local joint distribution of squared radius and droplet-sampled supersaturation, where here $s$ denotes the scalar value of the microscopic supersaturation sampled by a droplet. Thus
\bes
\int_0^\infty\int_{-\infty}^{\infty} \mathcal{H}_{\x,t}(R,s) \, \d s \, \d R = 1.
\ees
For any microphysical observable $\Psi(R,s)$, we define
\bes
\langle \Psi\rangle_{\x,t} = \int_0^\infty\int_{-\infty}^{\infty} \Psi(R,s)\mathcal{H}_{\x,t}(R,s) \, \d s \, \d R.
\ees
The local droplet-size distribution is the $R$-marginal,
\bes
h_{\x,t}(R) = \int_{-\infty}^{\infty} \mathcal{H}_{\x,t}(R,s) \, \d s.
\ees
Therefore, for observables depending only on $R$, i.e.~$\Psi(R,s)=F(R)$, we have
\bes
\langle F(R)\rangle_{\x,t} = \int_0^\infty F(R)h_{\x,t}(R) \, \d R.
\ees
In particular, for $F(R)=R$ and $\Psi(R,s)=s$, 
\bea\label{Eq_def_chi_phi}
&\phi(\x,t) = \langle R\rangle_{\x,t}, \\
& \chi(\x,t) = \langle s\rangle_{\x,t} = \int_0^\infty\int_{-\infty}^{\infty} s \mathcal{H}_{\x,t}(R,s) \, \d s \, \d R.
\eea

\section{Functional Jacobian and Turbulent Cascade Regularization}
\label{App_Jacobian_Regularization}
{The correction used in the Main Text can be obtained directly from the spatially regularized Stratonovich equation, without manipulating a singular noise-to-field inverse. Write the noise {as a $Q$-Wiener process \citep{lord2014introduction}, expanded} in modes as $\eta\,\d t=\sum_j e_j(\x)\,\d W_j(t)$, with
\be
\sum_j e_j(\x)e_j(\y)=\rho_{\eta_K}(\x-\y),\qquad
\sum_j e_j(\x)^2=\rho_{\eta_K}(0).
\ee
The covariance must be positive semidefinite; for example one may take $\rho_{\eta_K}=h_{\eta_K}*\check h_{\eta_K}$ with a smooth normalized spatial filter. For $B_j(\chi)=\sigma\chi e_j$, the Stratonovich-to-It\^o drift is
\be\label{Eq_trace}
\frac12\sum_j DB_j(\chi)[B_j(\chi)]
=\frac{\sigma^2}{2}\rho_{\eta_K}(0)\chi.
\ee
Define $V_c=\rho_{\eta_K}(0)^{-1}\propto\eta_K^d$. At fixed $\phi$, the equivalent It\^o bath equation is therefore
\bea
\d\chi&=\big[D_\chi\nabla^2\chi-\Gamma^*(\phi)\chi+J(\phi)\big]\d t
+\sigma\chi\,\d W_{\eta_K},\\
\Gamma^*&=\Gamma-\frac{\sigma^2}{2V_c}.
\eea
This is the spatial counterpart of Eq.~\eqref{Eq_Gas_Ito}. Its It\^o-discretized MSR action has the drift correction
\be
\Delta\Action_{\rm Jac}=-\int\d t\,\d\x\,\frac{\sigma^2}{2V_c}\bar\chi\chi.
\ee
A midpoint formulation describes the same process when its prescription-dependent measure terms are retained consistently \citep{Aron_al2010,aron2016dynamical,cugliandolo2017rules,cugliandolo2019covariant}. We use the equivalent It\^o action after this conversion, rather than adding the correction to an otherwise unchanged midpoint action.

To parameterize the unresolved mixing fluctuations, let $\zeta_K(t)$ be their local rate fluctuation and identify its white-time intensity with $2\int_0^\infty\langle\zeta_K(t)\zeta_K(0)\rangle\,\d t$. Writing this integral as $\sigma_S^2/\tau_K$ defines the dimensionless forcing-strength parameter $\sigma_S^2$ and gives the calibration
\be\label{Eq_sigma_scaling}
\sigma^2=2\sigma_S^2\frac{V_c}{\tau_K},\qquad
\Delta\Action_{\rm Jac}=-\int\d t\,\d\x\,\frac{\sigma_S^2}{\tau_K}\bar\chi\chi.
\ee
Thus the cancellation of $V_c$ holds under this calibration. It is not a universal independence from the spatial cutoff at fixed continuum noise amplitude. Nor does this relation identify $\sigma_S^2$ with the stationary supersaturation variance, which depends additionally on the bath relaxation, forcing and condensation coupling.}

\section{Turbulence Cascade and Clouds}
\label{Sec_Turb_Cascade}

Although the macroscopic thermodynamic shift derived above is structurally independent of the numerical value of $\eta_K$, the Kolmogorov scale remains essential for the physical validity of the Eulerian stochastic field theory. It sets the microscopic cutoff at which the turbulent cascade ceases to provide distinct fluid environments. The Kolmogorov length and time scales,
\be
\eta_K=\left(\frac{\nu^3}{\varepsilon}\right)^{1/4},
\qquad
\tau_K=\left(\frac{\nu}{\varepsilon}\right)^{1/2},
\ee
are fixed by the kinematic viscosity $\nu$ and the turbulent kinetic-energy dissipation rate $\varepsilon$. In typical stratocumulus and non-convective cloud boundary layers, the relevant dissipation rates place this cutoff in the millimeter range, $\eta_K\sim0.5\text{--}2.5{\rm mm}$.

This fluid cutoff remains several orders of magnitude larger than unactivated aerosols and twilight-cloud droplets, whose radii are typically in the range $r\sim1\text{--}10,\mu{\rm m}$. As a result, their Kolmogorov-scale Stokes number is very small, $\mathcal S_K\ll1$ \citep{santos2026brownian}. The droplets therefore behave, to leading order, as passive tracers of the local turbulent flow rather than as inertially detached particles. They do not appreciably cross streamlines by inertia, but are advected with the dissipative eddies that carry the local thermodynamic environment \citep{santos2026brownian}.

This tracer property is the physical bridge between the Lagrangian droplet picture and the Eulerian stochastic field theory used in the Main Text. The corresponding mathematical bridge is the local-bath approximation derived in {\it  Supplementary Note 5, Step 3}, where the droplet-sampled supersaturation average is written as a droplet-number-weighted Eulerian filter,
$\langle s\rangle_{\x,t}=\overline{n_d s}(\x,t)/\overline{n_d}(\x,t)$,
and then approximated by the resolved supersaturation bath $\chi(\x,t)\simeq\overline{s}(\x,t)$ when the sampling covariance $\tau_{s n_d}/\overline{n_d}$ is negligible. Since small-Stokes droplets remain locked to the local turbulent fluid motion at the Kolmogorov scale, this approximation has a direct physical meaning: the droplets sample supersaturation fluctuations that can be represented as localized Eulerian bath fluctuations in $\chi(\x,t)$. Thus the fluctuating supersaturation field integrated out in the MSR construction is not an artificial Eulerian device, but the coarse-grained representation of the thermodynamic forcing experienced by droplets as they are swept through the turbulent cascade. The scale separation $\eta_K\gg r$ therefore justifies treating droplet growth as a local stochastic response to a turbulent supersaturation bath, while the finite turnover time $\tau_K$ supplies the temporal regularization that survives in the dressed drift correction.

\section{Effective Action and Induced Macroscopic Noise: Derivation}\label{App_Effective_Action}

To perform the  Gaussian functional integration over the fast bath fluctuations referenced in Section \ref{Sec_Integrating_out_Fast}, we separate the quadratic action (Eq.~\eqref{Eq_S_quad}) into three distinct components based on their dependence on the fluctuation field $\delta\chi$ and the response field $\bar{\chi}$:
\be
\Action_{quad} = \Action_{MF} + \Action_{lin} + \Action_{var},
\ee
where:
\bea
\Action_{MF} &= \int \d t \int \d \x \left[ -2D \bar{\phi} {\chi_s} \right] \\
\Action_{lin} &= \int \d t \int \d \x \left[ \bar{\chi} \mathcal{L} \delta\chi - 2D \bar{\phi} \delta\chi \right] \\
\Action_{var} &= \int \d t \int \d \x \left[ -{\frac12} V(\phi) \bar{\chi}^2 \right].
\eea
Here, $\mathcal{L} = \partial_t - D_\chi \nabla^2 + \Gamma^*(\phi)$ is the linear differential operator of the renormalized bath, and $V(\phi) = {2}\sigma_S^2 V_c (\chi_0^*(\phi))^2/\tau_K$ represents the turbulently dressed noise variance.

We evaluate the nested path integral $\int \mathcal{D}[\bar{\chi}] \int \mathcal{D}[\delta\chi]$ in two stages. Crucially, because neither $\Action_{MF}$ nor $\Action_{var}$ depend on $\delta\chi$, they passively factor out of the inner integral. The inner path integral therefore solely isolates the terms linear in $\delta\chi$:
\be
\int \mathcal{D}[\delta\chi] \exp\left( -\Action_{lin} \right).
\ee
To evaluate this, we integrate the bilinear coupling ($\bar{\chi}\mathcal{L}\delta\chi$) by parts over space and time. This defines the formal adjoint operator, $\mathcal{L}^T = -\partial_t - D_\chi \nabla^2 + \Gamma^*(\phi)$, allowing us to factor out $\delta\chi$:
\be
\int \mathcal{D}[\delta\chi] \exp\left( \int \d t \d \x \Big[ \delta\chi \big( \mathcal{L}^T \bar{\chi} - 2D\bar{\phi} \big) \Big] \right).
\ee
Because the fluctuation field $\delta\chi$ appears purely linearly (lacking any Gaussian $\delta\chi^2$ components), this inner integral acts as a  functional Fourier transform. To intuitively grasp this mechanism, we can discretize the continuous spatiotemporal domain into a lattice with grid volumes $\Delta \x \Delta t$. The functional measure $\mathcal{D}[\delta\chi]$ becomes an infinite product of standard differentials, and the exponential of the integral sum translates into a product of individual local exponentials:
\be
\prod_{\x,t} \int \d(\delta\chi_{\x,t}) \exp\left( \delta\chi_{\x,t} \Big[ \mathcal{L}^T \bar{\chi} - 2D\bar{\phi} \Big]_{\x,t} \Delta \x \Delta t \right).
\ee
{Integrating the real fluctuation $\delta\chi$ against the imaginary response constraint} thus amounts to performing a 1D Fourier-style integral at every single point in space and time simultaneously. This operation yields a continuous product of local delta functions for every point, which rigorously defines the functional Dirac delta, $\delta(\mathcal{L}^T \bar{\chi} - 2D\bar{\phi})$. This strictly forces the deterministic constraint $\mathcal{L}^T \bar{\chi} = 2D\bar{\phi}$ everywhere across the macroscopic field.

To solve for the constrained response field $\bar{\chi}$, we invert the adjoint operator using the transposed Green's function:
\be
\bar{\chi} = (\mathcal{L}^T)^{-1} (2D\bar{\phi}) = \mathcal{G}^T (2D\bar{\phi}).
\ee
Mathematically, the transposed Green's function acts as the advanced propagator, defined by swapping the spatiotemporal arguments of the standard retarded Green's function:
\be
\mathcal{G}^T(\mathbf{y},s; \x,t) = \mathcal{G}(\x,t; \mathbf{y},s).
\ee
Applying this to our constraint yields the slaved response field explicitly in integral form:
\be
\bar{\chi}(\mathbf{y},s) = \int \d t \int \d \x \, \mathcal{G}(\mathbf{y},s; \x,t) \big(2D \bar{\phi}(\x,t)\big).
\ee

In the second stage of our nested integration, we substitute this strictly constrained response field back into the variance term $\Action_{var}$ that was waiting outside the inner integral. Because the slaved field $\bar{\chi}$ is itself an integral over space and time, squaring it requires introducing two independent sets of integration coordinates, $(\x,t)$ and $(\x',t')$:
\bea
\bar{\chi}(\mathbf{y},s)^2 = &\left( \int \d t \d \x \, \mathcal{G}(\mathbf{y},s; \x,t) 2D\bar{\phi}(\x,t) \right) \times  \\
&\left( \int \d t' \d \x' \, \mathcal{G}(\mathbf{y},s; \x',t') 2D\bar{\phi}(\x',t') \right).
\eea

Plugging this expanded square back into $\Action_{var}$ yields a massive nested integral. To extract the macroscopic physics, we apply Fubini's theorem to exchange the order of integration, pulling the macroscopic response fields $\bar{\phi}$ and their respective coordinates to the outside, while grouping the fast bath's internal variables ($s, \mathbf{y}$) on the inside:
\begin{widetext}
\be
\Action_{var} = -\frac{1}{2} \int \d t \d \x \int \d t' \d \x' \, \big(2D\bar{\phi}(\x,t)\big) \underbrace{\left[ \int \d s \d \mathbf{y} \, V\big(\phi(\mathbf{y},s)\big) \mathcal{G}(\mathbf{y},s; \x,t) \mathcal{G}(\mathbf{y},s; \x',t') \right]}_{\mathcal{C}(\x,t; \x',t')} \big(2D\bar{\phi}(\x',t')\big).
\ee
\end{widetext}
This rearrangement defines the non-local correlation function $\mathcal{C}$ presented in Eq.~\eqref{Eq_Corr_SpaceTime}. 

\section{Emergent Noise Variance: Physical Constraint}\label{Sec_Emergent_Noise_Constraint}

{The Gaussian residual bath used in the closure has a stationary covariance only when $\Gamma^*(\phi)>0$. A sufficient uniform condition over $\phi>0$ is
\be\label{Eq_cond_validity_noise}
\frac{\sigma_S^2}{\tau_K}<\lambda.
\ee
This condition concerns the retained Gaussian bath. It is not a sufficient second-moment condition for the original multiplicative system, nor a necessary bound at a fixed positive background $\phi$.

To see the distinction, freeze $\phi$ and neglect spatial transport. The original It\^o bath takes the form
\be
\d\chi=(J-\Gamma^*\chi)\d t+b_K\chi\d W_t,
\qquad b_K^2=\frac{\sigma^2}{V_c}=\frac{2\sigma_S^2}{\tau_K}.
\ee
Its moments satisfy
\bea
\frac{\d}{\d t}\mathbb E\chi&=J-\Gamma^*\mathbb E\chi,\\
\frac{\d}{\d t}\mathbb E\chi^2&=2J\mathbb E\chi-(2\Gamma^*-b_K^2)\mathbb E\chi^2.
\eea
For nonzero $J$, a finite stationary second moment requires $2\Gamma^*>b_K^2$, or $\Gamma(\phi)>2\sigma_S^2/\tau_K$. Its variance is
\be
\operatorname{Var}\chi=\frac{b_K^2(\chi_0^*)^2}{2\Gamma^*-b_K^2}.
\ee
The Gaussian residual approximation instead gives $b_K^2(\chi_0^*)^2/(2\Gamma^*)$. Their agreement requires $b_K^2/(2\Gamma^*)\ll1$ in this local test. Spatial transport modifies the full moment problem, but this calculation identifies why positive mean relaxation alone does not control the discarded multiplicative fluctuations.

With $\tau_m=\lambda^{-1}$, Eq.~\eqref{Eq_cond_validity_noise} reads
\be\label{Eq_cond_validity_noise2}
\sigma_S^2<\frac{\tau_K}{\tau_m}.
\ee
A separate requirement is that the slow-to-fast rate ratio remain small. As $\Gamma^*$ decreases, the bath memory grows and the adiabatic expansion eventually fails. The divergent noise intensity predicted by the Gaussian closure signals this breakdown; it is not, by itself, a demonstration of runaway condensation or a universal upper bound on observed supersaturation variance.}

}

\end{appendices}

\bibliographystyle{ieeetr}
\bibliography{Polydisperse_paper_Stat_Phys}
\end{document}